\documentclass[11pt, a4paper]{article}
\usepackage{jheparxiv}
\usepackage[utf8]{inputenc}
\usepackage{amsmath}
\usepackage{amsfonts}
\usepackage{amssymb}
\usepackage{latexsym}
\usepackage{mathrsfs}
\usepackage{braket}		
\usepackage{graphicx}
\usepackage{color}
\usepackage{xcolor}
\usepackage{slashed}
\usepackage{twistor}
\usepackage[all]{xy}
\usepackage{tikz-cd}
\usepackage{mathtools}

\newcommand{\sa}{\mathsf{a}}

\newcommand{\sh}{\mathsf{h}}

\newcommand{\ba}{\mathbf{a}}

\newcommand{\m}{\mathrm{m}}

\newcommand{\dt}[1]{\dot{\tilde{#1}}}

\renewcommand{\d}{\mathrm{d}}

\newcommand{\qed}{\hfill \ensuremath{\Box}}

\newcommand{\eps}{\epsilon}
\newcommand{\veps}{\varepsilon}

\newcommand{\msf}[1]{\mathsf{#1}}
\newcommand{\F}{\mathcal{F}}
\newcommand{\scF}{\mathscr{F}}

\newcommand{\sF}{\mathsf{F}}

\newcommand{\nbar}{\bar{\nabla}}
\newcommand{\bigma}{\boldsymbol{\sigma}}
\newcommand{\scG}{\mathscr{G}}
\newcommand{\sX}{\mathsf{X}}

\title{Graviton scattering in self-dual radiative space-times}

\author[a]{Tim Adamo,}
\author[b]{Lionel Mason}
\author[c]{\& Atul Sharma}

\affiliation[a]{School of Mathematics and Maxwell Institute for Mathematical Sciences \\
        University of Edinburgh, EH9 3FD, United Kingdom}
        
\affiliation[b]{The Mathematical Institute \\
		University of Oxford, OX2 6GG, United Kingdom}
	
\affiliation[c]{Center for the Fundamental Laws of Nature and Black Hole Initiative \\
        Harvard University, Cambridge, MA, 02138, United States of America}	

\emailAdd{t.adamo@ed.ac.uk}
\emailAdd{lmason@maths.ox.ac.uk}
\emailAdd{atulsharma@fas.harvard.edu}

\abstract{The construction of amplitudes on curved space-times is a major challenge, particularly when the background has non-constant curvature. We give formulae for all tree-level graviton scattering amplitudes in curved self-dual radiative space-times; these are chiral, source-free, asymptotically flat spaces determined by free characteristic data at null infinity. Such space-times admit an elegant description in terms of twistor theory, which provides the powerful tools required to exploit their underlying integrability. The tree-level S-matrix is written in terms of an integral over the moduli space of holomorphic maps from the Riemann sphere to twistor space, with the degree of the map corresponding to the helicity configuration of the external gravitons. For the MHV sector, we derive the amplitude directly from the Einstein-Hilbert action of general relativity, while other helicity configurations arise from a natural family of generating functionals and pass several consistency checks. The amplitudes in self-dual radiative space-times exhibit many novel features that are absent in Minkowski space, including tail effects. There remain residual integrals due to the functional degrees of freedom in the background space-time, but our formulae have many fewer such integrals than would be expected from space-time perturbation theory. In highly symmetric special cases, such as self-dual plane waves, the number of residual integrals can be further reduced, resulting in much simpler expressions for the scattering amplitudes.}

\begin{document}

\maketitle

\section{Introduction}
\label{sec:intro}

Despite the perturbative non-linearity the Einstein-Hilbert action~\cite{DeWitt:1967yk,DeWitt:1967ub,DeWitt:1967uc,tHooft:1974toh}, the scattering amplitudes of general relativity (GR) and supergravity (SUGRA) in Minkowski space-time are surprisingly simple. At tree-level, there are now remarkably compact formulae for the \emph{entire} tree-level S-matrix of GR in flat space that are totally divorced from the usual perspective of space-time Feynman rules~\cite{Cachazo:2012kg,Cachazo:2013hca}. Indeed, a variety of modern `on-shell' techniques (e.g., double copy, twistor string theory and unitarity) have enabled higher-loop calculations in gravity that would be impossible with the standard perturbative quantum field theory (QFT) toolkit (see~\cite{Cheung:2017pzi,Bern:2019prr,White:2019ggo} for some recent reviews).  

Given these remarkable achievements, it is natural to ask if analogous results can be found for gravitational scattering amplitudes in \emph{curved} space-times; that is, for graviton scattering in a background metric which solves the Einstein equations. This background is treated classically but non-perturbatively as an exact solution, neglecting back-reaction, so that one is looking at graviton scattering, including mutual non-linearities, in the `strong' gravitational background field. Of course, in a generic space-time the S-matrix will not exist due to lack of unitarity or particle creation effects from event horizons or singularities (cf., \cite{DeWitt:1975ys,Birrell:1982ix}) or because the space-time is not asymptotically flat. But one can always consider scattering at large distances (i.e., probing weakly curved regions far from horizons) or the natural analogues of scattering amplitudes in asymptotically (anti-)de Sitter space-times (i.e., boundary correlation functions~\cite{Witten:1998qj} or in-in correlators~\cite{Witten:2001kn,Strominger:2001pn}).

From the perspective of space-time perturbation theory, such amplitudes (or their analogues) are computed using the background-field formalism applied to the Einstein-Hilbert action (cf., \cite{DeWitt:1967ub,Abbott:1981ke}). There are two broad reasons for being interested in such observables. The first is practical: they capture non-linear strong field effects which are invisible in Minkowski space and physically relevant (e.g., time-delay, the memory effect, tails, pair production). Their importance ranges from gravitational non-gaussianities in early universe cosmology~\cite{Mukhanov:1990me,Maldacena:2002vr} to encoding infinite resummations of small momentum-transfer scattering in flat space~\cite{tHooft:1987vrq,Kabat:1992tb,Adamo:2021rfq}. The second reason is conceptual: scattering amplitudes in curved backgrounds are a theoretical playground where perturbative and non-perturbative physics meet. They test the robustness of on-shell techniques developed in Minkowski space, many of which simply break down in the presence of space-time curvature. For instance, in a generic space-time, momentum space Feynman rules and unitarity techniques fail, as Fourier transforms are obstructed by the background curvature and even tree-level scattering amplitudes are not rational functions of the kinematic data.

This tension between the importance and difficulty of computing scattering amplitudes in curved space-times means that, although they are often studied, nothing close to the all-multiplicity formulae available at tree-level in Minkowski space exists in even the simplest cases. For instance, state-of-the-art for tree-level graviton scattering in plane wave space-times is 3-points~\cite{Adamo:2017nia}, while in AdS backgrounds -- where CFT methods in the dual boundary theory can be used -- it is 4-points (cf., \cite{DHoker:1999kzh,Raju:2012zs,Rastelli:2016nze,Caron-Huot:2018kta,Alday:2020dtb}), with the exception of a particular $R$-symmetry sector of supergravity in AdS$_5\times S^5$, where it is 5-points~\cite{Goncalves:2019znr}\footnote{There are some results in AdS which are not generic graviton boundary correlators but are nevertheless all-multiplicity. These include `maximal U$(1)$-violating' correlators of components of the graviton multiplet in type IIB superstring theory on AdS$_5$~\cite{Green:2020eyj,Dorigoni:2021rdo}, and integral kernels for the bulk dynamics of gravity in AdS$_4$~\cite{Adamo:2013tja,Adamo:2015ina,Adamo:2021bej}. The latter do not explicitly encode boundary contributions to the AdS amplitudes and their relationship to position or momentum space expressions is obscure.}. The gap between these low-point examples and the entire tree-level graviton S-matrix is profound and begs the question: is graviton scattering in curved space-time so complicated that we simply cannot hope to recover all-multiplicity results?

\medskip

In this paper, we find all-multiplicity expressions for tree-level graviton scattering in a large class of four-dimensional, (almost everywhere) asymptotically flat \emph{complex} vacuum space-times. These space-times have a self-dual Weyl tensor determined by characteristic data which is a free function of three variables; as such, we refer to them as \emph{self-dual radiative space-times}. The functional degrees of freedom in the curvature of such space-times means that there is no hope of obtaining high-multiplicity scattering amplitudes using standard space-time perturbation theory, but self-duality allows us to reformulate the perturbative expansion using twistor theory, which manifests the underlying integrability of the self-dual sector. The integrability means that the self-dual sector itself has trivial scattering.  This can be seen perturbatively from the vanishing of the all positive helicity and one negative helicity tree amplitudes.  It can also be seen directly from the twistor theory for the space-times, which naturally identifies past and future null infinity together with the characteristic data there.  This ensures that linearised solutions used to build the perturbative S-matrix on a background can be taken to be asymptotically momentum eigenstates at both $\scri^\pm$ and that we have crossing symmetry. This is of course a very special feature of these integrable backgrounds.

Using twistor theory, we derive an all-multiplicity expression for the `maximal helicity violating (MHV)' tree-level graviton scattering amplitude on \emph{any} self-dual radiative space-time; this is the scattering configuration with two negative helicity and arbitrarily many positive helicity external gravitons. This confirms an earlier conjecture for these amplitudes in the special case of self-dual plane wave space-times~\cite{Adamo:2020syc}. Furthermore, we also provide a conjecture for the tree-level graviton S-matrix in any helicity configuration on any self-dual radiative space-time. While conjectural away from the MHV sector, this general formula has its origin in a natural family of generating functionals and passes several consistency tests.

\medskip

Our all-multiplicity formulae for scattering amplitudes in curved backgrounds generalize the well-known flat background formulae on Minkowski space in the forms due to Hodges~\cite{Hodges:2012ym} at MHV and Cachazo-Skinner~\cite{Cachazo:2012pz,Cachazo:2012kg} at higher MHV degree, but have many new terms, reflecting new physical features.  Two of the most prominent of these are residual integrals arising from the background curvature (which would reduce to momentum conserving delta functions in flat space) and tail terms coming from the interaction between gravitons and the background; these reflect the violation of Huygens' principle~\cite{Friedlander:2010eqa,Noonan:1989,Wunsch:1990,Harte:2013dba}. The precise formulae are given by equation \eqref{MHV2} at MHV and \eqref{NkMHV} beyond, but schematically, we show that a tree-level, $n$-point N$^{d-1}$MHV amplitude (i.e., with $d+1$ negative helicity and $n-d-1$ positive helicity external gravitons) takes the form:
\begin{equation}\label{introForm}
\cM_{n,d}\:= \sum_{\substack{\mathrm{\#\: of\: tails,\: }t\\ \mathrm{mult.\: of\: tails,\: }p}} \int\d\mu_{d,n}\:\mathcal{I}_{n,t}\,\prod_{i=1}^{n}h_{i}\,\prod_{\m=1}^{t}\partial^{p-2}N_{\m}\,.
\end{equation}
Here, there are sums over the number $t$ of tail contributions to the amplitude (running from zero to an upper bound dictated by $n$ and $d$) and the `multiplicity' of each of these tails -- roughly how many times it is involved in interactions with the external gravitons or other tails. The amplitude is then computed in terms of an integral over the moduli space of holomorphic, rational maps from the Riemann sphere $\P^1$ to twistor space $\PT$ of degree $d$ with $n$ marked points; this moduli space comes with a natural holomorphic measure $\d\mu_{d,n}$. Kinematic data is packaged into an `integrand' $\mathcal{I}_{n,t}$, in the form of a reduced determinant that also depends on the background geometry.  The $h_i$ are twistor wavefunctions for the external gravitons that can be taken to be asymptotic momentum eigenstates. Finally, there are explicit insertions of (derivatives of) the news function $N$ of the self-dual radiative space-time, one for each tail contribution.

Our main tool for obtaining these formulae is the chiral twistor sigma model of \cite{Adamo:2021bej}.  Our earlier formulae \cite{Adamo:2020yzi} on self-dual plane-wave backgrounds were based on curved background extensions of Skinner's twistor string for Einstein gravity~\cite{Skinner:2013xp}, but this had awkward extra gauge dependences that were hard to eliminate.  The chiral twistor sigma model was derived in order to find a more basic description that avoids this problem, and is more natural to express on a nonlinear background.

There are already some important features to be pointed out even at the schematic level of \eqref{introForm}. In Minkowski space, the sums over tails and insertions of the news function are absent, and the integrand and wavefunctions are such that the moduli integral is totally localised by delta functions, leaving a rational function of the kinematic data with four momentum conserving delta functions. In a self-dual radiative background this is not the case: there are $4d$ integrals which are not localized by delta functions, and no momentum conserving delta functions. When there are residual symmetries this can change; for instance, on a self-dual plane wave there are three momentum conserving delta functions and only $2d-1$ residual integrals.  

It is interesting to compare and contrast these results with analogous formulae for gluon scattering amplitudes of pure Yang-Mills theory in strong, self-dual radiative \emph{gauge field} backgrounds (on Minkowski space)~\cite{Adamo:2020yzi}. There, we also used twistor theory to find explicit formulae for the complete tree-level S-matrix (which were proved in the MHV sector), and these also have similar moduli counting and residual integrals due to the self-dual radiative background. However, non-linear effects such as tails are much less explicit there, being encoded implicitly in holomorphic frames associated with the background gauge field.

\medskip

The paper is structured as follows. Section~\ref{sec:twistor} provides a definition of the self-dual radiative class of space-times and their description in terms of twistor theory. We also show how gravitational perturbations to these space-times are captured by the twistor framework. The constructions are illustrated at each stage by two classes of explicit solutions, the self-dual plane waves, and the analytic continuation to split signature (or the complex) of the Gibbons-Hawking solutions with just one symmetry (i.e., depending on free data depending on two variables).  In Section~\ref{sec:mhv} we derive the tree-level MHV graviton scattering amplitudes on self-dual radiative space-times directly from general relativity by lifting the space-time generating functional for these amplitudes to twistor space. Section~\ref{sec:nmhv} generalizes the MHV formula to all helicity sectors, providing a conjecture for the full tree-level S-matrix of gravity in any self-dual radiative space-time. Section~\ref{sec:discuss} concludes, while Appendix~\ref{app:prl} shows the equivalence between the MHV formula and our earlier (less compact) expression in the literature~\cite{Adamo:2020syc}, and Appendix~\ref{app:grav} includes some explicit tests of our formulae against space-time perturbation theory at low multiplicity.


\section{Twistor theory of self-dual radiative space-times}
\label{sec:twistor}

In this section we introduce the curved space-times on which we will construct scattering amplitudes. For scattering from past to future null infinity, $\scri^-\rightarrow \scri^+$, one usually requires Lorentz signature; however, self-dual fields are necessarily complex in Lorentzian signature. Thus, we consider complex self-dual space-times holomorphic on regions of $\C^4$  that are small perturbations of flat space defined on some neighbourhood of a real $\R^4$ contour, but on this contour the metric will necessarily be complex. They will be determined by their self-dual asymptotic data in such a way that they admit both a $\scri^+$ and $\scri^-$ that compactify the $\R^4$ contour.
 
After introducing such backgrounds, we explain in \S\ref{sec:radtwistor} how these are constructed via twistor theory from twistor data that is simply related to the asymptotic data at null infinity. This provides the necessary technology not only to derive the underlying self-dual background metrics, but also to obtain formulae for perturbations which are asymptotic momentum eigenstates, as shown in \S\ref{sec:perts}. These will provide the linear fields (and momenta) from which the S-matrix will be built. Throughout, we introduce the two families of examples in which we can express our twistor formulae particularly explicitly: the self-dual plane waves and analytically continued Gibbons-Hawking spaces. Both of these examples admit symmetries which lead to violations of asymptotic flatness, but only in isolated directions which still suffices for our purposes. In \S\ref{sec:twistor-exs} the twistor theory is adapted to these examples where the results can be expressed in terms of straightforward integral formulae.
Much of this is a review of standard material but adapted to the radiative space-times in which we work.

A four-dimensional space-time will be said to be purely radiative if it is a solution of the vacuum Einstein equations, asymptotically flat with a smooth topologically $\R\times S^2$ past and future null infinity ($\scri^{-}$ or $\scri^+$) and completely determined by characteristic free data on $\scri^\pm$ obeying suitable regularity assumptions~\cite{Friedrich:1986rb,Friedrich:2013jua,Chrusciel:2013xha}. Thus $\scri^{\pm}$ are Cauchy surfaces for radiative space-times.\footnote{Such space-times are also known as \emph{asymptotically simple}~\cite{Penrose:1986uia}.  A general asymptotically flat space-time will only possess one of $\scri^{\pm}$, depending on whether final or initial characteristic data is specified~\cite{Friedrich:1981at,Friedrich:1981wx} because singularities are expected to form from general data (cf., \cite{Kehrberger:2021uvf}). However, with sufficiently small data, there is a reasonable class of space-times with both a complete past and future null infinity, $\scri=\scri^{-}\cup\scri^+$ \cite{Christodoulou:1993}.} For such space-times the natural, diffeomorphism-invariant semi-classical observable for generally nonlinear massless fields propagating in the space-time is the tree-level S-matrix, given by the scattering map from the characteristic data at $\scri^-$ to that at $\scri^+$. 

This characteristic data is composed of two functions at $\scri^{\pm}$, encoded in the leading piece of the Weyl tensor in the asymptotic peeling expansion, which correspond to the self-dual and anti-self-dual degrees of freedom in the radiative gravitational field. By complexifying the data and $\scri^\pm$, it is possible to set one of the characteristic functions to zero while keeping the other non-zero, resulting in a \emph{self-dual} radiative solution of the vacuum Einstein equations. The self-dual sector is classically integrable, and any self-dual space admits a construction in terms of analytic data on twistor space~\cite{Mason:1991rf}; in the case of self-dual radiative space-times, the twistor construction can be expressed directly in terms of the non-vanishing characteristic data at $\scri^{\pm}$, originally formulated in terms of Newman's \emph{good cuts} of $\scri$~\cite{Newman:1976gc}.  Integrability makes the scattering map for the space-time itself trivial: $\scri^\pm$ are identified together canonically along with the asymptotic data.  Thus, scattering on such backgrounds is crossing symmetric.


\subsection{Radiative metrics and self-duality}
\label{sec:sdrad}

We first introduce a homogeneous coordinatization of $\scri$ that will allow us to work with manifest Lorentz invariance.  It will also expedite the geometric correspondence between $\scri$ and twistor space following a framework~\cite{Eastwood:1982} that connects Newman's $\cH$-space~\cite{Newman:1976gc} with Penrose's non-linear graviton construction~\cite{Penrose:1976js}. 

\paragraph{Homogeneous description of $\scri^+$:} An asymptotically flat space-time can be described in terms of retarded Bondi coordinates $(u,r,\zeta,\bar{\zeta})$, where $u$ is retarded Bondi time, $r$ is a parameter along the outgoing null geodesics of constant-$u$ hypersurfaces and $(\zeta,\bar{\zeta})$ are stereographic coordinates on the celestial sphere. These coordinates are not global on the celestial sphere, so do not manifest Lorentz-invariance; this is remedied by instead working with homogeneous Bondi coordinates defined modulo the rescalings~\cite{Sparling:1990,Eastwood:1982,Adamo:2014yya,Geyer:2014lca,Adamo:2015fwa,Adamo:2021dfg}:
\be\label{homcoords}
(u,r,\lambda_{\alpha},\bar{\lambda}_{\dot\alpha})\sim(|b|^2u,\,|b|^{-2}r,\,b\lambda_{\alpha},\,\bar{b}\bar{\lambda}_{\dot\alpha})\,,\quad \forall\; b\in\C^*\,.
\ee
Here $\alpha,\dot\alpha$ are spinor indices of $\mathfrak{sl}(2,\C)$, and the affine coordinates on the celestial sphere are simply recovered by going to a stereographic coordinate patch:
\be
\lambda_\al = \frac{2^{1/4}}{\sqrt{1+|\zeta|^2}}\,(1,\zeta)\,.
\ee
Choosing a time-like unit vector $t^{\alpha\dot\alpha}$ so that
\be\label{affineBC}
x^{\alpha\dot\alpha}=\frac{u\,t^{\alpha\dot\alpha}}{\la\lambda\,\hat{\lambda}\ra}+r\,\lambda^{\alpha}\,\bar{\lambda}^{\dot\alpha}\,, \qquad \hat{\lambda}_{\alpha}:=t_{\alpha\dot\alpha}\,\bar{\lambda}^{\dot\alpha}\,,
\ee
packages the (non-homogeneous) Bondi coordinates in spinor form $x^{\alpha\dot\alpha}$; the choice of $t^{\alpha\dot\alpha}$ is equivalent to a choice of conformal factor $\la\lambda\,\hat{\lambda}\ra^{-1}$ on the unit sphere (see also section 2.1 of \cite{Adamo:2020yzi} for further discussion). We use standard spinor helicity notation for SL$(2,\C)$-invariant spinor contractions: $\la a\,b\ra:=\epsilon^{\alpha\beta}\,a_{\beta}\,b_{\alpha}$, $[\tilde{a}\,\tilde{b}]:=\epsilon^{\dot\alpha\dot\beta}\,\tilde{a}_{\dot\beta}\,\tilde{b}_{\dot\alpha}$, with $\epsilon^{\alpha\beta}$, $\epsilon^{\dot\alpha\dot\beta}$ the 2-dimensional Levi-Civita symbols.

In this homogeneous formalism, spin- and conformal-weight are formulated as homogeneity under the projective rescalings \eqref{homcoords} of the coordinates. Define the  line bundles $\cO(p,q)$ over space-time (or subsets thereof), whose sections transform as 
\begin{equation}
f(|b|^2u,\,|b|^{-2}r,\,b\lambda,\bar{b}\bar{\lambda})=b^{p}\bar{b}^{q}\,f(u,r,\lambda,\bar{\lambda})\, .
\end{equation}
The usual notions of spin weight ($s$) and conformal weight ($w$) then correspond to $s = \frac{p-q}{2}$ and $w = \frac{p+q}{2}$. Thus, spin/conformally-weighted functions correspond to functions valued in the line bundles $\cO(p,q)$ in this framework.

\medskip

An asymptotically flat metric in Bondi-Sachs gauge admits a large-$r$ expansion in these coordinates~\cite{Sachs:1961zz,Bondi:1962px,Sachs:1962wk,Exton:1969im,Penrose:1986uia,Madler:2016xju}:
\begin{multline}\label{AFgr1}
\d s^2=2\,\d u\,\d r-r^2\,\left( \D\lambda\,\D\bar{\lambda} -\frac{\bigma^0}{r}\,\D\lambda^2 - \frac{\bar{\bigma}^0}{r}\,\D \bar{\lambda}^2\right)
+  
\left(\frac{1}{\la\lambda\,\hat{\lambda}\ra^2}-2\,\frac{m_B}{r}\right)\d u^2  \\
 -\d u\left(\eth\bar{\bigma}^{0}\,\D\bar{\lambda}+\bar{\eth}\bigma^0\,\D\lambda\right) + O(r^{-1})\,,
\end{multline}
where $\D\lambda:=\la\lambda\,\d\lambda\ra$, $\D\bar{\lambda}:=[\bar{\lambda}\,\d\bar{\lambda}]$, and $\eth,\bar{\eth}$ are the spin-weighted covariant derivatives on the sphere~\cite{Goldberg:1966uu,Eastwood:1982}. The quantities $m_B$, $\bigma^0$, $\bar{\bigma}^0$ are spin- and conformally-weighted functions of $(u,\lambda,\bar{\lambda})$. In particular, for the line element to be homogeneous of degree zero, $m_B$ must be valued in $\cO(-3,-3)$, so is a spin-weight $0$, conformal-weight $-3$ quantity; this is the Bondi mass aspect, which generalizes the Schwarzschild mass parameter for generic asymptotically flat space-times. Similarly, $\bigma^0$ and $\bar{\bigma}^0$ are valued in $\cO(-3,1)$ and $\cO(1,-3)$, so both have conformal weight $-1$ and spin-weights $-2$ and $2$, respectively. These are the asymptotic shear optical scalars of the null hypersurfaces of constant $u$~\cite{Jordan:1961,Sachs:1961zz,Newman:1961qr}. The (retarded) time evolution of the mass aspect $m_{B}$ is controlled by $\bigma^0$ (and its complex conjugate) through the asymptotic Bianchi identities (cf., \cite{Newman:1961qr,Adamo:2009vu}).

Performing a conformal rescaling with conformal factor $R=r^{-1}$ gives 
\begin{equation}\label{AFgr1-proj}
\d \hat{s}^2:=R^2 \d s^2=-2\,\d u\,\d R-\D\lambda\,\D\bar\lambda  +
R\,\bigma^0\,\D\lambda^2+R\,\bar{\bigma}^0\,\D \bar{\lambda}^2 +O(R^2) \,.
\end{equation}
The future null conformal boundary $\scri^+$ is defined by $R=0$, where $\scri^+\cong\R\times S^2$ with degenerate conformal metric (cf., \cite{
Penrose:1962ij,Penrose:1964ge,Penrose:1965am,Penrose:1986uia,Frauendiener:2000mk})
\be\label{scrimet}
\d \hat s^{2}_{\scri^+}=0\times \d u^2+\D\lambda\,\D\bar{\lambda}\,.
\ee
The $(u,\lambda_{\alpha},\bar{\lambda}_{\dot\alpha})$ provide homogeneous coordinates on $\scri^+$ viewed as the total space of the line bundle $\cO_{\R}(1,1)\rightarrow\P^1$, whose sections are real-valued functions that obey $f(b\lambda, \bar{b}\bar{\lambda})=|b|^2\,f(\lambda,\bar{\lambda})$. The BMS asymptotic symmetry group~\cite{Bondi:1962px,Sachs:1962zza} acts via M\"obius transformations on the $\P^1$ base and supertranslations $u\to u+f$ for any $f$ in $\cO_{\R}(1,1)$, so there is no preferred choice of origin for the retarded time coordinate\footnote{This affine line bundle description of $\scri$ provides a formulation of the Carrollian geometry of~\cite{Figueroa-OFarrill:2021sxz,Herfray:2021qmp,Donnay:2022aba}.}. 

\medskip


\paragraph{Self-dual radiative metrics:}\label{sec:SDRadMets} The 2-spinor formalism decomposes the Weyl tensor $C_{abcd}$ on $M$ into self-dual (SD) and anti-self-dual (ASD) parts as~\cite{Penrose:1984uia} 
\be\label{Cdecom}
C_{\alpha\dot\alpha\beta\dot\beta\gamma\dot\gamma\delta\dot\delta} = \tilde\Psi_{\dot\alpha\dot\beta\dot\gamma\dot\delta} \,\eps_{\alpha\beta}\,\eps_{\gamma\delta}+\Psi_{\alpha\beta\gamma\delta}\, \eps_{\dot\alpha\dot\beta}\,\eps_{\dot\gamma\dot\delta}\,,
\ee
with SD part $C^{+}_{abcd}$ corresponding to $\widetilde\Psi_{\dot\alpha\dot\beta \dot\gamma \dot\delta}$ and ASD part $C^{-}_{abcd}$ to $\Psi_{\alpha\beta\gamma\delta}$. Both $\Psi_{\alpha\beta\gamma\delta}$ and $\tilde{\Psi}_{\dot\alpha\dot\beta\dot\gamma\dot\delta}$ are totally symmetric. Near $\scri^+$, they exhibit the fall-offs
\be
\Psi_{\alpha\beta\gamma\delta} =
o_\al\, o_\beta \, o_\gamma \, o_\delta\, \frac{\psi_4}{r} + O(r^{-2})\,,\qquad\tilde\Psi_{\dot\alpha\dot\beta\dot\gamma\dot\delta} = \tilde 
o_{\dal}\,\tilde o 
_{\dot\beta}\,\tilde o 
_{\dot\gamma}\,\tilde o
_{\dot\delta}\, \frac{\bar\psi_4}{r} + O(r^{-2})\,,
\ee
where $o_\al \tilde o_{\dal}=\nabla_{\alpha\dot\alpha} u$,  the outgoing null vector along constant  $u$.  
The coefficients $\psi_{4},\bar\psi_{4}$ are functions of $(u,\lambda,\bar{\lambda})$ and provide the \emph{radiation data}
given explicitly by \cite{Newman:1961qr,Newman:1981,Penrose:1986uia}:
\be\label{AFgr2}
\psi_4 =-\partial^{2}_{u}{\bigma}^0\,, \qquad \bar\psi_4=-\partial^{2}_{u}\bar \bigma^0\,,
\ee
with $\psi_4$ valued in $\cO(-5,-1)$ and $\bar{\psi}_4$ valued in $\cO(-1,-5)$. Thus, the complex shear $\bigma^0$ serves as a key ingredient in any definition of the radiative phase space of asymptotically flat space-times~\cite{Sachs:1962zzb,Bondi:1962px,Newman:1962cia,Ashtekar:1981bq,Ashtekar:1981hw,Ashtekar:1987tt,Herfray:2020rvq,Herfray:2021xyp}. For instance, the \emph{news function}
\be\label{AFgrnews}
N(u,\lambda ,\bar{\lambda })\vcentcolon=-\partial_u\bar\bigma^0 (u,\lambda ,\bar{\lambda })\,,
\ee
which takes values in $\cO(0,-4)$ on $\scri^+$, encodes the energy-momentum radiated through any interval of retarded time on $\scri^+$ through the Bondi mass-loss theorem~\cite{Bondi:1962px,Newman:1962cia}. 

To understand why $\bigma^0$ defines the radiation data through $\psi_4$, it is instructive to first consider the case of a linear perturbation of complexified Minkowski space, $\M\equiv\C^4$. The whole of $\M$ is easily reconstructed from $\scri^+$ in terms of light cone cuts: a point $x\in\M$ is described at $\scri^+$ by the cut
\be\label{lcc}
u = x^{\al\dal}\lambda_\al\bar\lambda_{\dal}\,,
\ee
which is the spherical cross section corresponding to the intersection between $\scri^+$ and the light cone with apex at $x$. This means that with suitable globality assumptions, it is easy to see that a linear gravitational perturbation on $\M$ is determined by the radiation data at $\scri^+$ by means of the Kirchoff-d'Adh\'emar integral formulae adapted to null infinity~\cite{Penrose:1980yx,Penrose:1984uia,Penrose:1986uia}. 

In particular, to linear order, the ASD Weyl spinor is given by the spin-2 field
\begin{equation}\label{K-dA}
\psi_{\alpha\beta\gamma\delta}(x)= \int_{\P^1}\lambda_{\alpha}\,\lambda_{\beta}\,\lambda_{\gamma}\,\lambda_{\delta}\,\D\lambda\wedge\D\bar{\lambda}\,\left. \frac{\p\psi_{4}}{\p u}\right|_{u=\la\lambda|x|\bar\lambda]}\,.
\end{equation}
By differentiating under the integral sign, it is easily checked that this solves the linearized field equation $\p^{\al\dal}\psi_{\al\beta\gamma\delta}=0$ on flat space. In this formula, the integral is taken over the spherical cut of $\scri^+$ corresponding to $x\in\M$ defined by \eqref{lcc}. Conversely, the field defined by \eqref{K-dA} is shown to give rise to the radiation data $\psi_4$ when the field is defined and differentiable over $\scri^+$ including the vertex $i^+$ (i.e., future time-like infinity)~\cite{Penrose:1984uia}.

For a generic asymptotically flat solution to the Einstein equations, the outgoing radiation field $\psi_4$ does not specify the full non-linear solution as one could include some constants of integration: although the evolution of the mass aspect $m_B$ in $u$ is determined by the asymptotic Bianchi identities at $\scri^+$, there is a constant of integration as $u\rightarrow\infty$ which encodes source contributions to the metric, such as a Schwarzschild mass or higher multipole moments. However, when such source terms are absent, $\psi_4$ -- and thus $\bigma^0$ -- does provide good Cauchy data for the Einstein equations with $\scri^+$ as the characteristic (final) data surface~\cite{Friedrich:1986rb,Friedrich:2013jua,Chrusciel:2013xha}. 

Such radiative space-times are therefore singled out by:
\begin{defn}
A \emph{radiative} four-dimensional space-time is almost everywhere asymptotically flat and characterized by $\bigma^0$, in particular,  $m_B\rightarrow 0$ as $u\rightarrow +\infty$.
\end{defn}
By `almost everywhere asymptotically flat,' we include also space-times that are singular on isolated generators of $\scri^+$ but otherwise admit a normal conformal compactification. This allows important examples such as (multi-)plane wave space-times to be included this radiative class.

\medskip

In Lorentzian-real signature the SD and ASD parts of the Weyl tensor are related by complex conjugation, but for complexified metrics (or in split and Euclidean-real signature) they are independent. Thus, for a complexified radiative space-time, $\bar{\bigma}\to\tilde{\bigma}^0$ which is independent free data no longer related to $\bigma^0$ by complex conjugation. The complexified data is viewed as living on a (partial) complexification of future null infinity, $\scri_{\C}^+$, for which Bondi time $u$ is complex.

A complex vacuum space-time is self-dual (SD) if $\Psi_{\alpha\beta\gamma\delta}=0$ and anti-self-dual (ASD) if $\tilde\Psi_{\dot\alpha\dot\beta\dot\gamma\dot\delta}=0$, and we can now define 
\begin{defn}
A \emph{self-dual (SD) radiative space-time} is a complex radiative space-time\footnote{For precise definitions of asymptotic flatness for complex radiative spacetimes, see~\cite{Ludvigsen:1981}.} determined by its radiative data $\tilde{\bigma}^0$, with $\bigma^0=0$. 
\end{defn}
These space-times were named \emph{Heavens} or $\mathcal{H}$-spaces~\cite{Newman:1976gc,Hansen:1978jz,Ko:1981}; they are tautologically holographic, with the `bulk' metric determined entirely by the free data $\tilde{\bigma}^0$ on $\scri^+_{\C}$. 

\subsubsection*{Examples}

It is useful to have some explicit examples of self-dual radiative space-times, and here we consider two particular families of such metrics. In what follows, these examples will provide concrete realizations of our more general constructions. 
 
\paragraph{\textit{Self-dual plane wave metrics:}} Several explicit SD radiative space-times have been constructed in the literature (e.g., \cite{Tod:1979tt,Sparling:1981nk}), but a key example is provided by self-dual plane wave (SDPW) metrics. These are SD metrics with a five-dimensional Heisenberg algebra of Killing vectors, with centre given by a covariantly constant null vector $n$. A SDPW metric in Einstein-Rosen coordinates\footnote{While Einstein-Rosen coordinates are not generally global due to null geodesic focusing in plane wave space-times~\cite{Penrose:1965rx}, these SD examples do not suffer from this issue.} takes the form
\be\label{sdpwgr}
\d s^2 = 2\left(\d x^+\,\d x^- - \d z\,\d\tilde z + f(x^-)\,\d\tilde z^2\right)\,, 
\ee
with $f(x^-)$ a free function of the lightfront coordinate $x^-$ determining the wave profile. The null vector $n^a\p_a=\p_{+}$ is easily seen to be a covariantly constant Killing vector. 

At this stage, we introduce a spinor dyad $o_\al = \tilde o_{\dal} = (1,0)$, $\iota_\al=\tilde\iota_{\dal} = (0,1)$, normalized so that $\la\iota\,o\ra=1=[\tilde{\iota}\,\tilde{o}]$. Define the coordinates $x^{\al\dal}$ to be
\be
x^{\al\dal} = \begin{pmatrix}x^+&\tilde z\\z& x^-\end{pmatrix}
\ee
so that the spinor components of $n^a$ become $n^{\al\dal} = \iota^\al\,\tilde\iota^{\dal}$. In terms of these, introduce a tetrad of 1-forms (i.e., a frame of the cotangent bundle):
\be\label{sdpwtetrad}
e^{\alpha\dot\alpha}(x) = \d x^{\alpha\dot\alpha} + f(x^-)\,o^\alpha\,\tilde\iota^{\dot\alpha}\,\d \tilde z\, , \qquad \d \tilde z=o_\beta\,\tilde\iota_{\dot\beta}\,\d x^{\beta\dot\beta}\,,
\ee
for which the SDPW metric \eqref{sdpwgr} can be reexpressed as
\be\label{orthods2}
\d s^2 = \eps_{\alpha\beta}\,\eps_{\dot\alpha\dot\beta}\,e^{\alpha\dot\alpha}\,e^{\beta\dot\beta}\,.
\ee
The dual tetrad (i.e., frame of the tangent bundle) is given by
\be\label{sdpwdual}
\nabla_{\alpha\dot\alpha} = \p_{\alpha\dot\alpha} + f(x^-)\,o_\alpha\,\tilde\iota_{\dot\alpha}\,\frac{\p}{\p z}\, ,  \qquad \frac{\p}{\p z}=-o^\beta\,\tilde\iota^{\dot\beta}\,\p_{\beta\dot\beta}\,,
\ee
satisfying $\nabla_{\beta\dot\beta}\,\lrcorner\, e^{\alpha\dot\alpha} = \delta^\alpha_\beta\,\delta^{\dot\alpha}_{\dot\beta}$. Dotted and undotted (equivalently, SD and ASD) spinor bundles can now be set up with respect to this tetrad. For instance, the spinor components $v^{\alpha\dot\alpha}$ of a vector field $v^a$ are defined as the coefficients in its expansion $v = v^{\alpha\dot\alpha}\,\nabla_{\alpha\dot\alpha}$ in this basis. 

To see that the metric \eqref{sdpwgr} is indeed SD, one may compute the ASD and SD spin connections, finding
\be\label{sdpwcon}
\tilde\Gamma_{\dot\alpha\dot\beta} = -\dot f(x^-)\,\tilde\iota_{\dot\alpha}\,\tilde\iota_{\dot\beta} \,\d \tilde z\,,\qquad \Gamma_{\alpha\beta} = 0\,,
\ee
respectively, with $\dot{f}(x^-):=\partial_{-}f(x^-)$. Thus the ASD part of the Riemann curvature 2-form is $R_{\alpha\beta} = \d\Gamma_{\alpha\beta} + \Gamma_{\alpha}{}^{\gamma}\wedge\Gamma_{\gamma\beta}=0$, and while the SD part is
\be\label{riemannspin}
\tilde R_{\dot\alpha\dot\beta} = \d\tilde\Gamma_{\dot\alpha\dot\beta} + \tilde\Gamma_{\dot\alpha}{}^{\dot\gamma}\wedge\tilde\Gamma_{\dot\gamma\dot\beta}= -\ddot{f}(x^-)\, \tilde\iota_{\dot\alpha}\,\tilde\iota_{\dot\beta} \,\d x^-\wedge\d\tilde z\,,
\ee
so that the Weyl curvature of the metric is given by 
\be\label{sdpwWeyl}
\tilde\Psi_{\dot\alpha\dot\beta\dot\gamma\dot\delta} = -\ddot f\,\tilde\iota_{\dot\alpha}\,\tilde\iota_{\dot\beta}\,\tilde\iota_{\dot\gamma}\,\tilde\iota_{\dot\delta}\,,\qquad\Psi_{\alpha\beta\gamma\delta} = 0\,.
\ee
From now on, we also assume the sandwich condition on our SDPW metrics, meaning that $\ddot f(x^-)$, and hence the curvature, is compactly supported in $x^-$. This ensures that the SDPW metric is asymptotically flat except in the $n$-direction and admits a well-defined S-matrix~\cite{Gibbons:1975jb,Adamo:2017nia}.

As the SDPW metric is Kerr-Schild, its SD Weyl curvature tensor is essentially a linear field on the Minkowski background so the complex conjugate of \eqref{K-dA} can be used to identify the radiation data giving rise to \eqref{sdpwWeyl}. The corresponding asymptotic data is distributional in nature:
\be\label{sdpwshear}
\tilde{\bigma}^0(u,\lambda ,\bar{\lambda})\,\D\bar \lambda=\frac{\la o\,\lambda \ra^2}{[\tilde{o}\,\bar{\lambda}]}\, \cF\!\left(\frac{u}{\la\lambda \,o\ra\,[\bar{\lambda }\,\tilde{o}]}\right)\, \bar\delta\left( \la\iota\,\lambda \ra \right)\,,
\ee
with $\cF(x^-) = \int^{x^-}f(y)\,\d y$. This $\tilde{\bigma}^0$ has distributional support along the generator of $\scri^+$ corresponding to $(\lambda ,\bar{\lambda })=(\iota,\tilde{\iota})$, and the associated news function is easily seen to be
\be\label{sdpwnews}
N(u,\lambda ,\bar{\lambda })=\frac{\la o\,\lambda\ra}{[\tilde{o}\,\bar{\lambda}]^{2}}\, f\!\left(\frac{u}{\la\lambda \,o\ra\,[\bar{\lambda }\,\tilde{o}]}\right)\, \bar\delta(\la \iota\,\lambda\ra)\,.
\ee

\medskip


\paragraph{\textit{Gibbons-Hawking metrics:}} While Gibbons-Hawking metrics are famous as (asymptotically locally Euclidean or flat) gravitational instantons in Riemannian signature~\cite{Hawking:1976jb,Gibbons:1978tef,Gibbons:1979xm}, in split-signature or as complex metrics they can be viewed as SD radiative metrics, encoding a solution to the wave equation in $3$-dimensions. These metrics have a self-dual Killing vector.  Introduce a constant 4-vector $Y^{\alpha\dot\alpha}$ on flat space normalized so that $Y^2=2$. The line element on this flat background is given by  
\be\label{GH1}
\d s^{2}=V(\mathbf{x})\,\d \mathbf{x}_{\alpha\beta}\,\d \mathbf{x}^{\alpha\beta}+\frac{(\d y+\omega(\mathbf{x}))^2}{V(\mathbf{x})} \,,
\ee
with  coordinates $y=Y_{\alpha\dot\alpha} x^{\alpha\dot\alpha}$ and 
\be\label{GHtc}
\mathbf{x}^{\alpha\beta}:=Y^{(\alpha|\dot\alpha}x^{\beta)}_{\dot\alpha}=\frac{1}{\sqrt{2}}\left(\begin{array}{cc}
                                                                                                                                        t+z & x \\
                                                                                                                                           x & t-z
                                                                                                                                        \end{array}\right)\,,
\ee
and $\omega(\mathbf{x}) = \omega_{\al\beta}(\mathbf{x})\,\d\mathbf{x}^{\al\beta}$ a 1-form in $\mathbf{x}$. This metric admits the Killing vector $\p_y$ and the function $V$ and 1-form $\omega$ are related by the $2+1$-dimensional abelian monopole equation: $\d V=*_{3}\d\omega$, where $*_{3}$ is the Hodge star in flat $\mathbf{x}$-space. Thus, $V(\mathbf{x})$ is a solution to the 3-dimensional wave equation.

The tetrad 1-forms associated with metrics of this form are
\be\label{GHtet}
e^{\alpha\dot\alpha}=\sqrt{V}\,Y^{\beta\dot\alpha}\,\d\mathbf{x}_{\beta}{}^{\alpha}+\frac{Y^{\alpha\dot\alpha}}{2\sqrt{V}}\,(\d y+\omega)\,,
\ee
Their dual vector fields are found to be
\be\label{GHdtet}
\nabla_{\al\dal} = \frac{1}{\sqrt{V}}\,Y^{\beta}{}_{\dal}\left(\p_{\al\beta} - \omega_{\al\beta}\,\p_y\right)+ \sqrt{V}\,Y_{\al\dal}\,\p_y\,.
\ee
From \eqref{GHtet}, it is straightforward to compute the ASD 2-forms
\be\label{GH2forms}
\Sigma^{\alpha\beta}=\d\mathbf{x}^{\alpha\beta}\wedge(\d y+\omega)+V\,\d\mathbf{x}^{\alpha\gamma}\wedge\d\mathbf{x}^{\beta}{}_{\gamma} \,.
\ee
It then follows that $\d\Sigma^{\alpha\beta}=0$ as a consequence of the monopole relation between $V$ and $\omega$, so the complex space-time is indeed SD. The conformal compactification will not be smooth due to the symmetry in the $y$-direction, but the metric is nevertheless almost everywhere asymptotically flat with appropriate fall-off in directions not parallel to $\p_y$. 

To see that these metrics are radiative, it suffices to observe that the free data is a solution to the 3-dimensional wave equation, $\d*_3\d V(\mathbf{x})=0$. Since we are considering complexified (or split-signature) metrics, we can consider solutions to the wave equations which are sufficiently `small' so as to be radiative; the metric inherits its radiative data directly from the radiative data for $V$. More specifically, solutions to the wave equation can be generated by the Whittaker integral formula~\cite{Whittaker:1903} which in our homogeneous coordinates becomes
\be\label{Whitt2}
V(\mathbf{x})=\oint_{\Gamma} \D\lambda\;\varphi(q,\lambda)\,, \qquad q:=\mathbf{x}^{\alpha\beta}\lambda_{\alpha}\lambda_{\beta}\,,
\ee
where the integral is taken along a closed contour on the Riemann sphere and the function $\varphi(q,\zeta)$ is free data for the solution, with $\varphi$ possessing overall weight $-2$ in $\lambda_\al$. Choosing 
 the contour  $\Gamma= \{Y^{\alpha\dot\alpha}\lambda_{\alpha}\bar{\lambda}_{\dot\alpha}=0\}$, this version of Whittaker's formula for $V$ 
 is recast as a Kirchoff-d'Adh\'emar integral formula in 4-dimensions with
\be\label{Whitt3}
V(\mathbf{x})=\int_{\P^1}\D\lambda\wedge\D\bar\lambda\left.\frac{\la\lambda|Y|\eta]}{[\bar\lambda\,\eta]}\,\delta(\la\lambda|Y|\bar{\lambda}])\,\varphi(u,\lambda,\bar{\lambda})\right|_{u=x^{\alpha\dot\alpha}\lambda_{\alpha}\bar{\lambda}_{\dot\alpha}}\,,
\ee
with the characteristic data now lifted to $\scri^+_{\C}$, where $\eta_{\dal}$ is a constant reference spinor.  On the support of the delta function $\delta(\la\lambda|Y|\bar{\lambda}])$, $u\to q$ and $\varphi(u,\lambda,\bar{\lambda})\to \varphi(q,\lambda)$ and the integrand is independent of $\eta_{\dot\alpha}$.


\subsection{Twistor theory of SD radiative metrics}
\label{sec:radtwistor}

The twistor space $\PT$ of (complexified) Minkowski space $\M$ is an open subset of the three-dimensional complex projective space $\P^3$. If $Z^A=(\mu^{\dot\alpha},\lambda_{\alpha})$ are homogeneous coordinates on $\P^3$, related to $\M$ by the incidence relations 
\be\label{flatinc}
\mu^{\dot\alpha}=x^{\alpha\dal}\,\lambda_{\alpha}\,,
\ee
then this open subset is given by
\be\label{flatPT}
\PT=\{Z^A\in\P^3\,|\,\lambda_{\alpha}\neq0\}\,.
\ee
The incidence relations give a non-local correspondence between $\PT$ and $\M$ which associates a holomorphic, linearly-embedded Riemann sphere $X\cong\P^1\subset\PT$ -- a twistor line -- to each point $x\in\M$. The twistor geometry captures the conformal structure of Minkowski space: any two twistor lines $X,Y\subset\PT$ intersect if and only if the corresponding points $x,y\in\M$ are null separated. 

A key result is that this correspondence can be deformed to give a twistor theory of any SD complexified four-dimensional space-time through Penrose's non-linear graviton construction:
\begin{thm}[Penrose~\cite{Penrose:1976js}]\label{NLGTheorem}
There is a one-to-one correspondence between:
\begin{itemize}
 \item suitably convex regions of Ricci-flat, self-dual  
 4-manifolds $(\CM, g_{ab})$, and
 
 \item 3-dimensional complex manifolds $\CPT$ that are complex deformations of a region in $\PT$ that contain a 
 holomorphic $\CP^1$ with normal bundle $\cO(1)\oplus\cO(1)$, have a holomorphic fibration
 \begin{equation*}
 \pi:\CPT\to\P^1
 \end{equation*} 
and have a holomorphic Poisson structure $I$ up the fibres of $\pi$ with values in the pullback of $\cO(-2)$ from $\P^1$.
\end{itemize}
\end{thm}
Such a deformed twistor space $\CPT$ exists for each SD 4-manifold $\CM$. As in the flat twistor correspondence, each $x\in\CM$ corresponds to a Riemann sphere $X$ in $\CPT$. The curved twistor space $\CPT$ encodes the conformal structure of the SD 4-manifold $(\CM,g_{ab})$ by the statement that two twistor curves $X,Y\subset\CPT$ intersect if and only if the corresponding points $x,y\in\CM$ are lie on a null geodesic of the conformal class $[g_{ab}]$. 

When $\CM$ is a SD radiative space-time, it is possible to formulate the twistor construction entirely in terms of the radiative data $\tilde{\bigma}^0$ following a method of Sparling, \cite{Eastwood:1982}. To do this, we first adapt homogeneous coordinates on $\CPT$ to describe Ricci-flat SD manifolds, as in the treatments of~\cite{Mason:2007ct,Adamo:2021bej}. Let $\lambda_\alpha$ be homogeneous coordinates on the $\P^1$ base of the fibration $\pi$, and choose $\mu^{\dot\alpha}$ up the fibres so that
\be\label{infinityboy}
I := \eps^{\dot\alpha\dot\beta}\,\frac{\p}{\p\mu^{\dot\alpha}}\wedge\frac{\p}{\p\mu^{\dot\beta}}\,,
\ee
is the weighted holomorphic Poisson structure.

The complex structure of $\CPT$ can be represented as a complex deformation of the $\dbar$-operator $\dbar $ of flat $\PT$ expressed  as $\nbar=\dbar+V$,
 where $\dbar$ is the flat complex structure on $\P^3$ in homogeneous coordinates $(\mu^{\dot\alpha},\lambda_\alpha)$ and $V\in\Omega^{0,1}(\PT,T_{\PT})$ for $T_{\PT}$ the holomorphic tangent bundle of $\PT$. Integrability of this almost complex deformation requires $\nbar^2=0$. For $\lambda_\alpha$ and $I$ to be holomorphic in the complex structure $\nbar$, $V$ must be Hamiltonian with respect to this Poisson structure:
\be\label{VHamil}
V = V^{\dot\alpha}\,\frac{\p}{\p\mu^{\dot\alpha}} = \eps^{\dot\alpha\dot\beta}\,\frac{\p \sh}{\p\mu^{\dot\beta}}\,\frac{\p}{\p\mu^{\dot\alpha}}\,,
\ee
for some Hamiltonian $\sh\in\Omega^{0,1}(\PT,\cO(2))$. Denoting the Poisson bracket induced by \eqref{infinityboy} by $\{\cdot,\cdot\}$, the Dolbeault operator on $\CPT$ is thus $\nbar=\dbar + \{\sh,\cdot\}$. With these additional structures, the integrability requirement $\nbar^2=0$ becomes
\be\label{h-int}
\dbar \sh + \frac{1}{2}\,\{\sh,\,\sh\} = 0\,,
\ee
with $\sh$ encoding the data of a SD asymptotically flat vacuum metric~\cite{Mason:2007ct,Adamo:2021bej}. As before, one can introduce complex conjugations on $\CPT$ appropriate to obtain real Euclidean signature~\cite{Atiyah:1978wi,Woodhouse:1985id,Ward:1990vs} or split-signature SD four-manifolds~\cite{Lebrun:2007}.  For SD radiative space-times, \eqref{h-int} is solved directly in terms of the asymptotic radiative data to manifest the holographic nature of the metric~\cite{Newman:1976gc,Hansen:1978jz,Eastwood:1982}. To do so, define the projection from twistor space to $\scri^+_{\C}$ by 
\begin{equation}
p:\CPT\rightarrow\scri^+_\C\, , \qquad (\mu^{\dot\alpha},\,\lambda_\alpha)\mapsto (u=\mu^{\dot\alpha}\bar\lambda_{\dot\alpha},\,\lambda_\alpha,\,\bar{\lambda}_{\dot\alpha})\, ,\label{scri-proj}
\end{equation}
and solve \eqref{h-int} by taking:
\begin{equation}\label{radata}
\frac{\p\sh}{\p\mu^{\dot\alpha}}=\bar\lambda^{\dot\alpha}\,\tilde \bigma^0\, \D\bar\lambda\, , \qquad \mbox{ so } \quad \sh= \int^u \bigma^0(u',\lambda,\bar\lambda) \,\d u'  \,\D\bar\lambda\,.
\end{equation}
Now \eqref{h-int} follows because $\sh$ is holomorphic in ${\mu}^{\dot\alpha}$ and proportional to $\D\bar\lambda$. 

\medskip

Now, characterize the holomorphic Riemann spheres $X\subset \CPT$ as sections of the fibration $\pi:\CPT\rightarrow \P^1$ by 
\be\label{curmap}
(\mu^{\dot\alpha}=F^{\dot\alpha}(x,\lambda ),\,\lambda _{\alpha}): \P^1\rightarrow \CPT\,,
\ee
where $F^{\dot\alpha}(x,\lambda)$ is homogeneous of weight 1 in $\lambda$ but \emph{not}, at this stage, holomorphic. Using \eqref{VHamil}, $X$ is holomorphic with respect to $\nbar$ if $\nbar(\mu^{\dot\alpha}-F^{\dot\alpha}(x,\lambda))|_{X}=0$ which gives
\be\label{Vsplit}
\dbar|_{X} F^{\dot\alpha}(x,\lambda)=\varepsilon^{\dot\alpha\dot\beta} \left.\frac{\partial \sh}{\partial\mu^{\dot\beta}}\right|_{X}\,,
\ee 
as the equation determining the holomorphic curves in twistor space.
Such a rational curve $X$
can now be projected to $\scri^+_\C$ by \eqref{scri-proj} to give a curved space analogue of a light cone cut \eqref{lcc}, described by the cut function $u=Z(x,\lambda):=F^{\dot\alpha}\bar\lambda_{\dot\alpha}$. The equation \eqref{Vsplit} can then be written entirely in terms of data at $\scri^+_\C$ as:
\be\label{good-cut}
\frac{\partial^2 Z}{\partial\bar{\lambda}^{\dot\alpha}\partial\bar{\lambda}^{\dot\beta}}=\bar\lambda_{\dot\alpha}\bar{\lambda}_{\dot\beta}\,\tilde{\bigma}^0\,,\qquad \Leftrightarrow\qquad \bar \eth^2 Z=\tilde{\bigma}^0(Z,\lambda\bar\lambda)\, ,
\ee
which is Newman's \emph{good cut equation} written in the homogeneous formalism~\cite{Newman:1976gc,Hansen:1978jz,Eastwood:1982}.

\medskip


\paragraph{Reconstruction of the half-flat metric on $\CM$:} Under our assumptions, standard theorems of Kodaira~\cite{Kodaira:1962,Kodaira:1963} guarantee the existence of solutions $F^{\dot\alpha}$ of \eqref{Vsplit} -- or equivalently, of solutions $Z(x,\lambda)$ to \eqref{good-cut} -- parametrized by $x\in\CM$. 

To reconstruct the SD radiative metric on $\CM$ from $\CPT$, 
observe that the vector fields 
\be\label{aholvfs} \frac{\partial}{\partial\bar{\lambda}^{\dot\alpha}}+\tilde{\bigma}^{0}\,\bar{\lambda}_{\dot\alpha}\,\bar{\lambda}^{\dot\beta}\,\frac{\partial}{\partial\mu^{\dot\beta}}\,,
\ee
are anti-holomorphic in the deformed complex structure given by $\nbar$. These vector fields annihilate the differential forms $\d\mu^{\dot\alpha}-\tilde{\bigma}^{0}\,\D\bar\lambda\,\bar{\lambda}^{\dot\alpha}$, so these two 1-forms along with $\D\lambda$ span $\Omega^{1,0}(\CPT)$. This allows us to construct a holomorphic top form $\Omega_{\CPT}\in\Omega^{3,0}(\CPT,\cO(4))$ as
\be\label{OmegaCPT}
\Omega_{\CPT} = \D\lambda\wedge(\d\mu^{\dot\alpha}-\tilde{\bigma}^{0}\,\D\bar\lambda\,\bar{\lambda}^{\dot\alpha})\wedge(\d\mu_{\dot\alpha} - \tilde{\bigma}^{0}\,\D\bar\lambda\,\bar{\lambda}_{\dot\alpha})\,.
\ee
Setting $\mu^{\dot\alpha}=F^{\dot\alpha}$ and using \eqref{Vsplit} we obtain the pullback of $\Omega_{\CPT}$ to $\PS=\CM\times \P^1$:
\be\label{Omegapull}
p^*\Omega_{\CPT} = \D\lambda \wedge\d_x F^{\dot\alpha}\wedge\d_x F_{\dot\alpha}\,,
\ee
where $\d_x$ denotes the exterior derivative on $\PS$ along $\CM$.

As $\D\lambda $ is trivially holomorphic, the 2-form $\d_x F^{\dot\alpha}\wedge\d_x F_{\dot\alpha}$ must be globally holomorphic along any rational twistor curve $X$, as is confirmed by a brief calculation:
\begin{equation*}
\begin{split}
\dbar|_{X}\left(\d_x F^{\dot\alpha}\wedge\d_x F_{\dot\alpha}\right) &= 2\,\d_x\left(\tilde{\bigma}^0|_{X}\,\bar{\lambda }^{\dot\alpha}\,\D\bar{\lambda}\right)\wedge\d_xF_{\dot\alpha} \\
 & = -2\,N|_{X}\,\D\bar{\lambda}\,\bar{\lambda }^{\dot\alpha}\,\bar{\lambda }^{\dot\beta}\,\d_x F_{\dot\beta}\wedge\d_xF_{\dot\alpha} =0\,,
\end{split}
\end{equation*}
where $N|_{X}$ denotes the news function $N=-\partial_u\tilde{\bigma}^0$ evaluated on the twistor curve $X$. Thus, $\d_x F^{\dot\alpha}\wedge\d_x F_{\dot\alpha}$ is a holomorphic 2-form of homogeneity 2 in $\lambda _{\alpha}$; by Liouville's theorem, there exists a triplet of 2-forms $\Sigma^{\alpha\beta}=\Sigma ^{(\alpha\beta)}$ on $\CM$ such that
\be\label{Omegapull1}
\d_x F^{\dot\alpha}\wedge\d_x F_{\dot\alpha}=\lambda _{\alpha}\,\lambda _{\beta}\,\Sigma ^{\alpha\beta}(x)\,, \qquad p^{*}\Omega_{\CPT}=\D\lambda \wedge\Sigma ^{\alpha\beta}(x)\,\lambda _\alpha\,\lambda _\beta\,.
\ee
It follows by construction that the 2-forms $\Sigma ^{\alpha\beta}$ obey $\d\Sigma ^{\alpha\beta}=0$ and $\Sigma^{(\alpha\beta}\wedge\Sigma ^{\gamma\delta)} = 0$.

This implies the existence of a tetrad $e^{\alpha\dot\alpha}$ on $\CM$ for which the $\Sigma ^{\alpha\beta}$ are a basis of ASD two forms \cite{Capovilla:1991qb}:
\be\label{simplicity}
\Sigma ^{\alpha\beta} = e^{\alpha\dot\alpha}\wedge e_{\dot\alpha}{}^\beta\,.
\ee
With this tetrad, the SD radiative metric on $\CM$ is recovered as $\d s^{2}=\epsilon_{\alpha\beta}\epsilon_{\dot\alpha\dot\beta}\,e^{\alpha\dot\alpha} e^{\beta\dot\beta}$; self-duality follows as a consequence of $\d \Sigma ^{\alpha\beta}=0$. Cartan's structure equations then give vanishing ASD (undotted) spin connection in this spin frame. Hence, the full ASD curvature vanishes -- including both the ASD Weyl tensor and Ricci tensor.

\medskip


\paragraph{Spin connection, Lax formulation \& Plebanski form:} The ASD spin connection is flat, but the SD (dotted) one is not. The SD spin bundle and its connection on $\CM$ are constructed (via the Ward transform) from the rank-two bundle $V_\pi(-1):=V_\pi\otimes\cO(-1)$ on $\CPT$, where $V_\pi$ is the bundle of vertical vectors up the fibres of $\pi:\CPT\rightarrow \P^1$. 

The object $\d_xF^{\dot\alpha} \,\p/\p\mu^{\dot\alpha}$ gives a 1-form on $\CM$ with values in $V_\pi$; contraction with a vector at $x$ determines a section of $V_\pi$ along $X$ so that  $V_\pi|_X =\cN_X$, the normal bundle to $X$ in $\CPT$. Now $\d_xF^{\dot\alpha}$ can be evaluated in a tetrad on $\CM$ to obtain
\be\label{incitetrad}
\d_x F^{\dot\alpha}(x,\lambda) = H^{\dot\alpha}{}_{\dot\beta}(x,\lambda )\,e^{\alpha\dot\beta}(x)\,\lambda _\alpha\,,
\ee
where the factor of $\lambda_\alpha$ follows from \eqref{Omegapull1}. This defines $H^{\dot\alpha}{}_{\dot\beta}(x,\lambda )\in\text{SL}(2,\C)$ which provides a holomorphic trivialization of $V_\pi(-1)$. The global sections $V_\pi(-1)|_X$ are thus precisely the SD (dotted) spinors at $x\in\CM$. 

The SD vacuum equations also admit a Lax description; let $\nabla_{\alpha\dot\alpha}$ be the dual tetrad to $e^{\alpha\dot\alpha}$, then \eqref{incitetrad} becomes
\be\label{frameformula}
\lambda _\alpha\,H^{\dot\alpha}{}_{\dot\beta} = \nabla_{\alpha\dot\beta}\,\lrcorner\,\d_x F^{\dot\alpha} = \nabla_{\alpha\dot\beta}F^{\dot\alpha}\,,
\ee
which gives the Lax description
\be\label{incidencelost}
\lambda ^\alpha\,\nabla_{\alpha\dot\alpha}F^{\dot\beta} = 0\,.
\ee
Integrability of these Lax operators is equivalent to the vacuum self-duality equations on $\CM$. This also gives a d-bar equation for $H^{\dot\alpha}{}_{\dot\beta}$ by acting with $\dbar|_X $ on both sides of \eqref{frameformula} and using \eqref{Vsplit} to give
\be\label{Heqn*}
\begin{split}
\lambda _\alpha\,\dbar|_X H^{\dot\gamma}{}_{\dot\beta} &= \nabla_{\alpha\dot\beta}\dbar|_X F^{\dot\gamma} = \nabla_{\alpha\dot\beta } \frac{\p \sh}{\p\mu_{\dot\gamma}}=\lambda _\alpha\,\frac{\p^2 \sh}{\p\mu^{\dot\delta}\p\mu_{\dot\gamma}}\biggr|_X\,H^{\dot\delta}{}_{\dot\beta} \\
&\implies \dbar|_XH^{\dal}{}_{\dot\beta} = \frac{\p^2 \sh}{\p\mu^{\dot\gamma}\p\mu_{\dal}}\biggr|_X\,H^{\dot\gamma}{}_{\dot\beta}\,.
\end{split}
\ee
Using the explicit form of $\sh$ in terms of the asymptotic radiative data yields
\be\label{Heqngr}
\dbar|_X H^{\dot\alpha}{}_{\dot\gamma}(x,\lambda )= \bar\lambda^{\dot\alpha}\bar\lambda_{\dot\beta}  \, H^{\dot\beta}{}_{\dot\gamma}(x,\lambda ) \, N|_X\,\D\bar\lambda\, ,
\ee
which acts as a d-bar equation determining holomorphic frames for dotted spinor bundles over the light cone cuts $u=Z(x,\lambda,\bar\lambda)$.

It is interesting to observe the close similarity between the twistor description of SD radiative metrics and that of SD radiative gauge fields (cf., \cite{Adamo:2020yzi}). Indeed, the global frame $H^{\dot\alpha}{}_{\dot\beta}$ is the gravitational equivalent of the global holomorphic frame appearing in the gauge theory setting, with the Ward bundle on twistor space replaced by $V_{\pi}(-1)$. This equivalence can be made precise, with the role of the gauge theory radiative data $\tilde{\cA}^0$ -- valued in $\cO(0,-2)$ tensored with the Lie algebra of the gauge group -- played by $\bar{\lambda}^{\dot\alpha}\bar{\lambda}_{\dot\beta} N$ in the gravitational case.

Finally, self-dual Ricci-flat manifolds in 4-dimensions are necessarily hyperk\"ahler. So the metric on $\CM$ has a local expression in terms of a scalar (K\"ahler) potential, $\Omega$, known as the first Plebanski form~\cite{Plebanski:1975wn}. This Plebanski form can be recovered from the holomorphic curves in $\CPT$ by choosing a spinor dyad $\{o_{\alpha},\iota_{\alpha}\}$ such that $\la \iota\,o\ra=1$ and writing
\be\label{lambexp}
\lambda_{\alpha}=\lambda_{1}\,\iota_{\alpha}+\lambda_{2}\,o_{\alpha}\,,
\ee 
for the fibre coordinates of $\CPT$. Then $F^{\dot\alpha}(x,\lambda)$ can be expanded around $\lambda_2=0$ and $\lambda_1=0$, respectively to give~\cite{Chakravarty:1991bt,Dunajski:2000iq,Adamo:2021bej}
\be\label{Fexp}
F^{\dot\alpha}(x,\lambda)=\lambda_1\,z^{\dot\alpha}+\lambda_{2}\,\Omega^{\dot\alpha}(x)+O(\lambda_2^2)\,, \qquad F^{\dot{\tilde\alpha}}(x,\lambda)=\lambda_2\,\tilde{z}^{\dot{\tilde\alpha}}-\lambda_1\,\Omega^{\dot{\tilde\alpha}}+O(\lambda_1^2)\,,
\ee
using the globality of the 2-form $\d_x F^{\dot\alpha}\wedge\d_x F_{\dot\alpha}$. Here, $(z^{\dot\alpha},\tilde{z}^{\dot{\tilde\alpha}})$ give local coordinates on $\CM$, and the objects $\Omega^{\dot\alpha}$, $\Omega^{\dot{\tilde\alpha}}$ are functions on $\CM$. Using \eqref{Omegapull1} in conjunction with these expansions allows us to identify
\be\label{2forms*}
 \Sigma^{12}=\d z^{\dot\alpha}\wedge\d \Omega_{\dot\alpha}=-\d\tilde{z}^{\dot{\tilde\alpha}}\wedge\d\Omega_{\dot{\tilde\alpha}}\,, 
\ee
from which it follows
\be\label{Plebform1}
\Omega_{\dot\alpha}=\frac{\partial\Omega}{\partial z^{\dot\alpha}}\,, \qquad \Omega_{\dot{\tilde\alpha}}=\frac{\partial\Omega}{\partial \tilde{z}^{\dot{\tilde\alpha}}}\,,
\ee
for some potential $\Omega(x)=\Omega(z,\tilde{z})$, a (pseudo-) Kahler scalar in the appropriate  signature. Defining
\be\label{Plebform2}
\Omega_{\dot\alpha\dot{\tilde\beta}}:=\frac{\partial^2\Omega}{\partial z^{\dot\alpha}\partial\tilde{z}^{\dot{\tilde\beta}}}\,,
\ee
and requiring that $\Sigma^{22}$ calculated from expansions around $\lambda_1=0$ and $\lambda_2=0$ agree, one obtains Plebanski's first heavenly equation~\cite{Plebanski:1975wn} or Monge-Amp\`ere equation:
\be\label{Heqn}
\Omega_{\dot\alpha\dot{\tilde\beta}}\,\Omega^{\dot\alpha}{}_{\dot{\tilde\gamma}}=\epsilon_{\dot{\tilde\beta}\dot{\tilde\gamma}}\,, \qquad \Leftrightarrow\qquad \det \frac{\p^2\Omega}{\p z^{\dot\alpha}\p \tilde z^{\dot{\tilde\alpha}}}=1\, .
\ee
This equation ensures that the tetrad
\be\label{Plebformtet}
e^{\alpha\dot\alpha}=\bigl(\d z^{\dot\alpha},\, \Omega^{\dot\alpha}{}_{\dot{\tilde\beta}}\,\d\tilde{z}^{\dot{\tilde\beta}}\bigr)\,,
\ee
is hyperk\"ahler, or equivalently, SD and Ricci-flat.

\medskip


\subsection{Gravitational perturbations} \label{sec:perts}
Metric perturbations $h_{\alpha\dot\alpha\beta\dot\beta}(x)$ on a SD radiative space-time $\CM$ have two on-shell degrees of freedom, corresponding to positive and negative helicity (i.e., SD and ASD) modes. However, the metric perturbation itself does \emph{not} split cleanly into SD and ASD parts (as it does in Minkowski space-time) due to the SD curvature of the background. The concept of a SD perturbation of the metric remains well-defined, but (asymmetrically) we only have a concept of an ASD perturbation of the Weyl curvature; for a generic solution to the linearized vacuum equations the condition that the corresponding perturbation of the ASD Weyl tensor should vanish is the condition that it is SD.

A negative helicity graviton can be characterized by $\nabla^{\alpha\dot\alpha}\psi_{\alpha\beta\gamma\delta}=0$, where $\psi_{\alpha\beta\gamma\delta}$ is the perturbation's linearised ASD Weyl spinor and $\nabla_{\alpha\dot\alpha}$ is just the dual tetrad on $\CM$ as the ASD spin connection is flat. The Penrose transform encodes such linearised fields on $\CM$ through cohomology on $\CPT$~\cite{Hitchin:1980hp,Ward:1990vs}:
\be\label{nhgrav1}
H^{0,1}_{\nbar}(\CPT,\cO(-6))\cong \left\{h_{ab} \mbox{ on } \CM\,|\,\nabla^{\alpha\dot\alpha}\psi_{\alpha\beta\gamma\delta}=0\right\}\,,
\ee
where $H^{0,1}_{\nbar}$ denotes Dolbeault cohomology with respect to the deformed complex structure $\nbar=\dbar+\{\sh,\cdot\}$ on $\CPT$. Given some $\tilde{h}\in H^{0,1}_{\nbar}(\CPT,\cO(-6))$, the field on $\CM$ is recovered from the integral formula
\be\label{psipenrose}
\psi_{\alpha\beta\gamma\delta}(x) = \int_X\D\lambda \wedge\lambda _\alpha\,\lambda _\beta\,\lambda _\gamma\,\lambda _\delta\,\tilde{h}|_X\,.
\ee
Since $\lambda^\alpha\nabla_{\alpha\dot\alpha}\tilde{h}|_X=0$ due to \eqref{incidencelost}, it follows that this solves the linear field equation $\nabla^{\alpha\dot\alpha}\psi_{\alpha\beta\gamma\delta}(x) = 0$. If we wish to find a metric perturbation that has  $\psi_{\alpha\beta\gamma\delta}$ as its ASD linearized curvature, we find that  setting the variation of the SD curvature to zero is obstructed by the SD background curvature; in particular it is inconsistent with diffeomorphism (gauge) invariance (cf., \cite{Mason:2008jy}).

Note that for a particular representative of $\tilde{h}$ within its Dolbeault equivalence class, \eqref{psipenrose} is in fact equivalent to the Kirchoff-d'Adh\'emar formula \eqref{K-dA}~\cite{Mason:1986}. This is achieved with the identification $\partial_{u}\psi_{4}\,\D\bar{\lambda}=\tilde{h}$, which is easily seen to be homogeneous of weight $-6$ on twistor space and $\nbar$-closed by virtue of being proportional to $\D\bar{\lambda}$ \cite{Mason:1986}.

\medskip

For positive helicity, the corresponding SD massless fields on a SD background are no longer consistent beyond spin 1, but a potential-modulo-gauge description of such fields still works. In particular, at spin 2 we can simply perturb the nonlinear graviton construction. The Penrose transform~\cite{Hitchin:1980hp,Ward:1990vs} for a positive helicity graviton (with suitable analyticity) is
\be\label{phgrav1}
H^{0,1}_{\nbar}(\CPT,\cO(2))\cong \left\{h_{ab} \mbox{ on } \CM\,|\,\psi_{\alpha\beta\gamma\delta}=0\right\}\,.
\ee
Given a representative $h\in H^{0,1}_{\nbar}(\CPT,\cO(2))$, note that on restriction to a rational curve $X$, $h|_X\in H^{0,1}(\P^1,\cO(2))$ since
\be\label{nbartrivial}
\dbar|_X\!\left(h|_X\right) = \left(\dbar h\right)\!|_X + \dbar|_XF^{\dot\alpha}\,\frac{\p h}{\p\mu^{\dot\alpha}}\biggr|_X = \left(\nbar h\right)\!|_X = 0\,,
\ee
having used \eqref{Vsplit}. But this cohomology group vanishes, so $h|_X$ must be $\dbar|_X$-exact:
\be\label{sdsplitgr}
h|_X = \dbar|_Xj(x,\lambda )\,,
\ee
for some function $j(x,\lambda )$ homogeneous of degree $2$ in $\lambda $. Since $h|_X$ is pulled back from $\CPT$, it follows that $\lambda ^\alpha\nabla_{\alpha\dot\alpha} h|_X = 0$ using \eqref{incidencelost}. As $\lambda^{\alpha}\nabla_{\alpha\dot\alpha}$ is holomorphic on $X$, \eqref{sdsplitgr} implies that $\lambda ^{\alpha}\nabla_{\alpha\dot\alpha} j$ is globally holomorphic in $\lambda$, so 
\be\label{thirdpot}
\lambda ^\alpha\nabla_{\alpha\dot\alpha}j(x,\lambda ) = \im\,\lambda ^\alpha\,\lambda ^\beta\,\lambda ^\gamma\,\varphi_{\dot\alpha\alpha\beta\gamma}(x)\,,
\ee
for some function $\varphi_{\dot\alpha\alpha\beta\gamma}$ on $\CM$ which is symmetric in its undotted indices. This is a potential for a metric perturbation $h_{ab}$~\cite{Penrose:1965am}:
\be\label{potchain}
h_{\alpha\dot\alpha\beta\dot\beta} = \D^\gamma{}_{(\dot\beta}\varphi_{\dot\alpha)\alpha\beta\gamma}\,,
\ee
where $\D_a$ is the Levi-Civita connection on $\CM$. It is straightforward to confirm that the linearised ASD Weyl spinor $\psi_{\alpha\beta\gamma\delta}$ associated to the metric perturbation \eqref{potchain} vanishes, and thus $h_{ab}$ defines a positive helicity graviton.

\medskip


\paragraph{Momentum eigenstates:} By virtue of their asymptotic flatness, we can consider graviton perturbations on SD radiative metrics which have asymptotic on-shell momentum $\kappa_{\alpha}\tilde{\kappa}_{\dot\alpha}$ at $\scri^{-}_{\C}$. Explicit representatives for such momentum eigenstates can be built using twistor representatives that are functionally identical to those on Minkowski space, thanks to the form of the twistor complex structure for SD radiative space-times. For a positive helicity graviton, the representative
\be\label{phgrrep}
h=\int_{\C^*}\frac{\d s}{s^3}\,\bar{\delta}^{2}(\kappa-s\,\lambda)\,\e^{\im\,s\,[\mu\,\tilde{\kappa}]}\,,
\ee
is in $H^{0,1}_{\nbar}(\CPT,\cO(2))$, with $\nbar$-closure following as it is a multiple of $\D\bar\lambda$, as is $\sh$, so that any potential curved correction term vanishes.  Restricting to a twistor curve $X\subset\CPT$ gives
\be\label{phgrrep2}
h|_X =  \frac{\la\xi\,\lambda\ra^3}{\la\xi\,\kappa\ra^3}\,\bar\delta\bigl(\la\lambda\,\kappa\ra\bigr)\,\e^{\im\phi} = \frac{1}{2\pi\im}\,\dbar|_X\left(\frac{\la\xi\,\lambda\ra^3}{\la\xi\,\kappa\ra^3\,\la\lambda\,\kappa\ra}\,\e^{\im\phi}\right),
\ee
where $\xi_{\alpha}$ is an arbitrary fixed reference spinor used to integrate out $s$ in \eqref{phgrrep} against one of the delta functions, setting $s=\la \xi\,\kappa\ra/\la\xi\,\lambda\ra$. The function $\phi(x)$ is defined by
\be\label{gravphase}
\phi(x):= [F(x,\kappa)\,\tilde\kappa] = F^{\dot\alpha}(x,\kappa)\,\tilde{\kappa}_{\dot\alpha}\,,
\ee
and it follows from \eqref{frameformula} that this solves the Hamilton-Jacobi equation:
\be\label{grHJ}
g^{ab}\,\p_a\phi\,\p_b\phi = 0\,,
\ee
with $g^{ab}$ the inverse metric on $\CM$ defined by the dual tetrad.

Comparing \eqref{phgrrep2} with \eqref{sdsplitgr} yields
\be\label{jsdgrav}
j(x,\lambda) = \frac{1}{2\pi\im}\,\frac{\la\xi\,\lambda\ra^3}{\la\xi\,\kappa\ra^3\,\la\lambda\,\kappa\ra}\,\e^{\im\phi}\,.
\ee
To descend to space-time via \eqref{thirdpot}, define the background-dressed momentum $K_{\alpha\dot\alpha} := \nabla_{\alpha\dot\alpha}\phi$, for which
\be\label{Kgrgen}
K_{\alpha\dot\alpha} = \nabla_{\alpha\dot\alpha}F^{\dot\beta}(x,\kappa)\,\tilde\kappa_{\dot\beta} = \kappa_\alpha\,\tilde K_{\dot\alpha}\,,\qquad\text{where}\;\tilde K_{\dot\alpha} := \tilde\kappa_{\dot\beta}\,H^{\dot\beta}{}_{\dot\alpha}(x,\kappa)\,,
\ee
having used \eqref{frameformula}. In other words, the holomorphic frame $H^{\dot\alpha}{}_{\dot\beta}$ serves to dress the dotted components of the graviton momentum as it traverses the SD radiative background. The potential on $\CM$ associated with \eqref{jsdgrav} is easily seen to be
\be\label{thirdpotrep}
\varphi_{\dot\alpha\alpha\beta\gamma} = \im\,\frac{\xi_\alpha\,\xi_\beta\,\xi_\gamma\,\tilde K_{\dot\alpha}}{\la\xi\,\kappa\ra^3}\,\e^{\im\phi}\,,
\ee
from which the positive helicity graviton perturbation follows via \eqref{potchain}:
\be\label{hplusgen}
h^{(+)}_{\alpha\dot\alpha\beta\dot\beta} = \nabla^{\gamma}{}_{(\dot\beta}\varphi_{\dot\alpha)\alpha\beta\gamma} - \tilde\Gamma^\gamma{}_{(\dot\beta\dot\alpha)}{}^{\dot\gamma}\,\varphi_{\dot\gamma\alpha\beta\gamma}\,,
\ee
where $\tilde\Gamma_{\dot\alpha\dot\beta}\equiv\tilde\Gamma_{\gamma\dot\gamma\dot\alpha\dot\beta}\,e^{\gamma\dot\gamma}$ is the background SD spin connection

For a negative helicity graviton, the representative
\be\label{nhgrrep}
\tilde{h}=\int_{\C^*}\d s\,s^5\,\bar{\delta}^{2}(\kappa-s\,\lambda)\,\e^{\im\,s\,[\mu\,\tilde{\kappa}]}\,,
\ee
is valued in $H^{0,1}_{\nbar}(\CPT,\cO(-6))$. Feeding this into \eqref{psipenrose} immediately gives
\be\label{nhgrrep2}
\psi_{\alpha\beta\gamma\delta}=\kappa_\alpha\,\kappa_\beta\,\kappa_\gamma\,\kappa_\delta\,\e^{\im\phi}\,,
\ee
which is easily seen to obey $\nabla^{\alpha\dot\alpha}\psi_{\alpha\beta\gamma\delta}=0$. Note that this differs from the expression for the linearised Weyl curvature of a negative helicity graviton on Minkowski space only by the substitution $\e^{\im k\cdot x}\rightarrow\e^{\im\phi}$; this is an expected consequence of the flatness of the undotted spin-connection that in turn follows from self-duality of the background.



\subsection{Explicit examples of the twistor construction}\label{sec:twistor-exs}

Although in general the twistor construction is implicit, for our two families of example self-dual radiative space-times the construction can be made explicit.

\medskip

\paragraph{\textit{SDPWs:}} The twistor construction of SDPW metrics was one of the first examples of the non-linear graviton construction~\cite{Ward:1978soq,Curtis:1978pw,Porter:1982uj}; our presentation differs from those in the literature by making use of a Dolbeault (rather than \v{C}ech) framework. The complex structure $\nbar=\dbar+\{\sh,\cdot\}$ is defined by
\be\label{twistorsdpwgr}
\sh = 2\pi\im\,\int_{\C^*}\frac{\d s}{s^3}\,\bar\delta^2(\iota-s\,\lambda)\,\scF(s[\mu\,\tilde\iota])\,,\qquad \mbox{ where }\;\scF(x^-) := \int^{x^-}\F(t)\,\d t\,,
\ee
and $\F(x^-) = \int^{x^-}f(t)\,\d t$ is the first antiderivative of $f(x^-)$. Observing that
\be\label{Vsdpw}
\left.\frac{\partial\sh}{\partial\mu_{\dot\alpha}}\right|_{X}=\dbar|_{X}\left(\frac{\la o\,\lambda\ra^2}{\la\iota\,\lambda\ra}\,\tilde{\iota}^{\dot\alpha}\,\F(x^-)\right)\,,
\ee
it follows from \eqref{Vsplit} that the holomorphic curves in $\CPT$ are given by
\be\label{SDPWcurve}
F^{\dot\alpha}(x,\lambda)=x^{\alpha\dot\alpha}\,\lambda_{\alpha}+\frac{\la o\,\lambda\ra^2}{\la\iota\,\lambda\ra}\,\tilde{\iota}^{\dot\alpha}\,\F(x^-)\,.
\ee
With the SDPW tetrad \eqref{sdpwtetrad}, equation \eqref{incitetrad} produces the associated holomorphic frame $H^{\dot\alpha}{}_{\dot\beta}$ for the SD spin bundle:
\be\label{sdpwframe}
H^{\dot\alpha}{}_{\dot\beta}(x,\lambda) = \delta^{\dot\alpha}_{\dot\beta} - \frac{\la o\,\lambda\ra}{\la\iota\,\lambda\ra}\,f(x^-)\,\tilde\iota^{\dot\alpha}\,\tilde\iota_{\dot\beta}\,.
\ee
It is easy to confirm that this $H^{\dot\alpha}{}_{\dot\beta}$ satisfies \eqref{frameformula}.

The solution to the Hamilton-Jacobi equations on a SDPW metric associated with incoming asymptotic momentum $\kappa_{\alpha}\tilde{\kappa}_{\dot\alpha}$ is thus
\be\label{phigr}
\phi(x) =
 \la\kappa|x|\tilde\kappa] + \frac{\la o\,\kappa\ra^2\,[\tilde\iota\,\tilde\kappa]}{\la\iota\,\kappa\ra}\,\F(x^-)\,.
\ee 
As expected, this is the SD truncation of the solution to the Hamilton-Jacobi equation on a general plane wave space-time~\cite{Ward:1987ws,Adamo:2017nia,Adamo:2020qru}. The null Killing vector $n^{\alpha\dot\alpha}=\iota^{\alpha}\tilde{\iota}^{\dot\alpha}$ of the SDPW metric enables us to impose lightfront gauge on graviton perturbations of both helicities. Applying \eqref{sdpwframe} to \eqref{hplusgen} gives the positive helicity graviton
\be\label{hplus}
h^{(+)}_{\alpha\dot\alpha\beta\dot\beta} = \frac{\xi_\alpha\,\xi_\beta}{\la\xi\,\kappa\ra^2}\left(\tilde K_{\dot\alpha}\,\tilde K_{\dot\beta} - \im\,\dot f\,\frac{[\tilde\iota\,\tilde\kappa]}{\la\iota\,\kappa\ra}\,\tilde\iota_{\dot\alpha}\,\tilde\iota_{\dot\beta}\right)\,\e^{\im\phi}\,,
\ee
where choosing the reference spinor $\xi_{\alpha}=\iota_{\alpha}$ imposes lightfront gauge, and the dressed graviton momentum is\footnote{Factors of $[\tilde\iota\,\tilde\kappa]/\la\iota\,\kappa\ra = 1$ are kept explicit to ensure uniform little group weight under the usual $\C^*$-scalings $(\kappa_\alpha,\tilde\kappa_{\dot\alpha})\mapsto(r\,\kappa_\alpha,r^{-1}\tilde\kappa_{\dot\alpha})$ for $r\in\C^*$.}
\be\label{sdpwdressgr}
K_{\alpha\dot\alpha}(x^-) = \kappa_{\alpha}\,\tilde{K}_{\dot\alpha}(x^-)\,, \qquad \tilde{K}_{\dot\alpha}=\tilde{\kappa}_{\dot\alpha}- \frac{\la o\,\kappa\ra\,[\tilde\iota\,\tilde\kappa]\,f(x^-)}{\la\iota\,\kappa\ra}\,\tilde\iota_{\dot\alpha}\,.
\ee
The explicit `tail' term in $h^{(+)}_{ab}$, proportional to $\dot{f}(x^-)$, is a result of the failure of Huygens' principle in the non-linear SD background~\cite{Mason:1989,Harte:2013dba,Adamo:2017nia}.

For negative helicity gravitons, the lightfront gauge condition is precisely what is required to consistently set the associated linearised SD Weyl spinor to zero. In particular, the negative helicity graviton associated with \eqref{nhgrrep2} is given by
\be\label{hminus}
h^{(-)}_{\alpha\dot\alpha\beta\dot\beta}=\frac{\kappa_{\alpha}\,\kappa_{\beta}}{[\tilde{\iota}\,\tilde{\kappa}]^2}\,\tilde{\iota}_{\dot\alpha}\,\tilde{\iota}_{\dot\beta}\,\e^{\im\,\phi}\,.
\ee
The combined de Donder-lightfront gauge is the only one in which it is possible to make this identification of the negative helicity graviton at metric level.


\paragraph{\textit{Gibbons-Hawking metrics:}} The twistor theory of metrics with SD Killing vectors, of which the Gibbons-Hawking form \eqref{GH1} is an example, has been well-studied~\cite{Tod:1979tt,Dunajski:2003gp}. The key idea is that the free data of the metric -- solutions to the abelian monopole equations $\d V=*_{3}\d\omega$ -- are encoded in terms of cohomology data on \emph{minitwistor space}, the space of oriented geodesics in 3-dimensions~\cite{Hitchin:1982gh}. This means that the complex structure on $\CPT$ is given by $\nbar=\dbar+\{\sh,\cdot\}$ with
\be\label{GIT1}
\sh(Z)=f(q,\lambda,\bar{\lambda})\,\D\bar{\lambda}\,, \qquad q:=Y^{\alpha}{}_{\dot\alpha}\,\mu^{\dot\alpha}\,\lambda_{\alpha}\,,
\ee
where $f$ is a function on $\PT$ of homogeneity weight $2$ holomorphically and weight $-2$ anti-holomorphically. 

The equation for holomorphic curves in twistor space is thus
\be\label{GIT2}
\dbar|_{X} F^{\dot\alpha}(x,\lambda)=\D\bar{\lambda}\,Y^{\alpha}{}_{\dot\alpha}\,\lambda_{\alpha}\,\left.\frac{\partial f}{\partial q}\right|_{X}\,,\qquad q|_{X}=\mathbf{x}^{\alpha\beta}\,\lambda_{\alpha}\lambda_{\beta}\,.
\ee
The usual arguments based on a natural extension of Liouville's theorem guarantee the existence of a splitting function $g(\mathbf{x},\lambda)$, homogeneous of degree zero holomorphically and anti-holomorphically, such that
\be\label{GIT3}
\dbar|_{X}g= -\D\bar{\lambda}\, \left.\frac{\partial f}{\partial q}\right|_{X}\,.
\ee
With this, it is straightforward to solve for the holomorphic curves as
\be\label{GIhc}
F^{\dot\alpha}(x,\lambda)=\left(x^{\alpha\dot\alpha}+g(\mathbf{x},\lambda)\,Y^{\alpha\dot\alpha}\right) \lambda_{\alpha}\,,
\ee
and $g$ can be related to $f$ by the integral formula
\be\label{GIT4}
g(\mathbf{x},\lambda)=\frac{-1}{2\pi\im}\int_{\P^1}\frac{\D\lambda'\wedge\D\bar{\lambda}'}{\la\lambda\,\lambda'\ra}\,\frac{\la\xi\,\lambda\ra}{\la\xi\,\lambda'\ra}\,\left.\frac{\partial f}{\partial q}\right|_{X'}\,.
\ee
Here, the choice of $\xi\in\P^1$ is just a boundary condition needed to specify the Cauchy kernel for $\dbar|_X$ acting on sections of $\cO\rightarrow\P^1$.

From \eqref{GIhc}, one can immediately compute
\be\label{GIT5}
\Sigma=\d_{x}F^{\dot\alpha}\wedge\d_{x}F_{\dot\alpha}=\lambda_{\alpha}\,\lambda_{\beta}\left(\d x^{\alpha\dot\alpha}\wedge\d x^{\beta}{}_{\dot\alpha}+2\,\d\mathbf{x}^{\alpha\beta}\wedge\d_{x}g\right)\,.
\ee
From \eqref{Omegapull1}, we expect this to be quadratic in $\lambda_\alpha$. This follows from the relation
\begin{equation}
\lambda^\beta\p_{\alpha\beta}g = \lambda_\alpha (V-1) + \lambda^\beta\omega_{\alpha\beta}
\end{equation}
which follows by differentiating \eqref{GIT4} under the integral sign giving
\be\label{GIscalar}
V(\mathbf{x})=1+\frac{1}{2\,\pi\,\im}\int_{\P^1}\D\lambda\wedge\D\bar{\lambda}\,\left.\frac{\partial^2 f}{\partial q^2}\right|_{X}\,,
\ee
and
\be\label{GI1form}
\omega_{\alpha\beta}(\mathbf{x})=\frac{-1}{\sqrt{2}\,\pi\,\im}\int_{\P^1}\frac{\D\lambda\wedge\D\bar{\lambda}}{\la\lambda\,\xi\ra}\,\left.\frac{\partial^2 f}{\partial q^2}\right|_{X}\,\lambda_{(\alpha}\,\xi_{\beta)}\,,
\ee
which provide standard versions of the Penrose transform integrals in the context of minitwistor theory~\cite{Adamo:2017xaf}.
Thus the ASD 2-forms on space-time take the form
\be\label{GIT6}
\Sigma^{\alpha\beta}=\d x^{\alpha\dot\alpha}\wedge\d x^{\beta}{}_{\dot\alpha}+\sigma^{\alpha\beta}(\mathbf{x})\,, 
\ee
in terms of a finite deformation $\sigma^{\alpha\beta}(\mathbf{x})$ of the flat structure that is independent of $\lambda_\al$. Using \eqref{GIT4} (and the Schouten identity), this deformation is given directly by
\be\label{GIform}
\sigma^{\alpha\beta}(\mathbf{x})=\frac{1}{\pi\im}\int_{\P^1}\frac{\D\lambda\wedge\D\bar{\lambda}}{\la\lambda\,\xi\ra}\,\left.\frac{\partial^2 f}{\partial q^2}\right|_{X}\,\lambda_{\gamma}\,\xi^{(\alpha}\,\d\mathbf{x}^{\beta)\delta}\wedge\d\mathbf{x}^{\gamma}{}_{\delta}\,,
\ee
in terms of the twistor data.

 Here, the choice of $\xi_{\alpha}$ for the Cauchy kernel manifests itself as a gauge choice for the 1-form $\xi^{\alpha}\omega_{\alpha\beta}=\xi_{\beta}(1-V)$. This choice drops out (as it must) upon computing the dual exterior derivative of the 1-form, which is easily seen to obey the abelian monopole equation $\partial^{\alpha\gamma}\omega_{\alpha\beta}=\partial_{\beta}{}^{\gamma} V$ upon differentiation under the integral sign in \eqref{GIscalar} and \eqref{GI1form}.

In other words, the complex structure \eqref{GIT1} encodes the abelian monopole data of the Gibbons-Hawking metric through the Penrose transform. A straightforward index manipulation from \eqref{GIform} using \eqref{GIscalar} and \eqref{GI1form} then identifies
\be\label{GIform2}
\sigma^{\alpha\beta}(\mathbf{x})=(V(\mathbf{x})-1)\,\d\mathbf{x}^{\alpha\gamma}\wedge\d\mathbf{x}^{\beta}{}_{\gamma}+\d\mathbf{x}^{\alpha\beta}\wedge\omega(\mathbf{x})\,,
\ee
from which the tetrad and Gibbons-Hawking metric are directly recovered. Finally, comparing \eqref{GIscalar} with \eqref{Whitt3} allows us to identify the twistor data $f(q,\lambda)$ in terms of the radiative data $\varphi(u,\lambda,\bar\lambda)$ for the potential $V(\mathbf{x})$.

To compute amplitudes of momentum eigenstates, the last ingredient one needs is the spin-frame $H^{\dal}{}_{\dot\beta}(x,\lambda)$. Recalling \eqref{frameformula}, this is found by acting with the dual tetrad \eqref{GHdtet} on the holomorphic curves \eqref{GIhc}. A short calculation then yields
\be
H^{\dal}{}_{\dot\beta}(x,\lambda) = \frac{1}{\sqrt{V}}\left(\delta^{\dal}_{\dot\beta} - \frac{Y^{\gamma\dal}\lambda_\gamma}{2\,\pi\,\im}\int_{\P^1}\frac{\D\lambda'\wedge\D\bar\lambda'}{\la\lambda\,\lambda'\ra}\;Y_{\dot\beta}^\beta\lambda'_\beta\left.\frac{\partial^2 f}{\partial q^2}\right|_{X'}\right)\,.
\ee
The integral formula \eqref{GIscalar} ensures that this satisfies $H_{\dal\dot\beta}H^{\dal}{}_{\dot\gamma} = \eps_{\dot\beta\dot\gamma}$, hence giving an $\SL(2,\C)$-valued spin-frame on the bundle of dotted spinors over a Gibbons-Hawking space-time. 

Finally, note that although we have assumed radiative data for the Gibbons-Hawking metrics, the general ansatz \eqref{GIT1} actually captures solutions with `large' data including the gravitational instantons in Riemannian signature.


\section{MHV scattering}
\label{sec:mhv}

The maximal helicity violating (MHV) configuration for graviton scattering involves two negative helicity and arbitrarily many positive helicity external gravitons at tree-level. A geometric framework for this configuration is provided by viewing the collection of positive helicity external gravitons as a self-dual radiative background; the generating functional for tree-level MHV scattering is then given by a 2-point function of negative helicity gravitons on this self-dual background~\cite{Mason:2008jy}. In~\cite{Adamo:2021bej}, this picture was used to provide a first-principles derivation of the Hodges formula for gravitational MHV scattering in Minkowski space by taking the SD background in the generating functional to be a superposition of positive helicity gravitons in flat space.

However, by instead considering a fixed, non-linear SD radiative background \emph{and} some number of positive helicity gravitons on this background, we easily obtain a generating functional for MHV graviton amplitudes on any SD radiative space-time. In this section, we describe this setup and its lift to twistor space, where the perturbative expansion of the generating functional is achieved using a matrix tree theorem, resulting in formulae for the MHV amplitudes at arbitrary multiplicity.


Here we review and extend~\cite{Mason:2008jy,Adamo:2021bej} to a SD space-time $\CM$ whose metric is given by a finite deformation of the metric on some other SD radiative space-time, $\cM$. In practice, this finite deformation will take the form of a superposition of positive helicity gravitons on $\cM$, so the situation we want to describe can be expressed heuristically as $\CM\simeq\cM\oplus(\mbox{gravitons})$. Since the gravitons on $\cM$ are themselves (linear) SD radiative perturbations, it follows that $\CM$ will also be a SD radiative space-time. Such a scenario is well-described by the twistor theory of SD radiative space-times, and the resulting K\"ahler potential (or Plebanski scalar) on $\CM$ can be computed explicitly via twistor theory.

In \S\ref{sec:sigma} we introduce the background field framework, and show how the K\"ahler scalar is obtained as the action of a chiral twistor sigma model for the holomorphic curves whose moduli space reconstructs the space-time in Penrose's nonlinear graviton construction.  In \S\ref{sec:mhv-gen} we review~\cite{Adamo:2021bej}, where it is shown that the integral of this Kahler scalar provides the second perturbation of the Einstein-Hilbert action needed to provide a generating function for the MHV amplitude.  These two frameworks are combined in \S\ref{sec:mhv-amp} to compute the MHV amplitude on some given background firstly as a tree-level correlation function \eqref{MHVexpand1} in the chiral twistor sigma model evaluated on the background.  We then explain how a matrix-tree theorem can be used to resum these diagrams into certain reduced determinants leading to formula \eqref{MHV2} with momentum eigenstates and made more explicit for self-dual plane waves in equation \eqref{MHVsdpw}.  Finally, in \S\ref{sec:mhv-cons} we show that the resulting formulae pass basic consistency tests.

\subsection{Self-dual deformations and twistor sigma models}
\label{sec:sigma}


\paragraph{Deformations of twistor space:} Assume that $\cM$ is an SD radiative space-time with twistor space $\CPT$. Let $\nbar = \dbar + \{\sh,\cdot\}$ be the complex structure on $\CPT$. We deform $\cM$ and its metric to a new SD Ricci-flat space-time $\CM$, thereby changing $\sh\mapsto\sh+h$ for some finite perturbation $h\in\Omega^{0,1}(\CPT,\cO(2))$, deforming the complex structure. The space-time $\cM$, its twistor space $\CPT$ and complex structure (specified by $\sh$) will be referred to as the `background.'

Let $x^{\al\dal}$ denote coordinates on $\cM$ which we continue to use as coordinates on $\CM$. Let $\sX:\mu^{\dal}=\sF^{\dal}(x,\lambda)$ be the background holomorphic twistor curves in $\CPT$. These solve
\be\label{curvepde}
\dbar|_x \sF^{\dal} = \frac{\p\sh}{\p\mu_{\dal}}\biggr|_{\sX}\,,
\ee
where $\dbar|_x$ denotes the antiholomorphic exterior derivative along the curve labeled by $x$. When the complex structure on $\CPT$ is deformed, the holomorphic curves are deformed accordingly. Denote the deformed curves by $X:\mu^{\dal} = F^{\dal}(x,\lambda)$, which we assume to be given by a finite perturbation
\be\label{pertcurve}
F^{\dal}(x,\lambda) = \sF^{\dal}(x,\lambda) + m^{\dal}(x,\lambda)\,.
\ee
These deformed curves must satisfy
\be\label{curvepde1}
\begin{split}
\dbar|_xF^{\dal} &= \dbar|_x\sF^{\dal} + \dbar|_xm^{\dal} =  \frac{\p\sh}{\p\mu_{\dal}}\biggr|_{X} +  \frac{\p h}{\p\mu_{\dal}}\biggr|_{X}\\
\implies \dbar|_xm^{\dal} &= \frac{\p h}{\p\mu_{\dal}}\biggr|_{X} + \frac{\p\sh}{\p\mu_{\dal}}\biggr|_{X} - \frac{\p\sh}{\p\mu_{\dal}}\biggr|_{\sX}\,,
\end{split}
\ee
having used \eqref{curvepde} to get the second line.

To fix a solution, impose the boundary conditions
\be\label{bdry}
F^{\dal}(x,\kappa_1)=\sF^{\dal}(x,\kappa_1)\equiv z^{\dal}\,,\qquad F^{\dt\al}(x,\kappa_2)=\sF^{\dt\al}(x,\kappa_2)\equiv \tilde z^{\dt\al}
\ee
at a pair of points $\kappa_1,\kappa_2$ along the curves. In writing this, we have assumed that the choice of $\kappa_1,\kappa_2$ avoids any singularities of $\sF^{\dal}$; this will turn out to be a reasonable assumption in the context of later physical calculations. The parameters $z^{\dal},\tilde z^{\dt\al}$ define complex coordinates on $\CM$. Using $\kappa_1,\kappa_2$ as a spinor basis, we can also write $\lambda_\al=\lambda_1\,\kappa_{1\,\al}+\lambda_2\,\kappa_{2\,\al}$. With the boundary conditions \eqref{bdry}, one can set
\be\label{mtoM}
m^{\dal}(x,\lambda) = \lambda_1\,\lambda_2\,M^{\dal}(x,\lambda)
\ee
for some field $M^{\dal}$ which is homogeneous of degree $-1$ in $\lambda$ and regular at $\lambda=\kappa_1,\kappa_2$.

As before, we have the equivalence
\be
\Sigma(x,\lambda) = \d_xF^{\dal}\wedge\d_xF_{\dal} = \lambda_1^2\,\Sigma^{11}(x)+2\,\lambda_1\,\lambda_2\,\Sigma^{12}(x) + \lambda_2^2\,\Sigma^{22}(x)\,,
\ee
where $\Sigma^{\al\beta}$ now denote the ASD 2-forms on the deformed SD space-time $\CM$. Using boundary conditions \eqref{bdry}, evaluating both sides at $\lambda=\kappa_1,\kappa_2$ yields
\be\label{sigma1122}
\Sigma^{11} = \d z^{\dal}\wedge\d z_{\dal}\,,\qquad\Sigma^{22} = \d\tilde z^{\dt\al}\wedge\d\tilde z_{\dt\al}\,.
\ee
From $F^{\dal}=\sF^{\dal}+\lambda_1\lambda_2M^{\dal}$, $\Sigma^{12}$ can be computed in two ways:
\be
\begin{split}
\Sigma^{12} = \frac{1}{2}\,\frac{\p\Sigma}{\p\lambda_2}\,\biggr|_{\lambda=\kappa_1}& = \frac{1}{2}\,\frac{\p\Sigma}{\p\lambda_1}\,\biggr|_{\lambda=\kappa_2}\\
\implies\Sigma^{12}=\d z^{\dal}\wedge\d(K_{\dal}+M_{\dal}(x,\kappa_1))& = -\d\tilde z^{\dt\al}\wedge\d(K_{\dt\al}-M_{\dt\al}(x,\kappa_2))\,,
\end{split}
\ee
where $K_{\dal} = \p K/\p z^{\dal}$, $K_{\dt\al} = \p K/\p \tilde z^{\dt\al}$ are first derivatives of a background K\"ahler potential $K$ on $\cM$, as follows from applying \eqref{2forms*}, \eqref{Plebform1} to the background. Equality of the two representations of $\Sigma^{12}$ provides the existence of a K\"ahler potential $\Omega$ on $\CM$, satisfying
\be\label{omegaders}
\frac{\p\Omega}{\p z^{\dal}} = \frac{\p K}{\p z^{\dal}} + M_{\dal}(x,\kappa_1)\,,\qquad\frac{\p\Omega}{\p\tilde z^{\dt\al}} = \frac{\p K}{\p\tilde z^{\dt\al}} - M_{\dt\al}(x,\kappa_2)\,,
\ee
with
\be\label{sigma12}
\Sigma^{12} = \Omega_{\dal\dt\al}\,\d z^{\dal}\wedge\d\tilde z^{\dt\al}\,.
\ee
The PDE determining $\Omega$ follows from the simplicity condition $\Sigma\wedge \Sigma=0$
\be
\Sigma^{11} = \Sigma\bigr|_{\lambda=\kappa_1} = \frac{1}{2}\,\frac{\p^2\Sigma}{\p\lambda_1^2}\biggr|_{\lambda=\kappa_2}\implies \Omega_{\dal\dt\al}\,\Omega^{\dal\dt\al} = 2\,.
\ee
This is precisely the first heavenly equation \eqref{Heqn}, now implicitly written in terms of a finite deformation around a non-trivial SD background.

\medskip

The PDE \eqref{curvepde1} governing the deformed holomorphic curves can be obtained as the equation of motion for a chiral sigma model governing rational maps from a Riemann sphere to twistor space $\CPT$:
\be
S_\Omega[m] = \frac{1}{\hbar}\int_{\P^1}\frac{\D\lambda}{\lambda_1^2\,\lambda_2^2}\left[\left[m\;\dbar|_x m\right] + 2\,h|_{X}+2\,\biggl(\sh|_X-\sh|_{\sX} - \frac{\p\sh}{\p\mu^{\dal}}\biggr|_{\sX}m^{\dal}\biggr)\right]\,,
\ee
where $\hbar$ is a formal book-keeping parameter. This model is a generalization of the `twistor sigma model of the first kind' introduced in~\cite{Adamo:2021bej} to a non-trivial SD background. By Taylor expanding $\sh|_X = \sh(\sF+m,\lambda)$ around $\mu^{\dal}=\sF^{\dal}$, this sigma model action can also be written as
\begin{multline}\label{pertac}
S_\Omega[m] = \frac{1}{\hbar}\int_{\P^1}\frac{\D\lambda}{\lambda_1^2\,\lambda_2^2}\left(\left[m\;\dbar|_x m\right] +2\,h|_X+\sum_{p=2}^\infty\frac{2}{p!}\,\frac{\p^p\sh}{\p\mu^{\dal_1}\cdots\p\mu^{\dal_p}}\biggr|_\sX\,m^{\dal_1}\cdots m^{\dal_p}\right)\,.
\end{multline}
Evaluated on-shell, the sigma model action computes the K\"ahler potential of $\CM$:

\begin{propn}\label{Kahlerprop}
The on-shell action $S_\Omega$ computes the K\"ahler potential $\Omega$ of $\CM$ as a perturbation of the K\"ahler potential $K$ of the background $\cM$,
\be\label{kpot}
\Omega = K - \frac{\hbar}{4\pi\im} S_\Omega[m]\bigr|_\text{\emph{on-shell}}\,,
\ee
where `on-shell' means evaluated on solutions of the equations of motion. 
\end{propn}
\proof Computing $\p S_\Omega/\p z^{\dal}$ gives
\begin{multline}
\hbar\,\frac{\p S_\Omega}{\p z^{\dal}} = \int_{X}\frac{\D\lambda}{\lambda_1^2\,\lambda_2^2}\left(\left[\frac{\p m}{\p z^{\dal}}\;\dbar|_x m\right] + \left[m\;\dbar|_x\frac{\p m}{\p z^{\dal}}\right]\right. \\
\left.+2\,\biggl(\frac{\p h}{\p\mu^{\dot\beta}}\biggr|_{X} + \frac{\p\sh}{\p\mu^{\dot\beta}}\biggr|_{X}-\frac{\p\sh}{\p\mu^{\dot\beta}}\biggr|_{\sX}\biggr)\,\frac{\p F^{\dot\beta}}{\p z^{\dal}}-2\,\frac{\p}{\p z^{\dal}}\biggl(\frac{\p\sh}{\p\mu^{\dot\beta}}\biggr|_\sX\biggr)\,m^{\dot\beta}\right)\,,
\end{multline}
having used $F^{\dal}=\sF^{\dal}+m^{\dal}$ to simplify the result. Evaluating this on a solution to the equation of motion \eqref{curvepde1}, we find
\begin{multline}
\hbar\,\frac{\p S_\Omega}{\p z^{\dal}} = \int_{X}\frac{\D\lambda}{\lambda_1^2\,\lambda_2^2}\left(\left[\frac{\p m}{\p z^{\dal}}\;\dbar|_x m\right] + \left[m\;\dbar|_x\frac{\p m}{\p z^{\dal}}\right]\right. \\
\left.-2\,\dbar|_x m_{\dot\beta}\left(\frac{\p \sF^{\dot\beta}}{\p z^{\dal}}+\frac{\p m^{\dot\beta}}{\p z^{\dal}}\right) + 2\,\dbar|_x\biggl(\frac{\p\sF_{\dot\beta}}{\p z^{\dal}}\biggr)\,m^{\dot\beta}\right)\,,
\end{multline}
having also used \eqref{curvepde} to simplify the last term.

Using the boundary conditions \eqref{mtoM} leaves
\be
\hbar\,\frac{\p S_\Omega}{\p z^{\dal}} = \int_{X}\D\lambda\left(\dbar|_x\!\left[M\;\frac{\p M}{\p z^{\dal}}\right] - \frac{2}{\lambda_1\,\lambda_2}\,\dbar|_x\biggl(\frac{\p \sF^{\dot\beta}}{\p z^{\dal}}\,M_{\dot\beta}\biggr)\right)\,,
\ee
on-shell. The first term is a total derivative and drops out, while the second can be computed using integration by parts and the fact that $\dbar|_x\,\lambda_i^{-1} = 2\pi\im\, \bar\delta(\lambda_i)$. We thus find
\be
\begin{split}
\frac{\hbar}{4\pi\im}\,\frac{\p S_\Omega}{\p z^{\dal}} &= \frac{\p \sF^{\dot\beta}}{\p z^{\dal}}\,M_{\dot\beta}\biggr|_{\lambda=\kappa_2} -  \frac{\p \sF^{\dot\beta}}{\p z^{\dal}}\,M_{\dot\beta}\biggr|_{\lambda=\kappa_1}\\
&= -M_{\dal}(x,\kappa_1) = \frac{\p K}{\p z^{\dal}} -  \frac{\p \Omega}{\p z^{\dal}}\,,
\end{split}
\ee
having applied \eqref{bdry} and \eqref{omegaders} to get the second line. One can similarly compute $\p S_\Omega/\p\tilde z^{\dt\al}$, thereby obtaining the K\"ahler potential \eqref{kpot}, as claimed. \qed


\subsection{MHV generating functional}\label{sec:mhv-gen}

In this subsection we review the derivation of the MHV generating functional \eqref{TliftMHV} from \cite{Adamo:2021bej}.
The computation of MHV amplitudes is facilitated by means of Plebanski's chiral formulation of general relativity~\cite{Plebanski:1977zz}. Let $\CM$ be the space-time manifold, with a coframe $e^{\al\dal}$ comprising a tetrad for its metric; the ASD 2-forms on $\CM$ are spanned by $\Sigma^{\al\beta} = e^{\al\dal}\wedge e^\beta{}_{\dal}$. Plebanski's action can be expressed in terms of $\Sigma^{\al\beta}$ and an ASD spin connection $\Gamma_{\al\beta}$:
\be\label{plebac}
S[e,\Gamma] = \frac{1}{\kappa^2}\,\int_\CM\Sigma^{\al\beta}\wedge(\d\Gamma_{\al\beta}+\Gamma_\al{}^{\gamma}\wedge\Gamma_{\gamma\beta})\,.
\ee
The equations of motion of this action state that the connection is Levi-Civita and the metric is Ricci-flat (cf., \cite{Plebanski:1977zz,Frauendiener:1990,Capovilla:1991qb,Krasnov:2009pu}). 

Following section~\ref{sec:sigma}, a generating functional for the MHV amplitude on a SD background $\cM$ is given by the 2-point function of ASD perturbations on $\CM$, viewed as incorporating an infinite number of positive helicity gravitons on $\cM$. Let $\Sigma_0^{\alpha\beta}$ be the ASD 2-forms of the background $\cM$ while $\Sigma^{\alpha\beta}$ be the ASD 2-forms on $\CM$; the ASD perturbations on $\cM$ are represented by perturbations $\gamma^{\alpha\beta}$ to the flat background ASD spin connection (i.e., to $\Gamma_0^{\alpha\beta}=0$). 

The generating functional of MHV amplitudes is then given by the interaction term in \eqref{plebac}~\cite{Mason:2008jy}:
\be\label{mhvgen}
\mathcal{G}(1,2) = \frac{1}{\kappa^2}\int_\CM\Sigma^{\al\beta}\wedge\gamma_{1\,\al}{}^\gamma\wedge\gamma_{2\,\gamma\beta}\,.
\ee
The MHV tree-amplitudes themselves are obtained by performing a classical perturbative expansion of this generating functional around $\cM$. One downside of this generating functional is that it is not manifestly diffeomorphism invariant, since the negative helicity gravitons are encoded by perturbations of the spin connection. This can be remedied by recasting the generating functional in terms of the K\"ahler potential of $\CM$, and the perturbative expansion is then operationalized by using the sigma model \eqref{pertac} introduced in the previous section. 


\paragraph{Lift to twistor space:} As discussed in section \ref{sec:radtwistor}, SD and ASD perturbations to the SD radiative background $\cM$ can be represented as momentum eigenstates with momenta $k_i^{\al\dal} = \kappa_i^\al\,\tilde\kappa_i^{\dal}$. The ASD spin connection perturbations are
\be\label{negfields1}
\gamma_i^{\alpha\beta}(x) = \frac{2\,\tilde b_i^{\dot\alpha}}{[\tilde b_i\,i]}\,\kappa_i^\alpha\,\kappa_i^\beta\,\e^{\im\phi_i}\,\d_x\sF_{\dal}(x,\kappa_i)\,,\qquad\phi_i(x) \equiv [\sF(x,\kappa_i)\,i]\,,
\ee
for $i=1,2$. The constant spinors $\tilde{b}_{i}^{\dot\alpha}$ encode residual gauge (diffeomorphism) freedom in the representation of the spin connection perturbations and drop out of invariant quantities like Weyl curvature perturbations. Applying \eqref{incitetrad}, the wavefunctions \eqref{negfields1} are seen to satisfy the required linearized field equations $\d\gamma_{i}^{\alpha\beta}=\psi_{i}^{\alpha\beta\gamma\delta}\Sigma_{\gamma\delta}$ with curvatures $\psi_i^{\al\beta\gamma\delta}$ of the form \eqref{nhgrrep2}. 

The expressions for the ASD perturbations are simplified by using the complex coordinates $(z^{\dal},\tilde z^{\dt\al})$ of \eqref{bdry}; these are adapted to the ASD perturbations by identifying the spinor basis $\kappa_1$, $\kappa_2$ in \eqref{bdry} with the undotted momentum spinors of the negative helicity gravitons. In these coordinates \eqref{negfields1} become
\be\label{negfields2}
\gamma_1^{\alpha\beta}(z,\tilde z) = 2\,\frac{[\tilde b_1\,\d z]}{[\tilde b_1\,1]}\,\kappa_1^\alpha\,\kappa_1^\beta\,\e^{\im[z\,1]}\,,\qquad \gamma_2^{\alpha\beta}(z,\tilde z) = 2\,\frac{[\tilde b_2\,\d\tilde z]}{[\tilde b_2\,2]}\,\kappa_2^\alpha\,\kappa_2^\beta\,\e^{\im[\tilde z\,2]}\,,
\ee
which take on a form that closely resembles negative helicity graviton perturbations in flat space. Without loss of generality, these states can be normalised so that $[\tilde b_1\,1]=1=[\tilde b_2\,2]$.

Inserting these into the MHV generating functional \eqref{mhvgen} and using \eqref{sigma1122} and \eqref{sigma12} leads to
\begin{align}
\label{PlebGen}
\mathcal{G}(1,2)&= \frac{1}{4\,\kappa^2}\int_{\CM} \d^2z\, \d^2\tilde z\,  \, \Omega_{\dot\alpha\dot{\tilde\alpha}}\, \tilde b_1^{\dot\alpha}\,\tilde b_2^{\dot{\tilde\alpha}}\, \e^{\im [z\, 1 ]+\im[\tilde z\, 2 ] } \nonumber \\
&=\frac{\im}{2\,\kappa^2}\int_{\CM} \d^2z\, \d^2\tilde z\,  \, \Omega_{\dot\alpha} \,\tilde b_1^{\dot\alpha}\, \e^{\im [z\, 1 ]+\im[\tilde z\, 2 ] }\nonumber \\
&=-\frac{\la 1 \,2\ra^4}{\kappa^2}\int_{\CM} \d^2z\, \d^2\tilde z\,  \, \Omega \, \e^{\im [z\, 1 ]+\im[\tilde z\, 2 ] } \,,
\end{align}
where the second and third lines follow upon integrating by parts with respect to $\tilde{z}$ and $z$, respectively. In the final line, explicit factors of $\la1\,2\ra$ (previously normalised to $\la1\,2\ra=1$) have been reinstated using little group scalings, and $\Omega$ is the K\"ahler potential on $\CM$ encoding both the background metric as well as SD gravitons. Using proposition~\ref{Kahlerprop} gives
 \be\label{TliftMHV}
\mathcal{G}(1,2) = \frac{\hbar\,\la1\,2\ra^{4}}{4\pi\im\,\kappa^2}\int_{\CM}\d^{2}z\,\d^{2}\tilde{z}\,\e^{\im\,[z\,1]+\im\,[\tilde{z}\,2]}\,S_{\Omega}\,,
\ee
where $S_\Omega$ is the twistor sigma model \eqref{pertac} evaluated on-shell. Thus, the MHV generating functional -- and its perturbative expansion -- is entirely controlled by the on-shell action of the twistor sigma model.


\subsection{MHV amplitude}\label{sec:mhv-amp}
In \cite{Adamo:2021bej} \eqref{TliftMHV} was expanded perturbatively about flat space to give the Hodges formula for the gravity MHV amplitude.  Here we extend that construction so as to perturbatively expand around a generic  SD radiative background space-time $\cM$. To recover the $n$-point tree-level MHV amplitude on  $\cM$, $\mathcal{G}(1,2)$ must be expanded to order $n-2$ in $h$. This corresponds to extracting the $n-2$-linear piece of the on-shell twistor sigma model action $S_{\Omega}$. Since on-shell actions are computed by tree-level Feynman graphs, this multi-linear contribution can be expressed in terms of a tree-level correlation function in the QFT on $\P^1$ defined by the twistor sigma model. Letting $h=\sum_{i}\epsilon_i\,h_i$, we can naturally express this $n-2$th term in the expansion of the on-shell action $S_\Omega$ as a \emph{tree-correlator}
\begin{multline}\label{MHVexpand1}
\la1\,2\ra^{2}\,\left\la\prod_{i=3}^{n}V_i\right\ra^{0}_{\mathrm{tree}}=\left(\prod_{i}\frac{\partial}{\partial\epsilon_i}\right) \int_{\P^1}\frac{\la1\,2\ra^2\,\D\lambda}{\la\lambda\,1\ra^2\,\la\lambda\,2\ra^2}\Bigg(\left[m\;\dbar|_x m\right] +2\,\sum_{i=3}^{n}\epsilon_i\,h_i(\sF+m,\lambda) \\
\left.\left.+\sum_{p=2}^\infty\frac{2}{p!}\,\frac{\p^p\sh(\sF,\lambda)}{\p\mu^{\dal_1}\cdots\p\mu^{\dal_p}}\,m^{\dal_1}\cdots m^{\dal_p}\right)\right|_{\epsilon_i=0} 
\,.
\end{multline}
Here we will see that the expectation value denotes a sum of connected tree diagrams (i.e., diagrams of $O(\hbar^0)$) in the ``background'' $\P^1$ theory
\be\label{QFT1}
S[m]=\frac{1}{\hbar}\int_{\P^1}\frac{\D\lambda}{\la\lambda\,1\ra^2\,\la\lambda\,2\ra^2}\left(\left[m\;\dbar m\right] +\sum_{p=2}^\infty\frac{2}{p!}\,\frac{\p^p\sh(\sF,\lambda)}{\p\mu^{\dal_1}\cdots\p\mu^{\dal_p}}\,m^{\dal_1}\cdots m^{\dal_p}\right)\,,
\ee 
with vertex operators given by
\be\label{vertexop}
V_{i}:=2\,\int_{\P^1}\frac{\D\lambda_i}{\la\lambda_i\,1\ra^2\,\la\lambda_i\,2\ra^2}\,h_{i}(\sF+m,\lambda_i)\,,
\ee
encoding the positive helicity gravitons. It will be assumed that each individual graviton is represented by a momentum eigenstate:
\be\label{vertexop2}
h_{i}(Z)=\int_{\C^*} \frac{\d s_i}{s_i^3}\,\bar{\delta}^2(\kappa_i -s_i\,\lambda)\,\e^{\im\,s_i\,[\mu\,i]}\,,
\ee
with on-shell momentum $\kappa_i^{\alpha}\tilde{\kappa}_i^{\dot\alpha}$.

The reduced sigma model action \eqref{QFT1} depends only on the fixed SD radiative background $\cM$ through $\sh$. However, there are infinitely many terms of higher and higher valence in $m^{\dot\alpha}$ appearing in this action, as curves corresponding to points in $\CM$ are expanded around curves corresponding to points in $\cM$. The quadratic term in this expansion simply defines the kinetic term of the action, while the contribution of all higher-order terms to the correlator \eqref{MHVexpand1} can be accounted for by absorbing them into \emph{background} vertex operators:
\be\label{bvo1}
U^{(p)}:=\frac{2}{p!}\,\int_{\P^1}\frac{\D\lambda}{\la\lambda\,1\ra^2\,\la\lambda\,2\ra^2}\,\frac{\p^p\sh(\sF,\lambda)}{\p\mu^{\dal_1}\cdots\p\mu^{\dal_p}}\,m^{\dal_1}\cdots m^{\dal_p}\,, \qquad p>2\,.
\ee
Taking into account the relationship \eqref{radata} between $\sh$ and the characteristic radiative data of $\cM$, these background vertex operators can be written equivalently as:
\be\label{bvo2}
U^{(p)}=\frac{2}{p!}\,\int_{\P^1}\frac{\D\lambda\wedge\D\bar{\lambda}}{\la\lambda\,1\ra^2\,\la\lambda\,2\ra^2}\,N^{(p-2)}(u,\lambda,\bar\lambda)\,[m\,\bar{\lambda}]^{p}\,,
\ee
where $u=[\sF\,\bar{\lambda}]$ and $N^{(k)}:=-\partial^{k+1}_{u}\tilde{\bigma}^0$ is the $k^{\mathrm{th}}$ $u$-derivative of the news function of $\cM$.

Since these background vertex operators do not introduce any new powers of $h$, arbitrarily many of them can be brought down into the correlation function \eqref{MHVexpand1}. Thus, the problem of computing the MHV amplitudes of $n-2$ positive helicity gravitons on $\cM$ is reduced to computing the connected tree-level correlation function
\be\label{Corr1}
\sum_{t=0}^{\infty}\sum_{p_1,\ldots,p_t}\left\la\prod_{i=3}^{n} V_{i}\,\prod_{\m=1}^{t}U^{(p_\m)}\right\ra^{0}_{\mathrm{tree}}\,,
\ee
in the free but background coupled QFT on $\P^1$ defined by
\be\label{QFT2}
S[m]=\int_{\P^1}\frac{\D\lambda}{\la\lambda\,1\ra^2\,\la\lambda\,2\ra^2}\left[m^{\dot\alpha}\,\left(\epsilon_{\dot\beta\dot\alpha}\dbar+\frac{\p^2\sh}{\p\mu^{\dal}\p\mu^{\dot\beta}}(\sF,\lambda)\right) m^{\dot\beta}\right]\,.
\ee
In \eqref{Corr1}, the second sum is over all $p_1,\ldots,p_t>2$; that is, over all possible `valences' of the background vertex operators. Physically, the role of these background vertex operators is clear: the $t>0$ terms in \eqref{Corr1} correspond to tail terms in the MHV amplitude which arise because of the failure of Huygens' principle for graviton scattering in any curved background~\cite{Friedlander:2010eqa,Harte:2013dba}.

\medskip

The computation of \eqref{Corr1} now proceeds making use of the OPE of the $m^{\dot\alpha}$ fields defined by the quadratic action \eqref{QFT2}:
\be\label{OPE}
m^{\dot\alpha}(\lambda_i)\,m^{\dot\beta}(\lambda_{j})\sim \frac{H^{\dot\alpha}{}_{\dot\gamma}(x,\lambda_i)\,H^{\dot\beta\dot\gamma}(x,\lambda_j)}{\la\lambda_{i}\,\lambda_{j}\ra}\,\la\lambda_i\,1\ra\,\la\lambda_{i}\,2\ra\,\la\lambda_j\,1\ra\,\la\lambda_{j}\,2\ra\,.
\ee
This is obtained from dressing the OPE of the $\sh=0$ free theory with factors of dotted spin-frames. Since the spin-frames satisfy \eqref{Heqn*}, this dressed OPE satisfies the equation of motion that follows from varying \eqref{QFT2}. The right-hand-side of this expression acts as a Green's function for the complex structure on $\CPT$ pulled back to $\sX$, and the appearance of the momentum spinors $\kappa_{1}^{\alpha}$, $\kappa_{2}^{\alpha}$ in the OPE represents a gauge choice for the inverse of the $\dbar$-operator acting on sections of $\cO(1)\rightarrow\P^1$.
While $m^{\dot\alpha}$ is valued in $\cO(1)$, the boundary conditions \eqref{mtoM} ensure that it has no zero modes, so all insertions of $m^{\dot\alpha}$ in the correlator must be Wick contracted away using \eqref{OPE} to obtain non-vanishing contributions.

\begin{figure}[t]
\centering
\includegraphics[scale=.85]{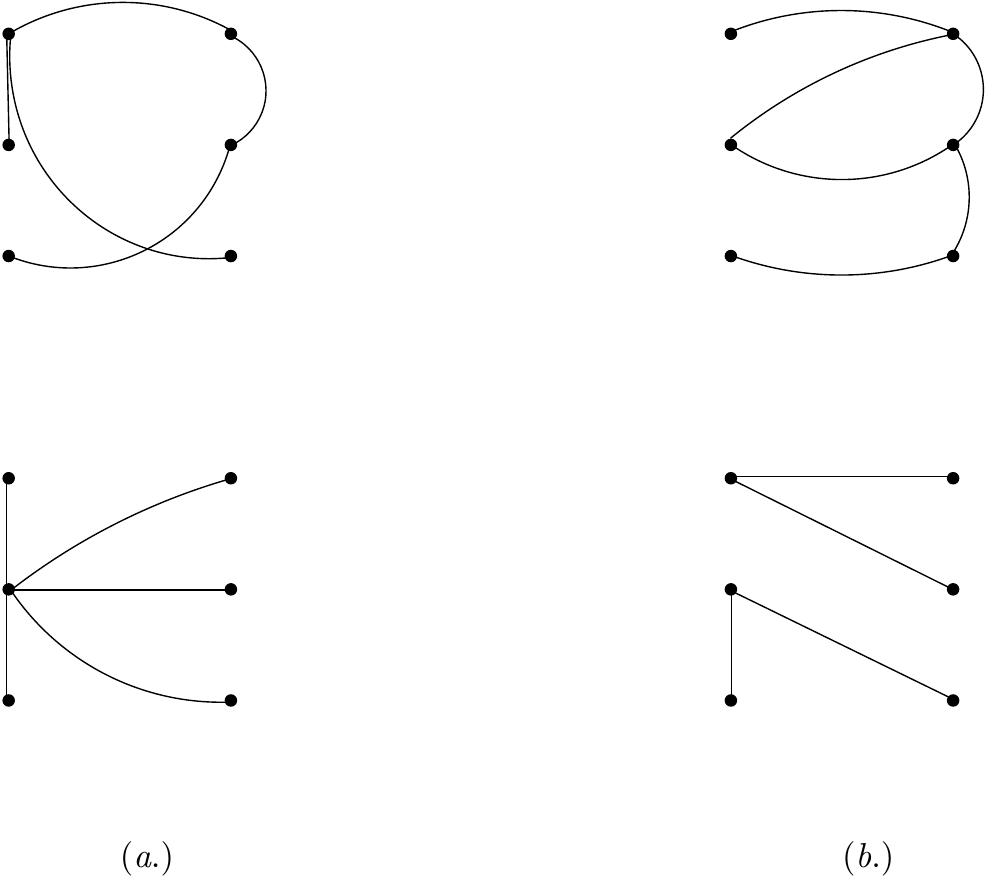}
\caption{Some examples of graphs included in the sum of connected trees on six vertices (\textit{a}.), and some graphs which are excluded from the sum (\textit{b}.) because they include loops and disconnected pieces respectively.}
\label{CTgraph}
\end{figure}

As a warm-up, it is illustrative to first consider the $t=0$ term in \eqref{Corr1}, which contains only external graviton vertex operators as would be the case on a flat background as in \cite{Adamo:2021bej}. In this case, the computation boils down to summing all connected tree graphs on $n-2$ vertices, with the weights assigned to each edge given by the OPE \eqref{OPE} acting between two vertex operators, say $i$ and $j$:
\be\label{GVOp}
-s_{i}s_{j}\,\frac{[\![i\,j]\!]}{\la\lambda_{i}\,\lambda_{j}\ra}\,\la\lambda_i\,1\ra\,\la\lambda_{i}\,2\ra\,\la\lambda_j\,1\ra\,\la\lambda_{j}\,2\ra\,,
\ee
which arises from \eqref{OPE} on computing the OPE of a pair of gravitons given by \eqref{vertexop2}. Here, the double-square bracket notation stands for contraction of the background dressed dotted momentum spinors defined in \eqref{Kgrgen}:
\be\label{bdshel}
[\![i\,j]\!]:=\tilde{K}^{\dot\alpha}_{i}\,\tilde{K}_{\dot\alpha\,j}=\tilde{\kappa}_{\dot\beta\,i}\,\tilde{\kappa}_{\dot\gamma\,j}\,H^{\dot\beta\dot\alpha}(x,\kappa_i)\,H^{\dot\gamma}{}_{\dot\alpha}(x,\kappa_j)\,.
\ee
Note that in the first instance, the OPE \eqref{OPE} means that the frames will be evaluated at $\lambda_i$ and $\lambda_j$, but the holomorphic delta functions in the momentum eigenstates \eqref{vertexop2} mean that they can be evaluated at $\kappa_i$ and $\kappa_j$, respectively.

Figure~\ref{CTgraph} illustrates some examples of graphs included and excluded from the sum of connected trees for the case $n=8$. The sum of these tree-level Feynman diagrams is then accomplished using the weighted matrix-tree theorem (cf., \cite{Stanley:1999,vanLint:2001,Stanley:2012}):
\be\label{MTT1}
\left\la\prod_{j=3}^{n} V_{j}\right\ra^{0}_{\mathrm{tree}}=\int_{(\P^1\times\C^*)^{n-2}}\left|\cL^{i}_{i}\right| \prod_{j=3}^{n}\frac{\bar\delta^2(\kappa_j-s_j\lambda_j)\,\e^{\im\,s_j\,[\sF(x,\lambda_j)\,j]}}{\la\lambda_j\,1\ra^2\,\la\lambda_j\,2\ra^2}\,\D\lambda_j\,\frac{\d s_j}{s_j^3}\,,
\ee
where $\cL$ is the weighted Laplacian matrix whose off-diagonal entries are given by \eqref{GVOp} and diagonal entries are
\be\label{wLap}
\cL_{ii}=s_{i}\,\sum_{j\neq i}s_{j}\,\frac{[\![i\,j]\!]}{\la\lambda_{i}\,\lambda_{j}\ra}\,\la\lambda_i\,1\ra\,\la\lambda_{i}\,2\ra\,\la\lambda_j\,1\ra\,\la\lambda_{j}\,2\ra\,,
\ee
and $|\cL^{i}_{i}|$ denotes the minor of $\cL$ obtained by removing the row and column corresponding to some external graviton $i$. The weighted matrix-tree theorem ensures that the equality \eqref{MTT1} is independent of the choice of $i\in\{3,\ldots,n\}$.

To make the connection to known formulae for graviton scattering in vacuum transparent, it is useful to rewrite \eqref{MTT1} by taking out a factor of $\la\lambda_i\,1\ra\,\la\lambda_{i}\,2\ra$ from each row and column of the determinant to give
\be\label{MTT2}
\left\la\prod_{j=3}^{n} V_{j}\right\ra^{0}_{\mathrm{tree}}=\int_{(\P^1\times\C^*)^{n-2}}\frac{|\HH^{i}_{i}|}{\la\lambda_i\,1\ra^2\,\la\lambda_i\,2\ra^2}\, \prod_{j=3}^{n} \bar\delta^2(\kappa_j-s_j\lambda_j)\,\e^{\im\,s_j\,[\sF(x,\lambda_j)\,j]}\,\D\lambda_j\,\frac{\d s_j}{s_j^3}\,,
\ee
where $\HH$ is the matrix with entries
\be\label{HHmat}
\HH_{ij}= -s_{i}s_{j}\,\frac{[\![i\,j]\!]}{\la\lambda_{i}\,\lambda_{j}\ra}\,, \quad i\neq j\,, \qquad \HH_{ii}=s_{i}\,\sum_{j\neq i}s_{j}\,\frac{[\![i\,j]\!]}{\la\lambda_{i}\,\lambda_{j}\ra}\,\frac{\la1\,\lambda_j\ra\,\la2\,\lambda_{j}\ra}{\la1\,\lambda_i\ra\,\la2\,\lambda_{i}\ra}\,,
\ee
which is a background-dressed version of the matrix appearing in Hodges' formula for MHV graviton scattering in flat space~\cite{Hodges:2012ym}. The $\lambda_j$ and $s_j$ integrals can also be performed explicitly against the delta functions. These simply set $\lambda_j=\kappa_j$ in the end.

\medskip

Now consider a generic term with fixed $t>0$ and $p_1,\ldots,p_t>2$ in \eqref{Corr1}. Each insertion of a background vertex operator $U^{(p_\m)}$ comes with at least 3 insertions of $m^{\dot\alpha}$, all of which must be Wick contracted for the correlator to be non-vanishing. These $m$-insertions can contract against the other background vertex operators or the external graviton vertex operators; it is easy to see that such Wick contractions cannot saturate the $m$-insertions via tree diagrams unless $t\leq n-4$, providing an upper bound on the number of background vertex insertions.

\begin{figure}[t]
\centering
\includegraphics[scale=.90]{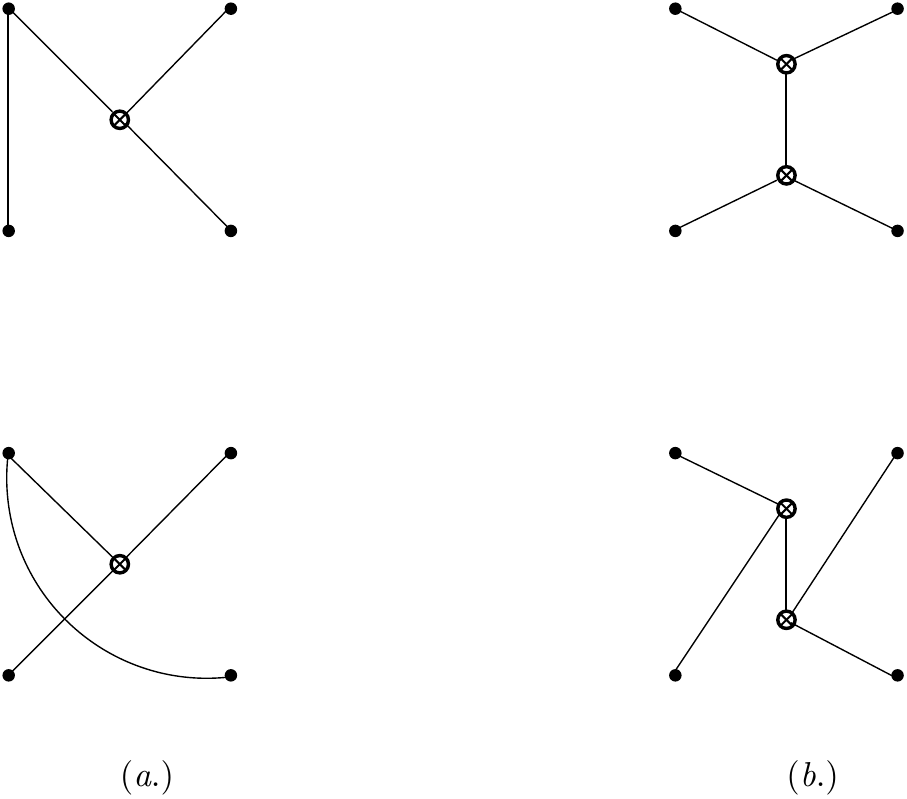}
\caption{Examples of connected tree graphs with background vertex operators contributing when $n-2=4$. In case (\textit{a}.), there is a single ($t=1$) background vertex operator (denoted by a crossed vertex) with valence $p_1=3$; in case (\textit{b.}) there are two background vertices ($t=2$) both with valence $p_1=p_2=3$.}
\label{BTgraph}
\end{figure}

With this in mind, we must now sum over all connected tree diagrams on $n+t-2$ vertices where the valence of the $t$ background vertices is fixed in each term in this sum. In particular, the vertex corresponding to $U^{(p_\m)}$ must have valence $p_\m$ in every spanning tree; any tree for which this vertex does not have this particular valence cannot contribute to the correlator, since there are either too many or too few $m^{\dot\alpha}$ insertions to saturate the corresponding Wick contractions. Some of the graphs contributing to the $t=1,2$ cases for $n=6$ are illustrated in Figure~\ref{BTgraph}. The weighted matrix tree theorem once again enables us to perform this sum, albeit in a slightly more subtle fashion.

First, consider the sum over all spanning trees without the fixed-valence restriction on the background vertex operators. After removing a factor of $\la\lambda_{A}\,1\ra\,\la\lambda_A\,2\ra$, for $A\in\{i=3,\ldots,n\}\cup\{\m=1,\ldots,t\}$, from every row and column of the weighted Laplacian matrix, this sum is computed by the reduced determinant $|\cH^{i}_{i}|$, where $\cH$ is the $(n+t-2)\times(n+t-2)$ matrix with block decomposition
\be\label{cHmat}
\cH=\left(\begin{array}{c c}
               \HH & \mathfrak{h} \\
               \mathfrak{h}^{\mathrm{T}} & \T
               \end{array}\right)\,,
\ee
and entries
\be\label{cHH}
\HH_{ij}= -s_{i}s_{j}\,\frac{[\![i\,j]\!]}{\la\lambda_{i}\,\lambda_{j}\ra}\,, \quad i\neq j
\ee
\begin{equation*}
 \HH_{ii}=s_{i}\,\sum_{j\neq i}s_{j}\,\frac{[\![i\,j]\!]}{\la\lambda_{i}\,\lambda_{j}\ra}\,\frac{\la1\,\lambda_j\ra\,\la2\,\lambda_{j}\ra}{\la1\,\lambda_i\ra\,\la2\,\lambda_{i}\ra}+s_i\,\sum_{\m=1}^{t}\varepsilon_\m\,\frac{[\![i\,\bar{\lambda}_\m]\!]}{\la\lambda_i\,\lambda_\m\ra}\,\frac{\la1\,\lambda_\m\ra\,\la2\,\lambda_\m\ra}{\la1\,\lambda_i\ra\,\la2\,\lambda_i\ra}\,,
\end{equation*}
\begin{equation*}
\mathfrak{h}_{i\m}=-s_i \,\varepsilon_\m\,\frac{[\![i\,\bar{\lambda}_\m]\!]}{\la\lambda_i\,\lambda_\m\ra}\,, \qquad \T_{\m\mathrm{n}}=-\varepsilon_\m \varepsilon_{\mathrm{n}}\,\frac{[\![\bar{\lambda}_\m\,\bar{\lambda}_{\mathrm{n}}]\!]}{\la\lambda_\m\,\lambda_{\mathrm{n}}\ra}\,, \quad \m\neq\mathrm{n}\,,
\end{equation*}
\begin{equation*}
\T_{\m\m}=\varepsilon_\m\,\sum_{i=1}^{n}s_i\,\frac{[\![\bar{\lambda}_\m\,i]\!]}{\la\lambda_\m\,\lambda_i\ra}\,\frac{\la1\,\lambda_i\ra\,\la2\,\lambda_i\ra}{\la1\,\lambda_\m\ra\,\la2\,\lambda_\m\ra}+\varepsilon_\m\,\sum_{\mathrm{n}\neq\m}\varepsilon_{\mathrm{n}}\,\frac{[\![\bar{\lambda}_\m\,\bar{\lambda}_{\mathrm{n}}]\!]}{\la\lambda_\m\,\lambda_{\mathrm{n}}\ra}\,\frac{\la1\,\lambda_{\mathrm{n}}\ra\,\la2\,\lambda_{\mathrm{n}}\ra}{\la1\,\lambda_\m\ra\,\la2\,\lambda_\m\ra}\,.
\end{equation*}
Here, a set of $t$ dummy parameters $\varepsilon_1,\ldots,\varepsilon_t$ have been introduced, whose role is to count the valence of the background vertices in each tree diagram. 

As it stands, $|\cH^{i}_{i}|$ over-counts the trees we are interested in, as there is no restriction on the valence of the background vertices. A generic term in the expansion of this determinant will be a homogeneous polynomial in the $\varepsilon_\m$, with the homogeneity in $\varepsilon_\m$ corresponding to the valence of the $\m^{\mathrm{th}}$ background vertex. Thus, the problem reduces to extracting the terms in $|\cH^i_i|$ proportional to $\prod_{\m=1}^{t}\varepsilon_{\m}^{p_\m}$; this is given simply by
\be\label{homogterm}
\left.\left(\prod_{\m=1}^{t}\frac{1}{p_\m!}\frac{\partial^{p_\m}}{\partial\varepsilon_\m^{p_\m}}\right)|\cH^{i}_{i}|\right|_{\varepsilon_1=\cdots=\varepsilon_t=0}\,.
\ee
Applying this mechanism to every term in \eqref{Corr1} and combining with \eqref{MHVexpand1}, \eqref{TliftMHV} enables the computation of the MHV amplitude with an arbitrary number of positive helicity external gravitons.

\medskip

Collecting all the pieces gives an all-multiplicity formula for the MHV amplitude:
\begin{multline}\label{MHV1}
\cM_{n,1}=\kappa^{n-2}\sum_{t=0}^{n-4}\sum_{p_1,\ldots,p_t}\int\d^{2}z\,\d^{2}\tilde{z}\,\frac{\la1\,2\ra^6}{\la1\,\lambda_i\ra^2\,\la2\,\lambda_i\ra^2}\,\e^{\im\,[z\,1]+\im\,[\tilde{z}\,2]}\,\left.\left(\prod_{\m=1}^{t}\frac{1}{p_\m!}\,\frac{\partial^{p_\m}}{\partial\varepsilon_\m^{p_\m}}\right)|\cH^{i}_{i}|\right|_{\varepsilon=0} \\
\times\prod_{j=3}^{n}\D\lambda_j\,\frac{\d s_j}{s_j^3}\,\bar{\delta}^{2}(\kappa_j-s_j\,\lambda_j)\,\e^{\im\,s_j\,[\sF(x,\lambda_j)\,j]}\,\prod_{\m=1}^{t}\D\lambda_{\m}\wedge\D\bar{\lambda}_\m\,N^{(p_\m-2)}_{\m}\,,
\end{multline}
where $N_\m:=N([\sF(x,\lambda_\m)\,\bar{\lambda}_\m],\lambda_\m,\bar{\lambda}_\m)$ denotes copies of the news function of the SD radiative background $\cM$ pulled back to the twistor curve. The choice of positive helicity graviton $i\in\{3,\ldots,n\}$ singled out in the first line is completely arbitrary, as a consequence of the weighted matrix-tree theorem. This formula can be simplified by performing the $2(n-2)$ integrals in $\D\lambda_j$ and $s_j$ against holomorphic delta functions, to give a formula for the MHV amplitude in momentum space:

\begin{propn}\label{prop:MHV}
The tree-level MHV amplitude for graviton scattering in a SD radiative space-time $(\cM,g)$ is given by 
\begin{multline}\label{MHV2}
  \cM_{n,1}=\kappa^{n-2}\sum_{t=0}^{n-4}\sum_{p_1,\ldots,p_t}\int\d^{4}x\,\sqrt{g}\,\frac{\la1\,2\ra^6}{\la1\,i\ra^2\,\la2\,i\ra^2}\,\left.\left(\prod_{\m=1}^{t}\frac{1}{p_\m!}\,\frac{\partial^{p_\m}}{\partial\varepsilon_\m^{p_\m}}\right)|\cH^{i}_{i}|\right|_{\varepsilon=0} \\
\times\exp\!\left(\im\sum_{j=1}^{n}\sF^{\dot\alpha}(x,\kappa_j)\,\tilde{\kappa}_{j\,\dot\alpha}\right)\,\prod_{\m=1}^{t}\D\lambda_{\m}\wedge\D\bar{\lambda}_\m\,N^{(p_\m-2)}_{\m}\,,
\end{multline}
where gravitons 1, 2 have negative helicity and all others have positive helicity. The entries of the matrix $\cH$ are
\be\label{cHHmom}
\HH_{ij}= -\frac{[\![i\,j]\!]}{\la i\,j\ra}\,, \quad i\neq j
\ee
\begin{equation*}
 \HH_{ii}=\sum_{j\neq i}\frac{[\![i\,j]\!]}{\la i\,j\ra}\,\frac{\la1\, j\ra\,\la2\, j\ra}{\la1\, i\ra\,\la2\, i\ra}+\sum_{\m=1}^{t}\varepsilon_\m\,\frac{[\![i\,\bar{\lambda}_\m]\!]}{\la i\,\lambda_\m\ra}\,\frac{\la1\,\lambda_\m\ra\,\la2\,\lambda_\m\ra}{\la1\,i\ra\,\la2\,i\ra}\,,
\end{equation*}
\begin{equation*}
\mathfrak{h}_{i\m}=- \varepsilon_\m\,\frac{[\![i\,\bar{\lambda}_\m]\!]}{\la i\,\lambda_\m\ra}\,, \qquad \T_{\m\mathrm{n}}=-\varepsilon_\m \varepsilon_{\mathrm{n}}\,\frac{[\![\bar{\lambda}_\m\,\bar{\lambda}_{\mathrm{n}}]\!]}{\la\lambda_\m\,\lambda_{\mathrm{n}}\ra}\,, \quad \m\neq\mathrm{n}\,,
\end{equation*}
\begin{equation*}
\T_{\m\m}=\varepsilon_\m\,\sum_{i=1}^{n}\frac{[\![\bar{\lambda}_\m\,i]\!]}{\la\lambda_\m\,i\ra}\,\frac{\la1\,i\ra\,\la2\,i\ra}{\la1\,\lambda_\m\ra\,\la2\,\lambda_\m\ra}+\varepsilon_\m\,\sum_{\mathrm{n}\neq\m}\varepsilon_{\mathrm{n}}\,\frac{[\![\bar{\lambda}_\m\,\bar{\lambda}_{\mathrm{n}}]\!]}{\la\lambda_\m\,\lambda_{\mathrm{n}}\ra}\,\frac{\la1\,\lambda_{\mathrm{n}}\ra\,\la2\,\lambda_{\mathrm{n}}\ra}{\la1\,\lambda_\m\ra\,\la2\,\lambda_\m\ra}\,,
\end{equation*}
and 
\be\label{newsinsert}
N^{(k)}_\m:=\left.\frac{\partial^{k}\,N(u,\lambda_\m,\bar{\lambda}_\m)}{\partial u^k}\right|_{u=[\sF(x,\lambda_\m)\,\bar{\lambda}_\m]}\,,
\ee
with $N(u,\lambda,\bar{\lambda})$ the news function of $\cM$.
\end{propn}

Note that in \eqref{MHV2}, we have reverted to `generic' coordinates $x^{\alpha\dot\alpha}$ from the complex coordinates $(z^{\dot\alpha},\tilde{z}^{\dot{\tilde{\alpha}}})$, with the plane wave exponentials $\e^{\im\,[z\,1]}$ and $\e^{\im\,[\tilde{z}\,2]}$ incorporated into the exponential factor on the second line.

\medskip

In this formula, the structure of the SD radiative background space-time $\cM$ appears through: the diffeomorphism-invariant integration measure (i.e., through $\sqrt{g}$); the function $\sF^{\dot\alpha}(x,\lambda)$ describing the holomorphic curves in the associated twistor space -- which appears as the argument of the exponential factor; the dressed momenta in the entries of $\cH$; and finally the insertions of (derivatives of) the background news function $N$ arising from tail contributions to the amplitude. The absence of any overall momentum conserving delta functions is an expected feature of scattering in a curved space-time, as the background gravitational field breaks Poincar\'e invariance. However, a striking feature of the formula is that there is only a \emph{single} residual space-time integral, regardless of the number of external positive helicity gravitons. Perturbatively, general relativity behaves as a cubic field theory (higher-point contact interactions can be re-absorbed into exchange diagrams built from cubic interactions, c.f.~\cite{Bern:2019prr}), so the na\"ive expectation is that a $n$-point tree amplitude in a curved background should entail $n-2$ space-time integrals.  

While this simplicity is remarkable from the perspective of perturbation theory based upon the Einstein-Hilbert or Plebanski actions, it is an expected feature of a \emph{MHV formalism}~\cite{Cachazo:2004kj} for general relativity, where all vertices are given by MHV interactions, linked by scalar propagators. While the most basic definition of such a formalism (based on all-line shifts) fails~\cite{Bjerrum-Bohr:2005xoa,Bianchi:2008pu}, there are several indications that a MHV formalism for gravity should exist. At least formally, MHV rules for GR can be defined indirectly by first identifying them in conformal gravity then restricting to Einstein degrees of freedom~\cite{Adamo:2013tja}, a truncation which is consistent at tree level~\cite{Maldacena:2011mk}. More recently, an off-shell description of general relativity in terms of an action functional on twistor space was found which has a structure compatible with an MHV vertex expansion~\cite{Sharma:2021pkl}. The formula \eqref{MHV2} can be viewed as another smoking gun for gravitational MHV rules, and using the twistor action it should be possible to give a them a precise formulation. 

\medskip


\paragraph{Evaluation on SDPWs:} The MHV amplitude \eqref{MHV2} simplifies considerably when the SD radiative background is a self-dual plane wave (SDPW), where the metric \eqref{sdpwgr} is controlled by a single function of lightfront time, $f(x^-)$. The background vertex operators are given on such a SDPW by  
\be\label{SDPWbVO}
U^{(p)}=\frac{2}{p!}\,\frac{f^{(p-2)}(x^{-})\,[m(\iota)\,\tilde{\iota}]^p}{\la\iota\,1\ra^2\,\la\iota\,2\ra^2}\,,
\ee
where $f^{(k)}\equiv\partial^{k}_{-}f(x^-)$ and $x^-:=x^{\alpha\dot\alpha}\iota_{\alpha}\tilde{\iota}_{\dot\alpha}$. Here, the $\P^1$ integral in the background vertex operator on a general background \eqref{bvo2} has been performed against the holomorphic delta function in $\sh$ which localises $\lambda=\iota$. One immediate consequence of the simplicity of \eqref{SDPWbVO} is that Wick contractions between background vertex operators in a SDPW vanish, dramatically simplifying the structure of the tail contributions. Each insertion of a background vertex operator in the correlator \eqref{Corr1} must be saturated by contractions with \emph{external} graviton vertex operators. This places a tighter bound on the number of tail contributions to the amplitude for fixed $n$, with $t\leq\frac{n-3}{2}$.

Using the expression for the twistor curves of a SDPW metric \eqref{SDPWcurve}, it is now straightforward to evaluate \eqref{MHV2} on this particular class of background. All but one of the four position space integrals can now be done analytically to obtain momentum conserving delta functions in the $x^+$, $z$ and $\tilde{z}$-directions, leaving
\begin{multline}\label{MHVsdpw}
\kappa^{n-2}\,\delta^{3}_{+,\perp}\!\left(\sum_{j=1}^{n}k_j\right)\,\frac{\la1\,2\ra^6}{\la1\,i\ra^2\,\la2\,i\ra^2}\sum_{t=0}^{\lfloor\frac{n-3}{2}\rfloor}\sum_{p_1,\ldots,p_t}\int_{-\infty}^{+\infty}\d x^{-}\,\left.\left(\prod_{\m=1}^{t}\frac{1}{p_\m!}\,\frac{\partial^{p_\m}}{\partial\varepsilon_\m^{p_\m}}\right)|\cH^{i}_{i}|\right|_{\varepsilon=0} \\
\times\,\e^{\im\,F_n(x^-)}\,\prod_{\m=1}^{t}f^{(p_\m-2)}(x^-)\,.
\end{multline}
The three momentum conserving delta functions allow the universal exponent appearing in the second line to be written in the form of a gravitational \emph{Volkov exponent}~\cite{Adamo:2020syc,Adamo:2020qru}
\be\label{grVolkov}
F_{n}(x^-):=\int^{x^-}\d s\,g^{ab}(s)\,\frac{\mathbb{K}_{a}(s)\,\mathbb{K}_{b}(s)}{2\,\la\iota|\mathbb{K}|\tilde{\iota}]}\,,
\ee
for $\mathbb{K}_{\alpha\dot\alpha}(x^-)$ given by the sum of any (distinct) $n-1$ of the $n$ dressed graviton momenta; for instance
\begin{equation*}
\mathbb{K}^{\alpha\dot\alpha}=\sum_{i=1}^{n-1}\kappa^{\alpha}_{i}\,\tilde{K}_{i}^{\dot\alpha}\,.
\end{equation*} 
The block structure \eqref{cHmat} of the matrix $\cH$ remains the same, but with the individual entries simplified:
\be\label{pwHHmom}
\HH_{ij}= -\frac{[\![i\,j]\!]}{\la i\,j\ra}\,, \quad i\neq j
\ee
\begin{equation*}
 \HH_{ii}=\sum_{j\neq i}\frac{[\![i\,j]\!]}{\la i\,j\ra}\,\frac{\la1\, j\ra\,\la2\, j\ra}{\la1\, i\ra\,\la2\, i\ra}+\sum_{\m=1}^{t}\varepsilon_\m\,\frac{[i\,\tilde{\iota}]}{\la i\,\iota\ra}\,\frac{\la1\,\iota\ra\,\la2\,\iota\ra}{\la1\,i\ra\,\la2\,i\ra}\,,
\end{equation*}
\begin{equation*}
\mathfrak{h}_{i\m}=- \varepsilon_\m\,\frac{[i\,\tilde{\iota}]}{\la i\,\iota\ra}\,, \qquad \T_{\m\mathrm{n}}=\delta_{\mathrm{mn}}\,\varepsilon_\m\,\sum_{i=1}^{n}\frac{[\tilde{\iota}\,i]}{\la\iota\,i\ra}\,\frac{\la1\,i\ra\,\la2\,i\ra}{\la1\,\iota\ra\,\la2\,\iota\ra}\,.
\end{equation*}
The formula \eqref{MHVsdpw} is an improvement on an earlier expression for MHV graviton scattering in a SDPW presented in~\cite{Adamo:2020syc}, in that it is manifestly diffeomorphism invariant, is separately symmetric in the positive and the negative helicity gravitons, and slightly more compact. Appendix~\ref{app:prl} demonstrates that \eqref{MHVsdpw} and the formula of~\cite{Adamo:2020syc} are equivalent.


\subsection{Consistency checks}
\label{sec:mhv-cons}
While the unitarity methods which can be used to prove tree-level amplitude formulae in vacuum no longer apply in the presence of a strong background, there are still some basic consistency checks which can be run on formula \eqref{MHV2}. The first of these is comparison with explicit Feynman diagram calculations using the background-coupled Einstein-Hilbert Lagrangian; of course, such computations are only tractable at low numbers of points and in highly symmetric SD radiative space-times where the Feynman rules in the background can be determined explicitly. In appendix~\ref{app:grav}, it is demonstrated that for a SDPW background, the formula \eqref{MHVsdpw} matches the amplitude computed with Feynman rules at 3- and 4-points.

Beyond this, the most straightforward all-multiplicity consistency check is the flat space limit, where the SD radiative background becomes (complexified) Minkowski space. In this case, the background news function $N(u,\lambda,\bar{\lambda})$ becomes trivial so only the $t=0$, no-tail term in \eqref{MHV2} contributes. For Minkowski space, the twistor curves become twistor lines:
\be\label{flatlim1}
\sF^{\dot\alpha}(x,\lambda)|_{\mathrm{flat}}=x^{\alpha\dot\alpha}\,\lambda_{\alpha}\,.
\ee
This enables all of the remaining integrals in \eqref{MHV2} to be performed immediately, with the result
\be\label{flatlim2}
\kappa^{n-2}\,\delta^{4}\!\left(\sum_{i=1}^{n}k_i\right)\,\frac{\la1\,2\ra^6}{\la1\,i\ra^2\,\la2\,i\ra^2}\,\left|\HH^{i}_{i}\right|\,,
\ee
where the entries of $\HH$ are now
\be\label{flatHH}
\HH_{ij}=-\frac{[i\,j]}{\la i\,j\ra}\,, \quad i\neq j\,, \qquad \HH_{ii}=\sum_{j\neq i}\frac{[i\,j]}{\la i\,j\ra}\,\frac{\la1\,j\ra\,\la2\,j\ra}{\la1\,i\ra\,\la2\,i\ra}\,,
\ee
defined in terms of un-dressed on-shell momenta. The expression \eqref{flatlim2} is precisely Hodges' formula~\cite{Hodges:2012ym} for gravitational MHV scattering in Minkowski space, so we do indeed produce the correct flat space limit.

\medskip

A more non-trivial consistency check is the behaviour of \eqref{MHV2} in the \emph{perturbative limit}, where the SD radiative background becomes weak and is treated like a single positive helicity graviton in flat space. In the perturbative limit, any SD radiative space-time will be well-approximated by a SDPW, so it suffices to work directly with \eqref{MHVsdpw} and isolate all contributions to the MHV amplitude which are linear in the background profile $f(x^-)$. This background-dependence enters \eqref{MHVsdpw} in three ways: through the matrix $\cH$, whose entries include dressed momentum spinors; through the Volkov exponent \eqref{grVolkov}; or through explicit tail factors when $t>0$. Clearly, only the $t=0,1$ terms can provide such linear contributions.

In the $t=0$ term, linear dependence on $f(x^-)$ arises only from the Volkov exponent or dressed momentum spinors. This is easily determined explicitly using \eqref{sdpwdressgr}, and representing the perturbative background by the Fourier mode
\be\label{pertlim1}
 f(x^-)=\kappa\,\e^{-\im\,\omega\,x^-}:=\kappa\,\e^{\im\, q\cdot x}\,, \qquad q_{\alpha\dot\alpha}=\omega\,\iota_{\alpha}\,\tilde{\iota}_{\dot\alpha}\,,
\ee
the contributions to the perturbative limit of the $t=0$ term are
\begin{multline}\label{pertlimt0}
\frac{\kappa^{n-1}\,\la1\,2\ra^6}{\la1\,i\ra^2\,\la2\,i\ra^2}\,\delta^{4}\!\left(q+\sum_{i=1}^{n}k_i\right)\left[\sum_{\substack{j,l=3 \\ j\neq l, \, j,l\neq i}}^{n}(-1)^{j+l}\,\frac{[\tilde{\iota}\,j]\,[\tilde{\iota}\,l]}{\la\iota\,j\ra\,\la\iota\,l\ra}\,|\HH^{ij}_{il}|-|\HH^{i}_{i}| \sum_{j=1}^{n} \frac{[\tilde{\iota}\,j]\,\la o\,j\ra^2}{\la\iota\,j\ra} \right. \\
\left. -\sum_{\substack{j=3 \\ j\neq i}}^n |\HH^{ij}_{ij}|\,\sum_{l\neq j}\frac{[\tilde{\iota}\,j]\,[\tilde{\iota}\,l]}{\la\iota\,j\ra\,\la\iota\,l\ra}\,\frac{\la1\,l\ra\,\la2\,l\ra}{\la1\,j\ra\,\la2\,j\ra}\right]\,,
\end{multline}
where $\HH$ is now understood to be the `flat' matrix given by \eqref{flatHH}. Using the Schouten identity and 4-momentum conservation, these terms can be re-written as
\begin{multline}\label{pertlim0*}
\frac{\kappa^{n-1}\,\la1\,2\ra^6}{\la1\,i\ra^2\,\la2\,i\ra^2}\,\delta^{4}\!\left(q+\sum_{i=1}^{n}k_i\right)\left[-|\HH^{i}_{i}|\,\sum_{j}\frac{[\tilde{\iota}\,j]}{\la\iota\,j\ra}\,\frac{\la1\,j\ra\,\la2\,j\ra}{\la1\,\iota\ra\,\la2\,\iota\ra} +\sum_{j,l, j\neq l}(-1)^{j+l}\,\frac{[\tilde{\iota}\,j]\,[\tilde{\iota}\,l]}{\la\iota\,j\ra\,\la\iota\,l\ra}\,|\HH^{ij}_{il}|\right. \\
\left. -\sum_{j} |\HH^{ij}_{ij}|\,\sum_{l\neq j}\frac{[\tilde{\iota}\,j]\,[\tilde{\iota}\,l]}{\la\iota\,j\ra\,\la\iota\,l\ra}\,\frac{\la1\,l\ra\,\la2\,l\ra}{\la1\,j\ra\,\la2\,j\ra}\right]\,,
\end{multline}
where the precise ranges of the various summations are implied by the structure of the summands.

The $t=1$ term only contributes to the perturbative limit through the explicit tail insertion itself; all other background-dependence (in the matrix $\cH$ and Volkov exponent) is set to zero when extracting those terms linear in $f(x^-)$. The resulting contribution is
\begin{multline}\label{pertlimt1}
\frac{\la1\,2\ra^6}{\la1\,i\ra^2\,\la2\,i\ra^2}\sum_{p=1}^{n-3}\sum_{j_1,\ldots,j_{p}}\left[\sum_{k,l}(-1)^{k+l}\,\left|\HH^{ikj_1\cdots j_p}_{ilj_1\cdots j_p}\right|\,\left(\prod_{\m=1}^{p}\frac{[\tilde{\iota}\,j_\m]}{\la\iota\,j_\m\ra}\,\frac{\la1\,\iota\ra\,\la2\,\iota\ra}{\la1\,j_\m\ra\,\la2\,j_\m\ra}\right)\,\frac{[\tilde{\iota}\,k]\,[\tilde{\iota}\,l]}{\la\iota\,k\ra\,\la\iota\,l\ra} \right. \\
\left.+\left|\HH^{ij_1\cdots j_p}_{ij_1\cdots j_p}\right|\,\left(\prod_{\m=1}^{p}\frac{[\tilde{\iota}\,j_\m]}{\la\iota\,j_\m\ra}\,\frac{\la1\,\iota\ra\,\la2\,\iota\ra}{\la1\,j_\m\ra\,\la2\,j_\m\ra}\right) \sum_{l}\frac{[\tilde{\iota}\,l]}{\la\iota\,l\ra}\,\frac{\la1\,l\ra\,\la2\,l\ra}{\la1\,\iota\ra\,\la2\,\iota\ra} \right]\,, \\
\end{multline}
where an overall factor of $\kappa^{n-1}$ and momentum conserving delta functions -- equivalent to those appearing in \eqref{pertlim0*} -- have been omitted and $\HH$ is once again understood as the matrix with `flat' entires. 

The $t=0$ and $t=1$ contributions combine auspiciously if we define a $(n-1)\times(n-1)$ matrix $\widehat{\HH}$ whose entries are equivalent to those of $\HH$, with the extra row and column corresponding to the background graviton with momentum $q_{\alpha\dot\alpha}$. Now, the minor $|\widehat{\HH}^{q}_{q}|$ differs from $|\HH|$ only by the diagonal entries of the two matrices, so
\be\label{pertHH}
|\widehat{\HH}^{q}_{q}|=|\HH|+\sum_{p=1}^{n-2}\sum_{j_1,\ldots,j_p}\left|\HH^{j_1\cdots j_p}_{j_1\cdots j_p}\right| \left(\prod_{\m=1}^{p}\frac{[\tilde{\iota}\,j_\m]}{\la\iota\,j_\m\ra}\,\frac{\la1\,\iota\ra\,\la2\,\iota\ra}{\la1\,j_\m\ra\,\la2\,j_\m\ra}\right)\,.
\ee
Applying this identiy to further reduced minors of $\HH$, it follows that the $t=0,1$ contributions to the perturbative limit combine to give
\be\label{pertlim}
\kappa^{n-1}\,\delta^{4}\!\left(q+\sum_{i=1}^{n}k_i\right)\,\frac{\la1\,2\ra^6}{\la1\,i\ra^2\,\la2\,i\ra^2}\,\left|\widehat{\HH}^{i}_{i}\right|\,,
\ee
which is the Hodges formula for MHV graviton scattering in Minkowski space with $(n-1)$ positive helicity gravitons whose momenta are $(k_3,\ldots,k_n,q)$. This confirms that \eqref{MHV2} has the correct perturbative limit, emerging only after a somewhat subtle collaboration between `pure scattering' (i.e., $t=0$) and tail (i.e., $t=1$) contributions to the amplitude.


\section{N$^k$MHV scattering}
\label{sec:nmhv}

A generic tree-level graviton amplitude will have an arbitrary number of both positive \emph{and} negative helicity external states: amplitudes with $(k+2)$ negative helicity gravitons and arbitrarily many positive helicity gravitons are in the N$^k$MHV helicity configuration. In Minkowski space, the Cachazo-Skinner formula extends Hodges' formula for MHV scattering to give every N$^k$MHV graviton scattering amplitude (i.e., the full tree-level S-matrix)~\cite{Cachazo:2012kg}. This formula presents the scattering amplitude in terms of integrals over the moduli space of holomorphic maps from the Riemann sphere to twistor space, with the degree $d$ of the map related to the helicity configuration by $d=k+1$, and arises as the genus zero worldsheet correlation functions in a twistor string theory~\cite{Skinner:2013xp}.

In \cite{Adamo:2021bej} we further showed how the Cachazo-Skinner formulae could be obtained from the perturbative expansion based on a higher degree version of the chiral twistor sigma model introduced earlier at MHV.
At degree one, we have seen that MHV amplitudes in self-dual radiative space-times are given in terms of a single space-time integral in \eqref{MHV2}. It is therefore natural to conjecture that a N$^k$MHV amplitude will be given by an integral over degree $d=k+1$ holomorphic maps from the Riemann sphere to the twistor space of the SD radiative background. 
This leads to a remarkably simple and compact formula for all tree-level graviton amplitudes in a SD radiative space-time. The formula contains $4(k+1)$ moduli integrals for the N$^k$MHV amplitude, as opposed to the $4(n-2)$ integrals na\"ively expected from the perturbative expansion of the background-coupled Einstein-Hilbert action. Thus, for any amplitude with $n>3$ external gravitons, our formula contains substantially fewer integrals, with further simplification possible upon restriction to SDPW backgrounds as a consequence of the additional symmetries.

In this section, we work with $\cN=8$ supergravity (SUGRA) to simplify the resulting formulae; since all expressions are at tree-level, specific components of the external graviton multiplets can be recovered by extracting the relevant terms in an expansion of the supermomenta (i.e., by algebraic Grassmann integration). Unlike the MHV amplitude \eqref{MHV2}, which was derived from first principles, these N$^k$MHV formulae are \emph{conjectural}. However, the starting point for our derivation is a well-motivated generating functional and the resulting expressions are easily seen to have the correct flat and perturbative limits.

In \S\ref{sec:nmhv-CS} we review the parts of \cite{Adamo:2021bej} that extend the the chiral twistor sigma model used in the previous sections to higher degree, leading to a reformulation of the Cachazo-Skinner formula for $\cN=8$ SUGRA amplitudes at N$^k$MHV degree. This allows us in \S\ref{sec:nmhv-background} to extend it to a version expanded around a background leading to formula \eqref{NkMHV}. This again is specialized to become more explicit for self-dual plane waves in equation \eqref{dsdpw}.


\subsection{$\cN=8$ supergravity \& the Cachazo-Skinner formula}\label{sec:nmhv-CS}

Supersymmetric twistor space $\PT$ is an open subset of $\P^{3|8}$ with homogeneous coordinates $Z^{I}=(\mu^{\dot\alpha},\lambda_{\alpha},\chi^{a})$, defined by $\lambda_{\alpha}\neq 0$, where the $\chi^{a}$ are Grassmann numbers and $a=1,\ldots,8$. In Minkowski space, the tree-level S-matrix of $\cN=8$ supergravity (SUGRA) can be obtained as a worldsheet correlation function in a twistor string theory governing holomorphic maps from $\P^1$ to $\PT$~\cite{Skinner:2013xp}; the resulting Cachazo-Skinner formula~\cite{Cachazo:2012kg} expresses N$^{d-1}$MHV amplitudes in terms of a moduli integral over the space of degree $d$ holomorphic maps to twistor space.

Rather than working with twistor string theory, the Cachazo-Skinner formula can also be derived using a higher-degree generalization of the generating functional \eqref{TliftMHV} for MHV amplitudes~\cite{Adamo:2021bej}. Working with explicit $\cN=8$ supersymmetry, all external states are given by on-shell $\cN=8$ graviton supermultiplets, represented on twistor space by $h_i\in H^{0,1}(\PT,\cO(2))$. For a $n$-point N$^{d-1}$MHV amplitude, these external graviton multiplets are partitioned into two disjoint sets: $\{1,\ldots,n\}=\tth\cup\mathtt{h}$, with $|\tth|=d+1$ and $|\mathtt{h}|=n-d-1$. The choice of this partition is entirely arbitrary, although it is natural to interpret the sets $\tth$ and $\mathtt{h}$ as indexing negative and positive helicity gravitons, respectively, upon extracting the multiplet components appropriate to N$^{d-1}$MHV scattering in general relativity.

A degree $d$ holomorphic map from $\P^1$ to $\PT$ can be written as
\be\label{curves}
Z^I(\sigma)=U_{\ba_1\cdots\ba_d}^{I}\,\sigma^{\ba_1}\cdots\sigma^{\ba_d}=:U^{I}_{\ba(d)}\,\sigma^{\ba(d)}\,, 
\ee 
where $\sigma^{\ba}=(\sigma^{\mathbf{0}},\,\sigma^{\mathbf{1}})$ are homogeneous coordinates on the underlying $\P^1$, $\ba(d)$ is a totally-symmetric rank $d$ multi-index, and the $4(d+1)$ bosonic and $8(d+1)$ fermionic parameters of $U^{I}_{\ba(d)}$ are the map moduli. In the presence of a finite deformation of the twistor space generated by Hamiltonian $h$, only the $\mu^{\dot\alpha}$ components of the map are modified, with $\mu^{\dot\alpha}=F^{\dot\alpha}(U,\sigma)$ for $F^{\dot\alpha}$ homogeneous of degree $d$ in $\sigma$ satisfying:
\be\label{dholc}
\dbar F^{\dot\alpha}(U,\sigma)=\frac{\partial h}{\partial\mu_{\dot\alpha}}(F,\sigma)\,.
\ee
Solutions to this equation can then be written in terms of a finite perturbation away from the undeformed curves \eqref{curves}:
\be\label{curves2}
F^{\dot\alpha}(U,\sigma)=U^{\dot\alpha}_{\ba(d)}\,\sigma^{\ba(d)}+m^{\dot\alpha}(\sigma)\,,
\ee
where $m^{\dot\alpha}$ takes values in $\cO(d)$ over $\P^1$. The moduli of $m^{\dot\alpha}$ are fixed by requiring that it has simple zeros at $d+1$ points on $\P^1$ indexed by $\tth$:
\be\label{curves3}
m^{\dot\alpha}(\sigma)=M^{\dot\alpha}(\sigma)\,\prod_{l\in\tth}(\sigma\,l)\,,
\ee
for $M^{\dot\alpha}$ valued in $\cO(-1)$ and regular at each $\sigma_l$, $l\in\tth$. Here and subsequently, the SL$(2,\C)$-invariant inner product on the homogeneous coordinates of $\P^1$ will be denoted with round brackets: $(\sigma\,\sigma'):=\epsilon^{\mathbf{ab}}\,\sigma_{\mathbf{b}}\sigma'_{\ba}$.  

There is an action principle for these higher-degree maps, given by:
\be\label{dAction}
S^{(d)}[m]=\int_{\P^1}\frac{\D\sigma}{\prod_{l\in\tth}(\sigma\,l)^2}\left([m\,\dbar m]+2\,h(F,\sigma)\right)\,,
\ee
where $\D\sigma:=(\sigma\,\d\sigma)$ is the weight $2$ holomorphic measure on $\P^1$. The equation of motion of this action is equivalent to \eqref{dholc} evaluated on \eqref{curves2}. The requirement that this variational principle is well-defined is equivalent to the boundary condition \eqref{curves3} on $m^{\dot\alpha}$. When $d=1$, M\"obius and scale invariance can be used to set $U_{\alpha\,\ba}=\epsilon_{\alpha\ba}$ so that the twistor coordinate $\lambda_{\alpha}$ is identified with the homogeneous coordinates on $\P^1$. In this case, it is easy to see that $S^{(1)}$ is equivalent to the action \eqref{pertac} on a trivial background (i.e., with $\sh=0$).

Continuing to follow the analogy with $d=1$, one defines a generating functional for the N$^{d-1}$MHV amplitudes where all positive helicity (or self-dual) gravitons emerge by a perturbative expansion of the classical action $S^{(d)}$. This definition is guided by several straightforward requirements: it should take the form of an integral over the moduli space of degree $d$ holomorphic curves in $\PT$, behave under degenerations of the map in accordance with unitarity, be linear in $S^{(d)}$ and homogeneous. These requirements are fairly restrictive, resulting in the generating functional
\be\label{dGenF}
\mathcal{G}^{(d)}=\int\frac{\d^{4|8(d+1)}U}{\mathrm{vol}\,\GL(2,\C)}\,\frac{\det^{\prime}(\HH^{\vee})}{|\tth|^2}\,S^{(d)}\,\prod_{l\in\tth}h_{l}(Z(\sigma_l))\,\D\sigma_l\,.
\ee
Here, division by the (infinite) volume of GL$(2,\C)$ is understood in the Fadeev-Popov sense, accounting for the $\C^*\times\SL(2,\C)$ redundancy in the parametrization of the map \eqref{curves}. The object $|\tth|$ appearing in the denominator is a Vandermonde determinant (not to be confused with the integer $d+1$):
\be\label{vdm1}
|\tth|:=\prod_{l<m\in\tth}(l\,m)\,.
\ee
The object $\det^{\prime}(\HH^{\vee})$ has a purely algebro-geometric definition as the \emph{resultant} of the map components $\lambda_{\alpha}(\sigma)$~\cite{Skinner:2013xp,Cachazo:2013zc}. This quantity is independent of the marked points on the Riemann sphere, and has the properties that it vanishes at points in the moduli space where $\lambda_{\alpha}(\sigma)=0$ (thereby ensuring that the generating function only has support when the image of the map lies in $\PT\subset\P^{3|8}$) and respects degeneration of the holomorphic map itself.

To obtain the $n$-point N$^{d-1}$MHV amplitude from this generating functional, the on-shell action $S^{(d)}$ must be perturbatively expanded to extract the graviton multiplets indexed by the set $\mathtt{h}$. This is equivalent to computing the connected, tree-level correlation function
\be\label{flatcorr1}
\left\la\prod_{j\in\mathtt{h}}V_j\right\ra^{0}_{\mathrm{tree}}\,, \qquad V_{j}=\int \frac{\D\sigma_j}{\prod_{l\in\tth}(l\,r)^2} \;h(F,\sigma_j)\,,
\ee
in the free theory on $\P^1$ with action
\be\label{flatcorr2}
\int_{\P^1} \frac{\D\sigma}{\prod_{l\in\tth}(\sigma\,l)^2}\,\epsilon_{\dot\beta\dot\alpha}\, m^{\dot\alpha}(\sigma)\,\dbar m^{\dot\beta}(\sigma)\,.
\ee
Using the corresponding OPE
\be\label{flatOPE}
m^{\dot\alpha}(\sigma)\,m^{\dot\beta}(\sigma')\sim \frac{\epsilon^{\dot\alpha\dot\beta}}{(\sigma\,\sigma')}\,\prod_{l\in\tth}(\sigma\,l)\,(\sigma'\,l)\,,
\ee
and the weighted matrix-tree theorem, it is easy to show that
\be\label{flatcorr3}
\left\la\prod_{j\in\mathtt{h}}V_j\right\ra^{0}_{\mathrm{tree}}=\int_{(\P^1)^{n-d-1}}\frac{|\HH^{i}_{i}|}{\prod_{l\in\tth}(i\,l)^{2}}\,\prod_{j\in\mathtt{h}}\,h_j(Z(\sigma_j))\,\D\sigma_j\,,
\ee
where $\HH$ is now a $(n-d-1)\times(n-d-1)$ matrix with entries indexed by $\mathtt{h}$:
\be\label{HHflatd}
\HH_{ij}=\frac{1}{(i\,j)}\,\left[\frac{\partial}{\partial\mu(\sigma_i)}\,\frac{\partial}{\partial\mu(\sigma_j)}\right]\,, \quad i\neq j\,, \qquad \HH_{ii}=-\sum_{j\neq i}\frac{1}{(i\,j)}\,\left[\frac{\partial}{\partial\mu(\sigma_i)}\,\frac{\partial}{\partial\mu(\sigma_j)}\right]\,\prod_{l\,\tth}\frac{(j\,l)}{(i\,l)}\,.
\ee
The choice of minor (labelled by $i\in\mathtt{h}$) in \eqref{flatcorr3} is completely arbitrary, as a consequence of the matrix-tree theorem.

Combining all of these ingredients, one obtains a formula for the tree-level, $n$-point N$^{d-1}$MHV amplitude:
\be\label{CaSk}
\cM_{n,d}=\int\frac{\d^{4|8(d+1)}U}{\mathrm{vol}\,\GL(2,\C)}\,\mathrm{det}^{\prime}(\HH^{\vee})\,\mathrm{det}^{\prime}(\HH)\,\prod_{i=1}^{n}h_{i}(Z(\sigma_i))\,\D\sigma_i\,,
\ee
where the reduced determinant
\be\label{HHred}
\mathrm{det}^{\prime}(\HH):=\frac{|\HH^{i}_{i}|}{|\tth\cup\{i\}|^2}\,.
\ee
This is precisely the Cachazo-Skinner formula for the tree-level S-matrix of $\cN=8$ SUGRA in Minkowski space~\cite{Cachazo:2012kg}. \emph{A priori}, this formula appears to contain a large number of moduli integrals, but their role is greatly clarified when evaluated on momentum eigenstates. A momentum eigenstate for the $\cN=8$ graviton multiplet is characterized by an on-shell supermomentum $(\kappa_{\alpha}\tilde{\kappa}_{\dot\alpha},\,\kappa_{\alpha}\eta_{a})$, represented in twistor space by (cf., \cite{Adamo:2011pv}):
\be\label{SUGRAmeig}
h_{i}(Z)=\int_{\C^*}\frac{\d s_i}{s^3_i}\,\bar{\delta}^{2}(\kappa_i-s_i\,\lambda)\,\e^{\im\,s_{i}\,([\mu\,i]+\chi^{a}\,\eta_{i\,a})}\,.
\ee
When \eqref{CaSk} is evaluated on these momentum eigenstates, the moduli of the $\mu^{\dot\alpha}$ and $\chi^a$ components of the map \eqref{curves} can be integrated out to obtain delta functions, leaving
\begin{multline}\label{CaSkme}
\cM_{n,d}=\int 
\frac{\d^{2(d+1)}\lambda}{\mathrm{vol}\,\mathrm{GL}(2,\C)}\,\delta^{2|8(d+1)}\!\left(\sum_{i=1}^{n}s_i\,(\tilde{\kappa}_i,\eta_i)\,\sigma_{i}^{\ba(d)}\right)\\  \times \mathrm{det}^{\prime}(\HH^{\vee})\,\mathrm{det}^{\prime}(\HH)\prod_{i=1}^{n}\frac{\d s_i\,\D\sigma_i}{s^3_i}\,\bar{\delta}^{2}(\kappa_i-s_i\,\lambda(\sigma_i))\,,
\end{multline}
where $\lambda_{\alpha}(\sigma)=\lambda_{\alpha\,\ba(d)}\sigma^{\ba(d)}$ and the entries of $\HH$ are now given by 
\be\label{meHH}
\HH_{ij}=-s_i s_j\,\frac{[i\,j]}{(i\,j)}\,, \quad i\neq j\,, \qquad \HH_{ii}=s_i\sum_{j\neq i}s_j\,\frac{[i\,j]}{(i\,j)}\,\prod_{l\in\tth}\frac{(j\,l)}{(i\,l)}\,.
\ee
The delta functions appearing in this formula are easily seen to imply $4|16$-dimensional supermomentum conservation, and all remaining moduli integrals $\{\lambda_{\alpha\ba(d)}, \sigma_i, s_i\}$ -- modulo GL$(2,\C)$ freedom -- are saturated by delta functions to give a residue sum.

Thus, there are actually \emph{no} residual integrals in the Cachazo-Skinner formula (when evaluated on momentum eigenstates), as expected for tree-level scattering amplitudes in Minkowski space. The veracity of the Cachazo-Skinner formula is easily established with unitarity methods~\cite{Cachazo:2012pz} or worldsheet factorization of the underlying twistor string~\cite{Adamo:2013tca}.


\subsection{Amplitudes in a SD radiative space-time} \label{sec:nmhv-background}

The generating functional \eqref{dGenF} extends naturally to N$^{d-1}$MHV scattering in any SD radiative space-time, closely following the calculation for MHV amplitudes. The difference from Minkowski space is that the degree $d$ maps are now written as a finite perturbation $\dbar+\sh\to\dbar+\sh+h$ away from holomophic maps into the twistor space $\CPT$ of the SD radiative manifold:
\be\label{dSDradcurve}
F^{\dot\alpha}(U,\sigma)=\sF^{\dot\alpha}(U,\sigma)+m^{\dot\alpha}(\sigma)\,,
\ee
where $\sF^{\dot\alpha}$ is homogeneous of degree $d$ in $\sigma$ and satisfies
\be\label{dbcurve}
\dbar\sF^{\dot\alpha}(U,\sigma)=\frac{\partial\sh}{\partial\mu_{\dot\alpha}}(\sF,\sigma)\,,
\ee
for $\sh$ defining the complex structure on $\CPT$. Theorems of Kodaira~\cite{Kodaira:1962,Kodaira:1963} ensure that there are the same number of map moduli $\{U^{I}_{\ba(d)}\}$ as in Minkowski space: $4|8(d+1)$. The field $m^{\dot\alpha}$ is valued in $\cO(d)$ and must solve
\be\label{dbcurve*}
\dbar m^{\dot\alpha}(\sigma)=\frac{\partial h}{\partial\mu_{\dot\alpha}}(\sF+m,\sigma)+ \frac{\partial \sh}{\partial\mu_{\dot\alpha}}(\sF+m,\sigma)-\frac{\partial\sh}{\partial\mu_{\dot\alpha}}(\sF,\sigma)\,,
\ee
subject to the boundary conditions \eqref{curves3}.

To obtain the background-coupled equation \eqref{dbcurve*}, the action $S^{(d)}$ must be modified with explicit background terms (once again mirroring the $d=1$ calculation):
\begin{multline}\label{dbAction}
S^{(d)}[m]=\int_{\P^1}\frac{\D\sigma}{\prod_{l\in\tth}(\sigma\,l)^2}\biggl([m\,\dbar m]+2\,h(\sF+m,\sigma) \\
 +\sum_{p=2}^{\infty}\frac{2}{p!}\,\frac{\partial^{p}\sh}{\partial\mu^{\dot\alpha_1}\cdots\partial\mu^{\dot\alpha_p}}(\sF,\sigma)\,m^{\dot\alpha_1}\cdots m^{\dot\alpha_p}\biggr)\,.
\end{multline}
With this definition, we conjecture that the generating functional for the N$^{d-1}$MHV amplitude on any SD radiative space-time is equivalent to \eqref{dGenF}. The non-trivial part of the calculation is now the computation of the correlation function
\be\label{dcorr1}
\sum_{t=0}^{\infty}\sum_{p_1,\ldots,p_t}\left\la\prod_{j\in\mathtt{h}}V_j\,\prod_{\m=1}^{t}U^{(p_\m)}\right\ra^{0}_{\mathrm{tree}}\,,
\ee
with background vertex operators defined by 
\be\label{dbVO}
U^{(p_\m)}=\frac{2}{p_\m!}\int_{\P^1}\frac{\D\sigma_\m}{\prod_{l\in\tth}(\m\,l)^2}\wedge\D\bar{\lambda}(\sigma_\m)\,[m\,\bar{\lambda}]^{p_\m}(\sigma_\m)\,N^{(p_\m-2)}(\sigma_\m)\,.
\ee
This tree-level, connected correlator is evaluated in the $\P^1$ theory defined by
\be\label{dcorr2}
\int_{\P^1}\frac{\D\sigma}{\prod_{l\in\tth}(\sigma\,l)^2}\,m^{\dot\alpha}\left(\epsilon_{\dot\beta\dot\alpha}\,\dbar+\frac{\partial^{2}\sh}{\partial\mu^{\dot\alpha}\partial\mu^{\dot\beta}}(\sF,\sigma)\right) m^{\dot\beta}\,,
\ee
whose OPE is
\be\label{dOPE}
m^{\dot\alpha}(\sigma_i)\,m^{\dot\beta}(\sigma_{j})\sim \frac{H^{\dot\alpha}{}_{\dot\gamma}(U,\sigma_i)\,H^{\dot\beta\dot\gamma}(U,\sigma_j)}{(i\,j)}\,\prod_{l\in\tth}(i\,l)\,(j\,l)\,.
\ee
The requirement that all $m^{\dot\alpha}$ insertions in the correlator be saturated by Wick contractions places an upper bound on the number of background vertex insertions in the correlator \eqref{dcorr1}: $t\leq n-d-3$.

\medskip

As in the MHV case, the computation of the correlator now boils down to counting (weighted) spanning tree graphs on the set of $n-d-1+t$ vertices, where the $n-d-1$ vertices indexed by $\mathtt{h}$ can have arbitrary valence while the $t$ vertices corresponding to insertions of the background vertex operators \eqref{dbVO} have their valence fixed by the value of $p_\m$. The final result is:
\begin{multline}\label{NkMHV}
\cM_{n,d}=\sum_{t=0}^{n-d-3}\sum_{p_1,\ldots,p_t}\int\frac{\d^{4|8(d+1)}U}{\mathrm{vol}\,\GL(2,\C)}\,\,\mathrm{det}^{\prime}(\HH^{\vee})\left(\prod_{\m=1}^{t}\frac{1}{p_\m!}\,\frac{\partial^{p_\m}}{\partial\varepsilon_{\m}^{p_\m}}\right)\,\mathrm{det}^{\prime}(\cH)\Big|_{\varepsilon=0} \\
\times \prod_{i=1}^{n} \frac{\D\sigma_i\,\d s_i}{s_i^3}\,\bar{\delta}^{2}(\kappa_i-s_i\,\lambda(\sigma_i))\,\e^{\im\,s_i\,[\sF(U,\sigma_i)\,i]}\,\prod_{\m=1}^{t}\D\sigma_\m\wedge\D\bar{\lambda}(\sigma_m)\,N^{(p_\m-2)}(\sigma_\m)\,.
\end{multline}
Here, the the matrix $\cH$ has the same block decomposition \eqref{cHmat} as before, with entries that are higher-degree generalisations of those appearing in the MHV amplitude \eqref{cHH}:
\be\label{cHHd}
\HH_{ij}= -s_{i}s_{j}\,\frac{[\![i\,j]\!]}{(i\,j)}\,, \quad i\neq j
\ee
\begin{equation*}
 \HH_{ii}=s_{i}\,\sum_{j\neq i}s_{j}\,\frac{[\![i\,j]\!]}{(i\,j)}\,\prod_{l\in\tth}\frac{(l\,j)}{(l\,i)}+s_i\,\sum_{\m=1}^{t}\varepsilon_\m\,\frac{[\![i\,\bar{\lambda}(\sigma_\m)]\!]}{(i\,\m)}\,\prod_{l\in\tth}\frac{(l\,\m)}{(l\,i)}\,,
\end{equation*}
\begin{equation*}
\mathfrak{h}_{i\m}=-s_i \,\varepsilon_\m\,\frac{[\![i\,\bar{\lambda}(\sigma_\m)]\!]}{(i\,\m)}\,, \qquad \T_{\m\mathrm{n}}=-\varepsilon_\m \varepsilon_{\mathrm{n}}\,\frac{[\![\bar{\lambda}(\sigma_\m)\,\bar{\lambda}(\sigma_{\mathrm{n}})]\!]}{(\m\,\mathrm{n})}\,, \quad \m\neq\mathrm{n}\,,
\end{equation*}
\begin{equation*}
\T_{\m\m}=\varepsilon_\m\,\sum_{i=1}^{n}s_i\,\frac{[\![\bar{\lambda}(\sigma_\m)\,i]\!]}{(\m\,i)}\,\prod_{l\in\tth}\frac{(l\,i)}{(l\,\m)}+\varepsilon_\m\,\sum_{\mathrm{n}\neq\m}\varepsilon_{\mathrm{n}}\,\frac{[\![\bar{\lambda}(\sigma_\m)\,\bar{\lambda}(\sigma_{\mathrm{n}})]\!]}{(\m\,\mathrm{n})}\,\prod_{l\in\tth}\frac{(l\,\mathrm{n})}{(l\,\m)}\,.
\end{equation*}
It is straightforward to check that the formula \eqref{NkMHV} is mathematically well-defined, in the sense that the integrand is a top-form on the moduli space and all of the projective integrals have homogeneity zero.

Although we have  emphasised that, unlike the Cachazo-Skinner formula or our earlier expression for the $d=1$ MHV sector, \eqref{NkMHV} is conjectural. This is because we do not have a first-principles derivation of the generating functional \eqref{dGenF} from general relativity when $d>1$, and the curved background spoils the unitarity techniques which enable proofs of such formulae in Minkowski space. Nevertheless, one can follow steps identical to the MHV case to show that \eqref{NkMHV} not only has the correct flat-space limit, but treating the background perturbatively then lands one back on the Cachazo-Skinner formula, and this is a highly non-trivial test.

While this formula for $\cM_{n,d}$ may seem quite complicated, it is remarkable that \emph{any} all-multiplicity, all-helicity expression for gravitational scattering in a curved space-time is available at all. And despite its somewhat intimidating appearance, \eqref{NkMHV} is actually much simpler that anything one might na\"ively expect from perturbative gravity. Indeed, from the perspective of `standard' space-time perturbation theory, the most optimistic bound on the diagrammatic complexity of gravity is given by viewing it as the double copy of gauge theory (cf., \cite{Bern:2019prr} for a review), which means that it behaves roughly like a field theory with cubic interactions. On a curved space-time, this means that one would expect $4(n-2)$ residual integrals in any $n$-point graviton amplitude at tree-level. Yet, by counting integrals and delta functions (and taking into account the GL$(2,\C)$ gauge-fixing which removes four integrals) we see that \eqref{NkMHV} has $4d$ residual integrals.\footnote{In our counting, we do not include integrals that come with insertions of the news function, as these are not really moduli integrals. Furthermore, in specific examples like the SDPW, these integrals can be performed explicitly.} Since $d\leq n-3$ for all $n>3$, it follows that our formula has much fewer remaining integrations than expected from space-time perturbation theory.

\medskip

\paragraph{Evaluation on SPDWs:} On a general self-dual radiative background, it is impossible to further simplify \eqref{NkMHV}, since the holomorphic curves in twistor space are only determined implicitly by \eqref{dbcurve}. However, for specific examples where the curves are known explicitly, some of the remaining moduli integrals can be done. As usual SDPW backgrounds provide an illustrative example. First, observe that the equation for holomorphic curves in the twistor space of a SDPW is
\be\label{dSDPWcur}
\dbar\sF^{\dot\alpha}(U,\sigma)=-2\pi\im\,\tilde{\iota}^{\dot\alpha}\,\bar{\delta}(\la\iota\,\lambda(\sigma)\ra)\,\la o\,\lambda(\sigma)\ra^2\,\cF\!\left(\frac{[\mu(\sigma)\,\tilde{\iota}]}{\la\lambda(\sigma)\,o\ra}\right)\,,
\ee
where we recall that $\cF(x^-)$ is the anti-derivative of the profile function $f(x^-)$ appearing in the metric \eqref{sdpwgr}, and we abuse notation slightly by letting $\mu^{\dot\alpha}$ stand for the homogeneous solution \eqref{curves} to $\dbar\sF^{\dot\alpha}=0$ on the right-hand-side of the equation. This can be re-written as an integral equation
\be\label{iSDPWcur}
\sF^{\dot\alpha}(U,\sigma)=\mu^{\dot\alpha}(U,\sigma)-\tilde{\iota}^{\dot\alpha}\,\int_{\P^1}\frac{\D\sigma'}{(\sigma\,\sigma')}\,\frac{(\sigma\,\xi)^{d+1}}{(\sigma'\,\xi)^{d+1}}\,\bar{\delta}(\la\iota\,\lambda(\sigma')\ra)\,\la o\,\lambda(\sigma')\ra^2\,\cF\!\left(\frac{[\mu(\sigma')\,\tilde{\iota}]}{\la\lambda(\sigma')\,o\ra}\right)\,,
\ee
where the choice of $\xi^{\ba}$ corresponds to fixing boundary conditions for the Green's function of $\dbar$ acting on sections of $\cO(d)\rightarrow\P^1$. As expected, the only moduli are those of the homogeneous solution $\mu^{\dot\alpha}$, which we denote by
\be\label{mumod}
\mu^{\dot\alpha}(\sigma)=\tilde{\iota}^{\dot\alpha}\,\tilde{m}_{\ba(d)}\,\sigma^{\ba(d)}+\tilde{o}^{\dot\alpha}\,m_{\ba(d)}\,\sigma^{\ba(d)}\,,
\ee
decomposed with respect to the dyad $\{\tilde{\iota}^{\dot\alpha},\tilde{o}^{\dot\alpha}\}$.

To proceed, we follow~\cite{Adamo:2020yzi} and make an auspicious parametrization for the $\lambda_{\alpha}(\sigma)$ components of the degree $d$ map to twistor space. Consider $\la\iota\,\lambda(\sigma)\ra$; locally this is just a homogeneous polynomial of degree $d$, so it can be parametrized by $d$ roots, say $\{b_1,\ldots,b_d\}$ and a scale:
\be\label{lambmap1}
\la\iota\,\lambda(\sigma)\ra=\la\iota\,b_0\ra\,\prod_{r=1}^{d}(\sigma\,b_r)\,,
\ee
where the additional parameter $b_0$ encodes the overall scale. This induces a natural basis of $H^0(\P^1,\cO(d))$, given by
\be\label{Odbasis}
\mathfrak{s}_0(\sigma)=\frac{\la\iota\,\lambda(\sigma)\ra}{\la\iota\,\lambda(b_0)\ra}\,, \qquad \mathfrak{s}_{r}(\sigma)=\frac{(\sigma\,b_0)}{(b_r\,b_0)}\,\prod_{s\neq0,r}\frac{(\sigma\,b_s)}{(b_r\,b_s)}\,, \quad r=1,\ldots,d\,,
\ee
in which we can expand 
\be\label{lambmap2}
\la o\,\lambda(\sigma)\ra=\nu_0\,\mathfrak{s}_0(\sigma)+\sum_{r=1}^{d}\nu_r\,\mathfrak{s}_r(\sigma)\,.
\ee
Note that this obeys $\la o\,\lambda(b_k)\ra=\nu_k$ for each $k=0,1,\ldots,d$, and the moduli of $\lambda_{\alpha}(\sigma)$ are given by the $2(d+1)$ parameters $\{b_0,b_1,\ldots,b_d,\nu_0,\nu_1,\ldots,\nu_d\}$.

These parametrizations allow us to evaluate \eqref{iSDPWcur} explicitly by performing the integral against the holomorphic delta function:
\be\label{SDPWcur}
\sF^{\dot\alpha}(U,\sigma)=\mu^{\dot\alpha}(U,\sigma)-\frac{\tilde{\iota}^{\dot\alpha}}{b_0}\,\sum_{r=1}^{d}\frac{\nu_r^2\,(\sigma\,\xi)^{d+1}}{(\sigma\,b_r)\,(b_r\,\xi)^{d+1}}\,\prod_{s\neq r}(b_r\,b_s)\,\cF\!\left(\frac{m_{\ba(d)}\,b_r^{\ba(d)}}{\nu_r}\right)\,.
\ee
Re-parametrizing $d+1$ of the $\mu^{\dot\alpha}$-moduli as
\be\label{mumod2}
y:=\frac{m_{\ba(d)}\,b_0^{\ba(d)}}{\la\iota\,\lambda(b_0)\ra}\,, \qquad x_r:=\frac{m_{\ba(d)}\,b_r^{\ba(d)}}{\nu_r}\,, \quad r=1,\ldots,d\,,
\ee
it follows that
\be\label{mumap1}
[\tilde{\iota}\,\sF(U,\sigma)]=y\,\la\iota\,\lambda(\sigma)\ra+\sum_{r=1}^d x_r\,\nu_r\,\mathfrak{s}_r(\sigma)\,,
\ee
\be\label{mumap2}
[\tilde{o}\,\sF(U,\sigma)]=-\tilde{m}_{\ba(d)}\,\sigma^{\ba(d)}+\frac{1}{b_0}\,\sum_{r=1}^{d}\frac{\nu_r^2\,(\sigma\,\xi)^{d+1}}{(\sigma\,b_r)\,(b_r\,\xi)^{d+1}}\,\prod_{s\neq r}(b_r\,b_s)\,\cF(x_r)\,.
\ee
An immediate consequence of this is that the $\P^1$ integrals associated with background vertex insertions in \eqref{NkMHV} can be performed explicitly. Indeed, for a SDPW space-time, the background vertex operator \eqref{dbVO} is
\be\label{sdpwbVO}
U^{(p_\m)}=\frac{2}{p_\m!}\,\int_{\P^1}\frac{\D\sigma_\m}{\prod_{l\in\tth}(\m\,l)^2}\,\la\lambda(\sigma_\m)\,o\ra^{3-p_\m}\,[m\,\bar{\lambda}]^{p_\m}(\sigma_\m)\,\bar{\delta}(\la\iota\,\lambda(\sigma_\m)\ra)\,f^{(p_\m-2)}\!\left(\frac{[\mu(\sigma_\m)\,\tilde{\iota}]}{\la\lambda(\sigma_\m)\,o\ra}\right)\,,
\ee
so making use of \eqref{lambmap1} and \eqref{lambmap2} the $\D\sigma_\m$ integral can be done against the holomorphic delta function to give
\be\label{sdpwbVO2}
U^{(p_\m)}=\frac{1}{b_0}\,\sum_{r=1}^{d}\frac{\nu_r^{3-p_\m}\,\bar{\nu}_{r}^{p_\m}\,[m(b_r)\,\tilde{\iota}]^{p_\m}}{\prod_{l\in\tth}(b_r\,l)^2}\,\prod_{s\neq r}(b_s\,b_r)\,f^{(p_\m-2)}(x_r)\,,
\ee
for each $\m=1,\ldots,t$.

Feeding all of this into \eqref{NkMHV}, the moduli integrals corresponding to $y$ and $\tilde{m}_{\ba(d)}$ (as well as all of the fermionic moduli) can now be performed to give delta functions, resulting in an expression for the N$^{d-1}$MHV amplitude on a SDPW space-time:
\begin{multline}\label{dsdpw}
\cM_{n,d}=\delta\!\left(\sum_{i=1}^{n}[\tilde{o}\,i]\,\la\iota\,i\ra\right)\sum_{t=0}^{\lfloor\frac{n-3}{2}\rfloor}\sum_{\substack{p_1,\ldots,p_t\\ r_1,\ldots,r_t}} \int\frac{\d^{d+1}b\,\d^{d+1}\nu\,\d^{d}x}{\mathrm{vol}\,\mathrm{GL}(2,\C)}\,J(b,\nu)\,\frac{\mathrm{det}'(\HH^{\vee})}{b_0^t} \\
\delta^{1|8(d+1)}\!\left(\sum_{j=1}^{n}s_j\,([\tilde{\iota}\,j],\eta_j)\,\sigma_j^{\ba(d)}\right) \left.\left(\prod_{\m=1}^{t}\frac{\nu_{r_\m}^{3-p_\m}}{p_\m!}\prod_{s\neq r_\m}(b_s\,b_{r_\m})\,f^{(p_\m-2)}(x_{r_\m})\,\frac{\partial^{p_\m}}{\partial\varepsilon_{\m}^{p_\m}}\right) \mathrm{det}'(\cH)\right|_{\varepsilon=0} \\
\times\,\prod_{i=1}^{n}\frac{\d s_i}{s_i^3}\,\D\sigma_i\,\bar{\delta}^{2}(\kappa_i-s_i\,\lambda(\sigma_i))\,\e^{\im\,\varphi_i}\,.
\end{multline}
Here, the sums over $\{r_1,\ldots,r_t\}$ runs from $1,\ldots,d$; the Jacobian
\be\label{Jacobian}
J(b,\nu)=\frac{b_{0}^{d+1}\,\nu_1\cdots \nu_{d}}{|b_0\,b_{1}\cdots b_{d}|}\,,
\ee
with $|b_0\,b_{1}\cdots b_{d}|$ the Vandermonde determinant, accounts for the reparametrization of the map moduli; and the exponents
\be\label{gravVolkov}
\varphi_i:=[\tilde{o}\,i]\,\sum_{r=1}^{d}x_r\,\nu_r\,\mathfrak{s}_r(\sigma_i)+\frac{[\tilde{\iota}\,i]}{b_0}\,\sum_{r=1}^{d}\frac{\nu_{r}^2\,(i\,\xi)^{d+1}}{(i\,b_r)\,(b_r\,\xi)^{d+1}}\,\prod_{s\neq r}(b_r\,b_s)\,\cF(x_r)\,,
\ee
are the gravitational Volkov exponents in this configuration. Note that the $\mathbb{T}$ block of the matrix $\cH$ becomes purely diagonal in the SDPW case:
\be\label{sdpwTblock}
\mathbb{T}=\mathrm{diag}\left(\varepsilon_\m\,\bar{\nu}_{r_\m}\,\sum_{i=1}^{n}s_i\,\frac{[\tilde{\iota}\,i]}{(b_{r_\m}\,i)}\,\prod_{l\in\tth}\frac{(l\,i)}{(l\,b_{r_\m})}\right)\,,
\ee
since background vertex operators have no contractions with each other.

In the SDPW background the number of residual integrations is substantially reduced, thanks to the additional symmetries in play. In particular, after accounting for the GL$(2,\C)$ quotient and the fact that the set of delta functions appearing in \eqref{dsdpw} imply 3-momentum conservation in the $x^+$, $z$ and $\tilde{z}$-directions, there are $2d-1$ residual integrals in the formula which are not saturated by delta functions.


\section{Discussion}
\label{sec:discuss}

In this work, we applied twistor theory to the problem of computing graviton scattering amplitudes in curved space-times. Focusing on self-dual radiative space-times, we derived formula \eqref{MHV2} for MHV amplitudes from first principles, and conjectured N$^k$MHV formulae \eqref{NkMHV} based on universal ingredients found in twistor sigma models and twistor strings. Although conjectural, we remark that by construction, the perturbative expansion of the background field in our formulae yields the correct flat space amplitudes~\cite{Cachazo:2012kg,Cachazo:2012pz}, so the formulae pass all perturbative checks. As our main example, we worked these out explicitly on self-dual plane wave space-times in \eqref{dsdpw}. At MHV degree, we obtained improved versions \eqref{MHVsdpw} of such formulae announced previously in~\cite{Adamo:2020syc}, with a detailed comparison provided in appendix \ref{app:prl}. Further computations with other explicit examples are possible, such as the analytically continued Gibbons-Hawking metrics discussed near the ends of sections \S\ref{sec:sdrad} and \S\ref{sec:radtwistor}.

While the structure of graviton scattering in a curved SD radiative space-time has many interesting features, the structure of tail contributions to the amplitude (encoded by the background news function) seems particularly noteworthy. Analogous `tail effects' also arise in the classical limit of gravitational scattering~\cite{Bonnor:1959,Bonnor:1966,Thorne:1980ru,Blanchet:1987wq,Blanchet:1993ec}, and have been the subject of study recently in the context of black hole scattering (e.g., \cite{Foffa:2011np,Galley:2015kus,Bern:2021yeh}). It would be interesting to establish what -- if any -- relationship there is between these notions of gravitational `tails', and if our all-multiplicity results can be of any use in the context of early-inspiral black hole physics.

More generally, combined with similar formulae for gluon scattering derived in~\cite{Adamo:2020yzi}, our results provide the first steps toward developing on-shell methods for perturbative computations in the presence of strong background fields. They can be heuristically viewed as pointing to the existence of a still mysterious MHV formalism for perturbative gravity. In gauge theory, the MHV formalism is well-understood~\cite{Cachazo:2004kj} and is crucial in explaining the fact that N$^k$MHV amplitudes on SD radiative gauge field backgrounds contain no more than $k+1$ position space integrals. Naively, the number of such integrals grows with the number of interaction vertices in the Feynman diagrams contributing to an amplitude. These give rise to momentum conserving delta functions in flat space but cannot be explicitly performed in generic curved space-times. The MHV formalism rescues the situation by trivializing many of these integrals irrespective of the choice of background, a fact that is owed to intricate integration-by-parts (IBP) identities that collapse sums of Feynman diagrams when doing standard perturbation theory. 

On the contrary, even in flat space the MHV rules of perturbative GR are only known in a rudimentary form~\cite{Bjerrum-Bohr:2005xoa} that fails at high multiplicity. So it is extremely interesting that our MHV amplitudes (and our conjecture for N$^k$MHV amplitudes) can nevertheless be written with the same number of leftover position space integrals as would follow from the existence of a gravitational MHV formalism in curved space-times! To further emphasize this magic, in appendix \ref{app:grav} we provide a direct Feynman diagram computation of the 4-graviton amplitude in a SD plane wave space-time and discuss the IBP identities responsible for its extremely simple form. 

These observations provide a variety of avenues for future work. A primary goal of studying scattering amplitudes in curved space-times is to understand which mathematical properties of scattering amplitudes are subordinate to the symmetries of flat space versus which properties are more universal in perturbative QFT. Our results show that the MHV formalism could be one such universal feature, and it would be extremely insightful to return to the pursuit of MHV rules for scattering in GR as well as its supersymmetric cousins. This is currently being investigated by means of a recently discovered twistor action that uplifts general relativity to twistor space~\cite{Sharma:2021pkl}, with the expectation that perturbation theory on twistor space would take the form of an MHV diagram expansion. On a similar note, it would be useful to extend other on-shell tools of scattering amplitudes, like BCFW recursion, generalized unitarity, etc., to curved backgrounds. Self-dual backgrounds provide a perfect setting to test how far these tools stretch, as twistor theory already provides us with at least one means of computing the tree amplitudes exactly.

In more practical directions, the results of our work pave the way for efficient computations of amplitudes in more commonly encountered strong backgrounds like cosmology, AdS/CFT, strong field QED/QCD, black holes, etc. For example, it is already possible to systematically treat scattering on Gibbons-Hawking instanton space-times, and in principle our calculations could also be extended to compute amplitudes in self-dual black hole backgrounds. Moving beyond self-duality brings its own set of hurdles. Generic non-self-dual backgrounds are not hyperk\"ahler, so our twistor sigma models can no longer capture their dynamics. Nonetheless, in recent years the tools to tackle such backgrounds have been provided by more general string theories \cite{Mason:2013sva} in \emph{ambitwistor space}: the space of null geodesics in space-time. At tree-level, these ambitwistor strings can be consistently coupled to general on-shell backgrounds in supergravity theories~\cite{Adamo:2014wea,Adamo:2018ege}. Worldsheet correlators of these models then compute amplitudes around such backgrounds. The feasibility of using ambitwistor strings to compute amplitudes has been demonstrated on plane wave backgrounds at three points~\cite{Adamo:2017sze}, and in (A)dS backgrounds at general multiplicity for certain theories~\cite{Roehrig:2020kck,Eberhardt:2020ewh,Gomez:2021qfd,Gomez:2021ujt}. It would be interesting to pursue this approach to build scattering equations and worldsheet formulae for amplitudes on other space-times of interest.

\acknowledgments

We would like to thank Harry Braden for interesting conversations about graph theory and Maciej Dunajski for many helpful discussions and comments. TA is supported by a Royal Society University Research Fellowship and by the Leverhulme Trust (RPG-2020-386). LJM  is grateful to the STFC for support under grant ST/T000864/1. AS is supported by a Mathematical Institute Studentship, Oxford and by the ERC grant GALOP ID: 724638.


\begin{appendix}

\section{Equivalence with past formulae}
\label{app:prl}

In~\cite{Adamo:2020syc}, we presented a formula for MHV graviton scattering on self-dual plane wave (SDPW) backgrounds which also passed all consistency tests but was not manifestly gauge invariant. We now prove its equivalence to the new formula \eqref{MHVsdpw}.

For any invertible $n\times n$ matrix $A = (A_{i,j}(\veps))$ whose entries depend on a parameter $\veps$, the $\veps$-derivative of its determinant is given by
\be\label{detder}
\p_\veps|A| = \sum_{i=1}^n\begin{vmatrix}
A_{1,\bullet}\\
\vdots\\
\p_\veps A_{i,\bullet}\\
\vdots\\
A_{n,\bullet}
\end{vmatrix}\,,
\ee
where one differentiates only the $i^\text{th}$ row in the $i^\text{th}$ term of the sum.  This formula is an easy consequence of the linearity of the determinant in its rows. Let's apply this to compute the derivatives
\be
\left.\left(\prod_{\m=1}^{t}\frac{1}{p_\m!}\,\frac{\partial^{p_\m}}{\partial\varepsilon_\m^{p_\m}}\right)|\cH^{i}_{i}|\right|_{\varepsilon=0} 
\ee
present in \eqref{MHVsdpw}.

The last $t$ rows of $\cH^i_i$ need to be differentiated first. The $(n-3+\m)^\text{th}$ row reads
\be
\begin{split}
&\begin{pmatrix}\mathfrak{h}_{\bullet\,\m}&&\mathbb{T}_{\m\,\bullet}\end{pmatrix}= \veps_\m\,\begin{pmatrix}\overline{\mathfrak{h}}_{\bullet\,\m}&&\overline{\mathbb{T}}_{\m\,\bullet}\end{pmatrix}\,,\\
&\text{where}\quad\overline{\mathfrak{h}}_{j\m} = -\frac{[j\,\tilde\iota]}{\la j\,\iota\ra}\,,\quad\overline{\mathbb{T}}_{\m\mathrm{n}} = \delta_{\m\mathrm{n}}\,\sum_{j=1}^{n}\frac{[\tilde{\iota}\,j]}{\la\iota\,j\ra}\,\frac{\la1\,j\ra\,\la2\,j\ra}{\la1\,\iota\ra\,\la2\,\iota\ra}\,,
\end{split}
\ee
with the rescaled matrices $\overline{\mathfrak{h}}$ and $\overline{\T}$ now being independent of all $\veps_\m$'s. Using the transformation of determinants under scalings of rows, we then find
\be
|\cH^i_i| = \left(\prod_{\m=1}^t\veps_\m\right)|\overline{\cH}^i_i|\,,\qquad\overline{\cH}=\left(\begin{array}{c c}
               \HH & \mathfrak{h} \\
               \overline{\mathfrak{h}}^{\mathrm{T}} & \overline{\T}
               \end{array}\right)\,.
\ee
It follows that
\be\label{hugeder}
\left.\left(\prod_{\m=1}^{t}\frac{1}{p_\m!}\,\frac{\partial^{p_\m}}{\partial\varepsilon_\m^{p_\m}}\right)|\cH^{i}_{i}|\right|_{\varepsilon=0} = \left.\left(\prod_{\m=1}^{t}\frac{1}{(p_\m-1)!}\,\frac{\partial^{p_\m-1}}{\partial\varepsilon_\m^{p_\m-1}}\right)|\overline{\cH}^{i}_{i}|\right|_{\varepsilon=0}\,.
\ee
The remaining derivatives will only act on the first $n-3$ rows of $\overline{\cH}^i_i$.

At this stage, we use the multinomial expansion of products of derivatives. The result is a sum over $t$ disjoint multi-indices $\sa_\m\subset\{3,\dots,n\}\,\backslash\,\{i\}$ of cardinalities $|\sa_\m| = p_\m-1\geq2$. Basically, the rows of $\overline{\cH}^i_i$ labeled by elements of $\sa_\m$ will be acted on by $(\p/\p\veps_\m)^{p_\m-1}$. To simplify life, let $\sa = \sa_1\cup\cdots\cup\sa_t$ (with the obvious ordering) and $\bar\sa = \{3,\dots,n\}\,\backslash\,(\sa\cup\{i\})$. By using the transformations of determinants under exchanging rows and columns, we first reorder $\overline{\cH}^i_i$ into
\be
|\overline{\cH}^i_i| =  \begin{vmatrix}
\HH_{\bar\sa\bar\sa}&&\HH_{\bar\sa\sa}&&\mathfrak{h}_{\bar\sa}\\
\HH_{\sa\bar\sa}&&\HH_{\sa\sa}&&\mathfrak{h}_\sa\\
\overline{\mathfrak{h}}_{\bar\sa}^T&&\overline{\mathfrak{h}}_\sa^T&&\overline{\T}
\end{vmatrix}\,.
\ee
Here, for two multi-indices $\sa,\msf{b}$, $\HH_{\msf{ab}}$ is the $|\sa|\times|\msf{b}|$ sub-matrix of $\HH$ with entries $(\HH_{\sa\msf{b}})_{jk} = \HH_{jk}$ for all $j\in\sa,k\in\msf{b}$. Similarly, $\mathfrak{h}_\sa$ is the $|\sa|\times t$ sub-matrix of $\mathfrak{h}$ with entries $(\mathfrak{h}_\sa)_{j\m} = \mathfrak{h}_{j\m}$ for $j\in\sa$, etc. 

Now note that $\mathfrak{h}_{\bar\sa} = 0$ if we set all $\veps_\m = 0$. Next, we compute the action of $(\p/\p\veps_\m)^{p_\m-1}$ on the rows of the sub-matrix $\begin{pmatrix}\HH_{\sa_\m\bar\sa}&&\HH_{\sa_\m\sa}&&\mathfrak{h}_{\sa_\m}\end{pmatrix}$. Any such row is only linearly dependent on $\veps_\m$. So each row receives one derivative in the determinant, with the derivatives getting homogeneously distributed in $(p_\m-1)!$ ways. Recalling \eqref{pwHHmom}, acting with $\p/\p\veps_\m$ on the row $j\in\sa_\m$ gives
\be
\frac{\p}{\p\veps_\m}\begin{pmatrix}(\HH_{\sa_\m\bar\sa})_{j\,\bullet}&&(\HH_{\sa_\m\sa})_{j\,\bullet}&&(\mathfrak{h}_{\sa_\m})_{j\,\bullet}\end{pmatrix} = \begin{pmatrix}0&&(\mathbb{D}_{\sa_\m\sa})_{j\,\bullet}&&(\tau_{\sa_\m})_{j\,\bullet}\end{pmatrix}\,,
\ee
with the new matrices
\be
(\mathbb{D}_{\sa_\m\sa})_{jk} =  \delta_{jk}\,\frac{[j\,\tilde{\iota}]}{\la j\,\iota\ra}\,\frac{\la1\,\iota\ra\,\la2\,\iota\ra}{\la1\,j\ra\,\la2\,j\ra}\,,\quad(\tau_{\sa_\m})_{j\mathrm{n}} = -\delta_{\m\mathrm{n}}\,\frac{[j\,\tilde\iota]}{\la j\,\iota\ra}\,,\quad j\in\sa_m\,,\,k\in\sa\,.
\ee
It is convenient to arrange these together as
\be
\mathbb{D} = \begin{pmatrix}
\mathbb{D}_{\sa_1\sa}\\
\vdots\\
\mathbb{D}_{\sa_t\sa}
\end{pmatrix}\,,\qquad
\tau = \begin{pmatrix}
\tau_{\sa_1}\\
\vdots\\
\tau_{\sa_t}
\end{pmatrix}\,.
\ee
With these definitions, the term in the expansion of \eqref{hugeder} in which the $\veps_\m$-derivatives act on the rows labeled by $\sa_\m$ is given by the determinant
\be
 \begin{vmatrix}
\mathbb{G}&&\mathfrak{g}&&0\\
0&&\mathbb{D}&&\tau\\
\overline{\mathfrak{h}}_{\bar\sa}^T&&\overline{\mathfrak{h}}_\sa^T&&\overline{\T}
\end{vmatrix}\,,
\ee
having set all $\veps_\m=0$ and defined $\mathbb{G} = \HH_{\bar\sa\bar\sa}|_{\veps=0}$, $\mathfrak{g} =  \HH_{\bar\sa\sa}|_{\veps=0}$.

We can simplify this further by adding multiples of rows of $(0\;\;\mathbb{D}\;\;\tau)$ to the rows of $(\overline{\mathfrak{h}}_{\bar\sa}^T\;\;\overline{\mathfrak{h}}_\sa^T\;\;\overline{\T})$. Using this fact, we can replace $(\overline{\mathfrak{h}}_{\bar\sa}^T\;\;\overline{\mathfrak{h}}_\sa^T\;\;\overline{\T})$ with $(\mathfrak{c}\;\;\mathfrak{t}\;\;\mathbb{S})$ expressed in terms of sub-matrices with entries
\begin{align}
\mathfrak{c}_{\m j} &= \overline{\mathfrak{h}}_{j\m}\,,\qquad j\in\bar\sa\,,\\
\mathfrak{t}_{\m j} &= \overline{\mathfrak{h}}_{j\m} + \sum_{k\in\sa_\m}\mathbb{D}_{kj}\,\frac{\la1\,k\ra\,\la2\,k\ra}{\la1\,\iota\ra\,\la2\,\iota\ra}\,\,,\qquad j\in\sa\,,\\
\mathbb{S}_{\m\mathrm{n}} &= \overline{\mathbb{T}}_{\m\mathrm{n}} +  \sum_{k\in\sa_\m}\tau_{k\mathrm{n}}\,\frac{\la1\,k\ra\,\la2\,k\ra}{\la1\,\iota\ra\,\la2\,\iota\ra}\,.
\end{align}
Explicitly, these modified combinations of rows are found to be
\be
\mathfrak{t}_{\m j} = \begin{cases}
-\frac{[j\,\tilde\iota]}{\la j\,\iota\ra}\quad &j\not\in\sa_\m\\
\;\;\;0\quad &j\in\sa_\m
\end{cases}\,,
\qquad \mathbb{S}_{\m\mathrm{n}} = \delta_{\m\mathrm{n}}\,\sum_{j\not\in\sa_\m}\frac{[\tilde{\iota}\,j]}{\la\iota\,j\ra}\,\frac{\la1\,j\ra\,\la2\,j\ra}{\la1\,\iota\ra\,\la2\,\iota\ra}\,.
\ee
Hence, defining the reduced matrices
\be
\cH^i_i[\sa_1,\dots,\sa_t] = \begin{pmatrix}
\mathbb{G}&&\mathfrak{g}&&0\\
0&&\mathbb{D}&&\tau\\
\mathfrak{c}&&\mathfrak{t}&&\mathbb{S}
\end{pmatrix}
\ee
and summing over all $(p_1-1)!\cdots(p_t-1)!$ permutations of distributing the derivatives, we finally reduce \eqref{hugeder} to
\be\label{hugeder1}
\left.\left(\prod_{\m=1}^{t}\frac{1}{p_\m!}\,\frac{\partial^{p_\m}}{\partial\varepsilon_\m^{p_\m}}\right)|\cH^{i}_{i}|\right|_{\varepsilon=0}  = \sum_{\sa_1,\dots,\sa_t}|\cH^i_i[\sa_1,\dots,\sa_t]|\,.
\ee
For any given $i$, the right hand side of this is exactly the kind of sums that occur in the old formula.

To prove the equivalence, recall the old formula:
\begin{multline}\label{oldformula}
-\kappa^{n-2}\,\delta^{3}_{+,\perp}\!\left(\sum_{j=1}^{n}k_j\right)\sum_{t=0}^{\lfloor\frac{n-3}{2}\rfloor}\sum_{p_1,\dots,p_t}\sum_{i=3}^n\frac{\la1\,2\ra^5\,[i\,\tilde\iota]}{\la1\,i\ra\,\la2\,i\ra^2\,[2\,\tilde\iota]}\\
\times\int_{-\infty}^{+\infty}\d x^{-} \sum_{\substack{\sa_1,\dots,\sa_t\\|\sa_\m|=p_\m}}|\cH^i_i[\sa_1,\dots,\sa_t]|\;\,\e^{\im\,F_n(x^-)}\,\prod_{\m=1}^{t}f^{(p_\m-2)}(x^-)\,.
\end{multline}
In this expression, the term with $t$ tails of valences $p_1,\dots,p_t$ contains the factor
\begin{multline}\label{eqstep}
\sum_{i=3}^n\frac{\la1\,2\ra^5\,[i\,\tilde\iota]}{\la1\,i\ra\,\la2\,i\ra^2\,[2\,\tilde\iota]}\sum_{\sa_1,\dots,\sa_t}|\cH^i_i[\sa_1,\dots,\sa_t]|\\
= \left.\left(\prod_{\m=1}^{t}\frac{1}{p_\m!}\,\frac{\partial^{p_\m}}{\partial\varepsilon_\m^{p_\m}}\right)\sum_{i=3}^n\frac{\la1\,2\ra^5\,[i\,\tilde\iota]}{\la1\,i\ra\,\la2\,i\ra^2\,[2\,\tilde\iota]}\,|\cH^{i}_{i}|\right|_{\varepsilon=0}\,,
\end{multline}
having applied \eqref{hugeder1}. Next, using the property
\be
\sum_{j=3}^n\cH_{ij}\,\la 1\,j\ra\,\la2\,j\ra + \sum_{\m=1}^t\cH_{i\m}\,\la1\,\iota\ra\,\la2\,\iota\ra = 0\,,
\ee
one can show that
\be
\frac{|\cH^{i}_{i}|}{\la1\,i\ra^2\,\la2\,i\ra^2} = \frac{|\cH^{j}_{j}|}{\la1\,j\ra^2\,\la2\,j\ra^2} 
\ee
for any $i,j\in\{3,\dots,n\}$. So, fixing a row $r\in\{3,\dots,n\}$, the factor of $|\cH^i_i|$ can be pulled out of the sum in \eqref{eqstep} to get
\be\label{eqstep1}
\begin{split}
\sum_{i=3}^n\frac{\la1\,2\ra^5\,[i\,\tilde\iota]}{\la1\,i\ra\,\la2\,i\ra^2\,[2\,\tilde\iota]}&\sum_{\sa_1,\dots,\sa_t}|\cH^i_i[\sa_1,\dots,\sa_t]|\\
 &= \left.\left(\prod_{\m=1}^{t}\frac{1}{p_\m!}\,\frac{\partial^{p_\m}}{\partial\varepsilon_\m^{p_\m}}\right)\frac{\la1\,2\ra^5}{\la1\,r\ra^2\,\la2\,r\ra^2\,[2\,\tilde\iota]}\,|\cH^{r}_{r}|\right|_{\varepsilon=0}\sum_{i=3}^n\la1\,i\ra\,[i\,\tilde\iota]\\
&= -\left.\left(\prod_{\m=1}^{t}\frac{1}{p_\m!}\,\frac{\partial^{p_\m}}{\partial\varepsilon_\m^{p_\m}}\right)\frac{\la1\,2\ra^6}{\la1\,r\ra^2\,\la2\,r\ra^2}\,|\cH^{r}_{r}|\right|_{\varepsilon=0}\,.
\end{split}
\ee
In the final step, we have used the residual momentum conserving delta functions present in \eqref{oldformula}. Substituting \eqref{eqstep1} back in \eqref{oldformula} proves that the latter is equal to \eqref{MHVsdpw}.


\section{$3$ and $4$-point amplitudes on SD plane waves}
\label{app:grav}

Direct calculation of the 3- and 4-graviton amplitudes from the Einstein-Hilbert action is, of course, impractically complicated. In practice, it is easier to work with the perturbative Lagrangian of~\cite{Cheung:2016say}, which utilizes field redefinitions that do not change the S-matrix to obtain a more tractable perturbative expansion. The details of interaction vertices on a SDPW background metric can be obtained directly from this background field Lagrangian. In Feynman gauge, the graviton propagator $\scG_{abcd}(x,y)$ solves 
\be\label{greengraviton}
\left(g_{ac}\,g_{bd}\,\D^2 - 2\,R_{acdb}\right)\scG^{cd}{}_{ef}(x,y) = g_{a(e}\,g_{f)b}\,\delta^4(x-y)\,,
\ee
where $g_{ab}$ is the SDPW metric \eqref{sdpwgr} with Riemann curvature tensor $R_{abcd}$. The corresponding Feynman propagator is given by
\be\label{hprop}
\scG_{abcd}(x,y) = \frac{\cN}{2\pi\im}\int\frac{\d^4k}{k^2+\im\,\varepsilon}\,\Delta_{abcd}(x^-,y^-)\,\e^{\im(\phi(x)-\phi(y))}\,,
\ee
with $\phi(x)$ given by \eqref{phigr} extended off-shell to a solution of the massive Klein-Gordon equation and $\cN$ a normalization factor. The tensor structure of the propagator (written in light-front variables) is:
\be\label{Deltasdpw}
\Delta_{abcd}(x^-,y^-):= D_{a(c|}\,D_{b|d)} - \frac{\im\,\Delta\dot f}{k_+}\,E_{a(c|}\,E_{b|d)}\,,
\ee
where
\be\label{spin1prop}
D_{ab}(x^-,y^-) = \frac{g_{ab}(x^-)+g_{ab}(y^-)}{2} + \frac{k}{k_+}\Delta f\,E_{ab}\,
\ee
for $\Delta f:= f(x^-)-f(y^-)$, $\Delta\dot f:= \dot f(x^-)-\dot f(y^-)$ and $E_{ab}:=\epsilon_{\alpha\beta}\,\tilde{\iota}_{\dot\alpha}\tilde{\iota}_{\dot\beta}$. The final term in \eqref{Deltasdpw} is the origin of gravitational wave tails created dynamically in the scattering process.

The tree-level 3-graviton amplitude on a general plane wave background was computed in~\cite{Adamo:2017nia}, where it was cast in a `double copy' format in terms of spin-1 polarization data. On a SDPW background, the amplitude is
\be\label{M3}
\begin{split}
\cM_3 = \frac{\im\,\kappa}{2}\,\delta_{+,\perp}^3\!\left(\sum_{i=1}^3 k_i\right)\int\d x^-\,&\e^{\im F_3}\biggl[(\cE_3\cdot\cE_1\,K_1\cdot\cE_2+\text{cyclic})^2\\
& - \im\,k_{1+}k_{2+}k_{3+}\dot f(x^-)\left(\cE_1\cdot\cE_2\,\frac{\eps_3}{k_{3+}}+\text{cyclic}\right)^2\biggr]\,.
\end{split}
\ee
where $F_n(x^-)$ is the gravitational Volkov exponent \eqref{grVolkov} and the $\cE_{i\,a}$ denote lightfront gauge spin-1 polarisations on an SDPW space-time:
\be\label{spin1polonM}
\cE_{i\,\alpha\dot\alpha}^{(+)}(x^-) = \frac{\iota_\alpha\,\tilde K_{i\,\dot\alpha}(x^-)}{\la\iota\,i\ra}\,,\qquad\cE_{i\,\alpha\dot\alpha}^{(-)} = \frac{\kappa_{i\,\alpha}\,\tilde\iota_{\dot\alpha}}{[\tilde\iota\,i]}\,.
\ee
The terms in the second line, proportional to $\dot f$, are the tail contributions at 3 points. The parameters $\eps_i$, $i=1,2,3$, are $1$ if the $i^\text{th}$-graviton is positive helicity and $0$ if it is negative helicity.

It is straightforward to show that $\cM_3$ vanishes in the all-positive and all-negative helicity configurations; in the MHV configuration with gravitons 1 and 2 of negative helicity, all tail terms are zero since $\cE_1\cdot\cE_2=\eps_1=\eps_2=0$. The remaining contributions to the MHV 3-point amplitude can be simplified using 3-momentum conservation and the lightfront gauge polarisations, leaving
\be\label{M3mhv}
\cM_3^\text{MHV} =  \frac{\im\,\kappa}{2}\,\delta_{+,\perp}^3\!\left(\sum_{i=1}^3 k_i\right)\frac{\la1\,2\ra^6}{\la2\,3\ra^2\,\la3\,1\ra^2}\,\int\d x^-\,\e^{\im F_3}\,.
\ee
This matches our general MHV formula \eqref{MHVsdpw} for $n=3$. This structure -- of the flat space 3-graviton MHV amplitude multiplying the leftover lightfront integral -- is special to the 3-point MHV configuration on the chiral background.

Indeed, we can also use \eqref{M3} to find the 3-point $\overline{\text{MHV}}$ amplitude with gravitons 1 and 2 positive helicity and particle 3 negative helicity:
\be\label{M3mhvbar}
\cM_3^{\overline{\text{MHV}}} =  \frac{\im\,\kappa}{2}\,\delta_{+,\perp}^3\!\left(\sum_{i=1}^3 k_i\right)\int\d x^-\,\e^{\im F_3}\biggl(\frac{[\![1\,2]\!]^6}{[\![2\,3]\!]^2\,[\![3\,1]\!]^2} - \im\,\dot f\,\frac{[\tilde\iota\,1]\,[\tilde\iota\,2]\,[\tilde\iota\,3]\,\la\iota\,3\ra^5}{\la\iota\,1\ra^3\,\la\iota\,2\ra^3}\biggr)(x^-)\,.
\ee
Note that $\cM_3^{\overline{\text{MHV}}}$ is not the parity conjugate of $\cM_{3}^{\text{MHV}}$ due to the chirality of the background, as was also observed in the gauge theory setting~\cite{Adamo:2020yzi}. Furthermore, $\cM_3^{\overline{\text{MHV}}}$ contains a tail term which is absent in the MHV configuration.

\medskip

The 4-graviton amplitude on a plane wave background has not been computed before in the literature; here we perform the computation with Feynman diagrams for a SDPW space-time. Let gravitons 1 and 2 to be negative helicity; for positive helicity gravitons 3 and 4 the reference spinors in the wavefunctions \eqref{hplus} can be set to $\xi_\alpha = \kappa_{1\,\alpha}$:
\be\label{polchoice4ptgr}
\cE_{j\,\alpha\dot\alpha\beta\dot\beta} = \frac{\kappa_{1\,\alpha}\,\kappa_{1\,\beta}}{\la1\,j\ra^2}\left(\tilde K_{j\,\dot\alpha}\,\tilde K_{j\,\dot\beta} - \im\,\dot f\,\frac{[\tilde\iota\,j]}{\la\iota\,j\ra}\,\tilde\iota_{\dot\alpha}\,\tilde\iota_{\dot\beta}\right)\,,\quad j=3,4\,.
\ee
The 4-point amplitude again recieves contributions from four Feynman diagrams,
\be\label{4ptdecomgr}
\cM_4 = \cM_\msf{s} + \cM_\msf{t} + \cM_\msf{u} + \cM_\text{cont}\,,
\ee
and with our gauge choices the contact contribution $\cM_{\text{cont}}$ vanishes.

In addition to the propagator \eqref{hprop}, the exchange contributions require the 3-point graviton vertices. For instance, we find
\be
\cM_\msf{s} = \int\d^4x\,\sqrt{|g(x)|}\,\d^4y\,\sqrt{|g(y)|}\,V_{12}^{ab}(x)\,\scG_{abcd}(x,y)\,V_{34}^{cd}(y)\,,
\ee
where, in transverse gauge $\D^a h_{i\,ab} = 0$ for the graviton perturbations $h_{i\,ab}$, the 3-point interaction reads 
\begin{multline}\label{S3varder}
V^{ab}_{ij} = \frac{\kappa}{4}\cdot\frac{1}{2}\left(\D^a h_{i\,cd}\,\D^b h_j^{cd} - 2\,\D_c h^a_{i\,d}\,\D^b h_j^{cd}+2\,\D_c h_i^{ad}\,\D_d h_j^{bc}\right.\\
\left.- 2\,h_i^{cd}\,\D_c\D_d h_j^{ab} + 2\,h_i^{ad}\,\D_c\D_d h_j^{bc} + 2\,h_i^{cd}\,\D_d\D^a h^b_{j\,c} + (i\leftrightarrow j)\right)\,.
\end{multline}
Here $\D_{a}$ is the Levi-Civita connection on the SDPW background. With these ingredients, the $\msf{s}$-channel contribution to \eqref{4ptdecomgr} is
\be\label{smhvgr}
\begin{split}
\cM_\msf{s} = \biggl(\frac{\kappa}{4}\biggr)^2\,&\delta_{+,\perp}^3\!\left(\sum_{i=1}^4k_i\right)\int\d^2\mu[\msf{s}]\,\frac{\la1\,2\ra^6}{\la1\,3\ra^2\la1\,4\ra^2}\,\frac{[\tilde\iota\,2]^2}{[\tilde\iota\,1]^2}\\
&\times\left([\![3\,4]\!]^2 - \im\,(k_3+k_4)_+\,\frac{[\tilde\iota\,3]}{\la\iota\,3\ra}\frac{[\tilde\iota\,4]}{\la\iota\,4\ra}\,\dot f\right)\!(y^-)+(x^-\leftrightarrow y^-)\,,
\end{split}
\ee
where the measure is defined by
\be\label{d2musgr}
\begin{split}
&\d^2\mu[\msf{s}] = \Theta(x^--y^-)\,\frac{\d x^-\;\d y^-}{(k_1+k_2)_+}\,\exp\!\left[\im\sum_{i=1,2}\int^{x^-}K_{i\,-}(s)\,\d s\right.\\
 &\left.+ \im\sum_{j=3,4}\int^{y^-}K_{j\,-}(s)\,\d s - \frac{\im\,(k_1+k_2)}{\,(k_1+k_2)_+}\int_{y^-}^{x^-}\left(\tilde k_1+\tilde k_2 + (k_1+k_2)\,f(s)\right)\d s\right]\,.
\end{split}
\ee
This can be simplified using an IBP relation. First, observe that
\be\label{KKder}
\D^2\!\left(K_3\cdot K_4\,\e^{\im\phi_3+\im\phi_4}\right) = -2\,\la3\,4\ra^2\left([\![3\,4]\!]^2 - \im\,(k_3+k_4)_+\frac{[\tilde\iota\,3]}{\la\iota\,3\ra}\frac{[\tilde\iota\,4]}{\la\iota\,4\ra}\dot f\right)\e^{\im\phi_3+\im\phi_4}\,,
\ee
and 
\be\label{newidstep}
\begin{split}
\frac{1}{\cN}\int&\d^4 x\;K_3\cdot K_4(y^-)\,\exp\!\left[\im\sum_{i=1}^4\phi_i(x)\right]\\
&= \frac{1}{\cN}\int\d^4 x\,\d^4y\;\delta^4(x-y)\,K_3\cdot K_4(x^-)\,\exp\!\left[\im\sum_{i=1,2}\phi_i(x) + \im\sum_{j=3,4}\phi_j(y)\right]\,.
\end{split}
\ee
Now, the propagator for a massless scalar in the SDPW background is
\be\label{scprop}
\scG(x,y) =\frac{\cN}{2\pi\im}\int\frac{\d^4k}{k^2+\im\,\varepsilon}\,\e^{\im(\phi(x)-\phi(y))}\,,
\ee
which obeys:
\be\label{greenscalar}
\D^2\scG(x,y) = \delta^4(x-y) = \D'^2\scG(x,y)\,,
\ee
where $\D'_a$ is the covariant derivative with respect to $y$. Using this in \eqref{newidstep}, the right-hand side becomes 
\begin{equation*}
\frac{1}{\cN}\int\d^4 x\,\d^4y\;\D'^2\scG(x,y)\,K_3\cdot K_4(y^-)\,\exp\!\left[\im\sum_{i=1,2}\phi_i(x) + \im\sum_{j=3,4}\phi_j(y)\right]\,.
\end{equation*}
Integration by parts and an application of \eqref{KKder} then produces the identity,
\be\label{newid1}
\begin{split}
\frac{1}{\cN}\int&\d^4 x\;K_3\cdot K_4(y^-)\,\exp\!\left[\im\sum_{i=1}^4\phi_i(x)\right]\\
&=-\int\d^2\mu[\msf{s}]\,\la3\,4\ra^2\left([\![3\,4]\!]^2 - \im\,(k_3+k_4)_+\frac{[\tilde\iota\,3]}{\la\iota\,3\ra}\frac{[\tilde\iota\,4]}{\la\iota\,4\ra}\dot f\right)\!(y^-)+(x^-\leftrightarrow y^-)\,,
\end{split}
\ee
which allows $\cM_{\msf{s}}$ to be simplified to
\be\label{smhvgr1}
\cM_\msf{s} = -\frac{1}{\cN}\left(\frac{\kappa}{4}\right)^2\delta_{+,\perp}^3\!\left(\sum_{i=1}^4k_i\right)\int\d x^-\,\e^{\im F_4}\,\frac{\la1\,2\ra^6\,[\![3\,4]\!](x^-)}{\la1\,3\ra^2\,\la1\,4\ra^2\,\la3\,4\ra}\,\frac{[\tilde\iota\,2]^2}{[\tilde\iota\,1]^2}\,,
\ee
containing only a single lightfront integral.

Similar, albeit slightly more involved, simplifications occur in the other channels. The $\msf{t}$-channel computation from Feynman diagrams gives a complicated result
\be\label{tmhvgr}
\begin{split}
&\cM_\msf{t} = \left(\frac{\kappa}{4}\right)^2\delta_{+,\perp}^3\!\left(\sum_{i=1}^4k_i\right)\frac{\la1\,2\ra^3}{\la4\,1\ra}\frac{[\tilde\iota\,3]^2}{(k_1+k_3)^2_+} \\
&\times \int\d^2\mu[\msf{t}]\left[\frac{\la\iota\,1\ra^2}{[\tilde\iota\,1]^2}\frac{[\tilde\iota\,4]^3}{[\tilde\iota\,2]}\left([\![1\,3]\!]^2-\im\,(k_1+k_3)_+\frac{[\tilde\iota\,1]}{\la\iota\,1\ra}\frac{[\tilde\iota\,3]}{\la\iota\,3\ra}\,\dot f\right)\!(x^-)\right.\\
&+\frac{\la\iota\,2\ra}{[\tilde\iota\,2]}\frac{[\tilde\iota\,3]^2\{\la\iota\,1\ra[\tilde\iota\,4]\la4\,2\ra+\la\iota\,2\ra[\tilde\iota\,2]\la1\,2\ra\}}{\la4\,1\ra[\tilde\iota\,1]^2}\left([\![2\,4]\!]^2 - \im\,(k_2+k_4)_+\frac{[\tilde\iota\,2]}{\la\iota\,2\ra}\frac{[\tilde\iota\,4]}{\la\iota\,4\ra}\,\dot f\right)\!(y^-)\\
&\left.+\frac{\la\iota\,1\ra[\tilde\iota\,3][\tilde\iota\,4]}{\la4\,1\ra[\tilde\iota\,1]^2[\tilde\iota\,2]}\left(\la\iota\,1\ra[\tilde\iota\,4]\la4\,2\ra+\la\iota\,2\ra[\tilde\iota\,2]\la1\,2\ra + \la\iota\,2\ra[\tilde\iota\,4]\la4\,1\ra\right)[\![1\,3]\!](x^-)[\![2\,4]\!](y^-)\right]\\
&+(x^-\leftrightarrow y^-).
\end{split}
\ee
where the measure $\d^2\mu[\msf{t}]$ can be obtained from $\d^2\mu[\msf{s}]$ by exchanging gravitons $2\leftrightarrow 3$ as usual. The first two terms in the integrand can be simplified via the $\msf{t}$-channel analogues of \eqref{newid1}, while the last term requires a new IBP identity. Starting with
\be\label{newidstep2}
\begin{split}
\frac{1}{\cN}\int&\d^4 x\;K_1\cdot K_3(x^-)\,\exp\!\left[\im\sum_{i=1}^4\phi_i(x)\right]\\
&= \frac{1}{\cN}\int\d^4 x\,\d^4y\;\delta^4(x-y)\,K_1\cdot K_3(x^-)\,\exp\!\left[\im\sum_{i=1,3}\phi_i(x) + \im\sum_{j=2,4}\phi_j(y)\right],
\end{split}
\ee
insert $\delta^4(x-y) = \D'^2\scG^F(x-y)$. The covariant derivatives with respect to $y$ do not hit $K_1\cdot K_3(x^-)$, so on integration by parts one obtains the identities:
\be\label{newid2}
\begin{split}
\frac{1}{\cN}\int&\d^4 x\;K_1\cdot K_3(x^-)\,\exp\!\left[\im\sum_{i=1}^4\phi_i(x)\right] = \frac{1}{\cN}\int\d^4 x\;K_2\cdot K_4(y^-)\,\exp\!\left[\im\sum_{i=1}^4\phi_i(x)\right]\\
&=-\int\d^2\mu[\msf{t}]\,K_1\cdot K_3(x^-)\,K_2\cdot K_4(y^-)+(x^-\leftrightarrow y^-)\,.
\end{split}
\ee
These identities, combined with 3-momentum conservation and applications of the Schouten identity bring $\cM_{\msf{t}}$ into a substantially simpler form:
\be\label{tmhvgr2}
\cM_\msf{t} = -\frac{1}{\cN}\left(\frac{\kappa}{4}\right)^2\delta_{+,\perp}^3\!\left(\sum_{i=1}^4k_i\right)\int\d x^-\,\e^{\im F_4}\,\frac{\la1\,2\ra^5[\![3\,4]\!](x^-)}{\la1\,3\ra\la1\,4\ra^2\la4\,2\ra}\,\frac{[\tilde\iota\,3]^2}{[\tilde\iota\,1]^2}\,.
\ee
The $\msf{u}$-channel contribution
\be\label{umhvgr}
\cM_\msf{u} = -\frac{1}{\cN}\left(\frac{\kappa}{4}\right)^2\delta_{+,\perp}^3\!\left(\sum_{i=1}^4k_i\right)\int\d x^-\,\e^{\im F_4}\,\frac{\la1\,2\ra^5[\![3\,4]\!](x^-)}{\la1\,3\ra^2\la1\,4\ra\la2\,3\ra}\,\frac{[\tilde\iota\,4]^2}{[\tilde\iota\,1]^2}\,,
\ee
is obtained by simply exchanging particles 3 and 4.

Adding all the contributions, the full MHV 4-graviton amplitude on the SDPW background is given by
\be\label{fullmhvgr}
\cM_4^\text{MHV} = -\frac{1}{\cN}\left(\frac{\kappa}{4}\right)^2\delta_{+,\perp}^3\!\left(\sum_{i=1}^4k_i\right)\int\d x^-\,\e^{\im F_4}\,\frac{\la1\,2\ra^7\,[\![3\,4]\!](x^-)}{\la1\,2\ra\,\la1\,3\ra\,\la1\,4\ra\,\la2\,3\ra\,\la2\,4\ra\,\la3\,4\ra}\,.
\ee
This matches the all-multiplicity MHV formula \eqref{MHVsdpw} at $n=4$; note that at 4-points the integrand of the amplitude retains the standard structure (now background dressed) of the 4-graviton amplitude in flat space (cf., \cite{Berends:1988zp}). This confirms -- by direct computation -- the prediction implicit in our general formula that tails appear in the MHV amplitude on a SDPW space-time only from 5-points. To see these tails directly from Feynman diagrams would therefore necessitate the (very complicated) computation of a 5-graviton amplitude in background perturbation theory.

\end{appendix}

\bibliography{sdpw1}
\bibliographystyle{JHEP}

\end{document}